\documentclass[aps,prb,twocolumn,superscriptaddress,longbibliography]{revtex4-2}
\usepackage{graphicx}
\usepackage{dcolumn}   
\usepackage{bm}        
\usepackage{amssymb}   
\usepackage{amsmath}
\usepackage{url}
\usepackage{layout}
\usepackage{color}
\usepackage{epsfig}
\usepackage{graphicx}
\usepackage{mathrsfs}\usepackage{bm}\usepackage{appendix}\usepackage{color}

\usepackage{cancel}

\usepackage[thinlines]{easytable}
\usepackage{makecell}
\usepackage{tikz,pgf}
\usepackage{booktabs}
\usepackage{float}
\usepackage{subfigure}
\usepackage{hyperref}
\hypersetup{colorlinks,allcolors=blue}

\usepackage{cancel}

\begin{document}

\title{Strong-coupling study of the pairing mechanism in pressurized La$_3$Ni$_2$O$_7$}
\author{Jia-Heng Ji}
\thanks{These four authors contributed equally to this work.}
\affiliation{School of Physics, Beijing Institute of Technology, Beijing 100081, China}
\author{Chen Lu}
\thanks{These four authors contributed equally to this work.}
\affiliation{School of Physics and Hangzhou Key Laboratory of Quantum Matter, Hangzhou Normal University, Hangzhou 311121, China}
\author{Zhi-Yan Shao}
\thanks{These four authors contributed equally to this work.}
\affiliation{School of Physics, Beijing Institute of Technology, Beijing 100081, China}
\author{Zhiming Pan}
\thanks{These four authors contributed equally to this work.}
\affiliation{Department of Physics, Xiamen University, Xiamen 361005, Fujian, China}
\author{Fan Yang}
\email{yangfan\_blg@bit.edu.cn}
\affiliation{School of Physics, Beijing Institute of Technology, Beijing 100081, China}
\author{Congjun Wu}
\email{wucongjun@westlake.edu.cn}
\affiliation{New Cornerstone Science Laboratory, Department of Physics, School of Science, Westlake University, Hangzhou 310024, Zhejiang, China}
\affiliation{Institute for Theoretical Sciences, Westlake University, Hangzhou 310024, Zhejiang, China}
\affiliation{Key Laboratory for Quantum Materials of Zhejiang Province, School of Science, Westlake University, Hangzhou 310024, Zhejiang, China}
\affiliation{Institute of Natural Sciences, Westlake Institute for Advanced Study, Hangzhou 310024, Zhejiang, China}

\begin{abstract}
Recently, the bilayer perovskite nickelate La$_3$Ni$_2$O$_7$ has been reported to exhibit high-temperature superconductivity near $80$ K under a moderate pressure of about $14$GPa. 
To investigate the underlying pairing mechanism and symmetry in this complex system, we propose and analyze a mixed spin-$1$ and spin-$\frac{1}{2}$ bilayer $t$-$J$ model in the strong coupling regime. 
This model explicitly incorporates the crucial role of strong Hund's coupling, which favors the formation of local spin-triplet states from the two onsite $E_g$ orbital electrons at half-filling.
We further investigate the model using both slave-particle mean-field theory and the density matrix renormalization group method. 
Our simulation results reveal that the dominate pairing channel is the interlayer one in the $3d_{x^2-y^2}$ orbital. 
The Hund's coupling is shown to enhance superconductivity within a reasonable physical range. 
Moreover, electron doping strengthens superconductivity by increasing carrier density; in contrast, hole doping weakens superconductivity. 
These findings offer critical insights into the unconventional superconductivity of pressurized La$_3$Ni$_2$O$_7$ and underline the important role of orbital-selective behavior and Hund's rule.
\end{abstract}

\maketitle

\section{Introduction}
The discovery of high-$T_c$ superconductivity (SC) with critical temperature $T_c$ $\approx 80\text{ K}$ in the bilayer perovskite nickelate La$_3$Ni$_2$O$_7$ under pressure ~\cite{Wang2023LNO, YuanHQ2023LNO,Wang2023LNOb,wang2023LNOpoly,wang2023la2prnio7,zhang2023pressure,zhou2025investigations,wang2024bulk,li2024pressure,Ueki2025pressure} has garnered widespread attention, both experimentally ~\cite{Fukamachi2001,Wang2022LNO,wang2023structure,chen2024electronic,Kakoi2024,xie2024neutron,feng2024unaltered,meng2024density,fan2024tunn,LI2024distinct,liu2024electronic,wang2024review,yang2024orbital,Li2024ele,zhang2024doping,liu2024growth,chen2024evidence,gupta2024anisotropic,xu2024pressure,li2024distinguishing,zhou2024revealing,su2024strongly,mijit2024local,ren2025resolving,khasanov2024pressure,zhao2024spin,chen2025unveiling,shi2025prerequisite,li2025ambient,huo2025low,zhang2025damage} and theoretically ~\cite{YaoDX2023,shilenko2023correlated,labollita2023ele,WangQH2023,Werner2023,oh2023type2,sui2023rno,Dagotto2023,huang2023impurity,ZhangGM2023DMRG,Yi_Feng2023,qin2023high,YangF2023,lechermann2023,Kuroki2023,HuJP2023,lu2023bilayertJ,liao2023electron,qu2023bilayer,jiang2023high,zhang2023trends,qu2023roles,jiang2023pressure,lu2023sc,kitamine2023,zhang2023strong,pan2023rno,sakakibara2023La4Ni3O10,lange2023mixedtj,yang2023strong,grusdt2023lno,lange2023feshbach,fan2023sc,cao2023flat,zhang2023structural,geisler2023structural,rhodes2023structural,zhang2023la3ni2o6,geisler2024optical,tian2023correlation,luo2023high,kaneko2023pair,WuWei2023charge,yang2024decom,Lu2024interplay,chen2023iPEPS,kakoi2023pair,Talantsev2024analysis,Ouyang2024absence,wu2024deconfined,zhang2024electronic,heier2023competing,oh2024high,wang2024self,xu2025competition,ryee2024quenched,chen2023critical,zheng2023twoorbital,liu2023dxy,wang2025mottness,kaneko2025tj,shi2025effect}.
More recently, the observation of ambient-pressure SC with $T_c$ $\approx 40\text{ K}$ in thin-film samples has attracted further attention \cite{ko2024signatures,zhou2025ambient,geisler2024fermi,liu2025superconductivity,bhatt2025resolving,yue2025correlated,shao2025band,hu2025electronic}.
These findings highlight the potential of nickelates as a new platform for exploring unconventional superconductors, 
particularly as analogs to the extensively studied cuprates ~\cite{tsuei2000pairing,lee2006doping,taillefer2010scattering,proust2019remarkable,botana2020similar,bhatta2025structural}.
In cuprates, doping introduces holes into the oxygen $2p$ orbitals, 
which form Zhang-Rice singlets with localized $3d_{x^2-y^2}$ orbital spins in the Cu$^{2+}$ ions \cite{zhang1988effective}. 
The suppression of long-range antiferromagnetic (AFM) order under doping leads to $d$-wave SC \cite{tsuei2000pairing,lee2006doping}. 
A similar scenario has been proposed for infinite-layer nickelates such as Nd$_{1-x}$Sr$_x$NiO$_2$ \cite{li2019sc,nomura2022sc,wang2024experimental}. 
However, La$_3$Ni$_2$O$_7$ presents a more complex challenge due to its bilayer structure, unusual orbital filling, and strong electron correlations. 
Moreover, the $T_c$ of the thin-film at ambient pressure is much lower than the one of the pressurized bulk La$_3$Ni$_2$O$_7$. 
Therefore, it is still necessary to understand the high $T_c$ above the boiling point of liquid nitrogen in the pressurized bulk La$_3$Ni$_2$O$_7$ by studying its pairing mechanism.

Recent theoretical and experimental studies have indicated that pressurized La$_3$Ni$_2$O$_7$ exhibits remarkable orbital-selective strong-correlation effects: 
According to first-principles calculations based on density-functional theory (DFT), the low-energy physics near the Fermi level are dominated by Ni-$3d_{z^2}$ and Ni-$3d_{x^2-y^2}$ orbitals, with occupancies of approximately half and one-quarter, respectively ~\cite{sui2023rno,YaoDX2023,Dagotto2023,cao2023flat,zhang2023structural,huang2023impurity,geisler2023structural,rhodes2023structural,zhang2023la3ni2o6,geisler2024optical}. 
A range of experiments have demonstrated the strongly-correlated characteristic of the bilayer nickelate material ~\cite{chen2024electronic,Kakoi2024,fan2024tunn,LI2024distinct,liu2024electronic,yang2024orbital,Li2024ele,li2024distinguishing,liu2024growth}. 
For instance, optical measurements report a notable reduction in electron kinetic energy, indicating proximity to the Mott phase ~\cite{liu2024electronic}; 
angle-resolved photoemission spectroscopy (ARPES) experiment reveals pronounced orbital-selective band renormalization ~\cite{yang2024orbital}; 
linear temperature dependence of resistivity points to ``strange-metal'' behavior~\cite{YuanHQ2023LNO}; 
the transport measurements of resistivity and magneto-resistance confirm Kondo-like scattering~\cite{liu2024growth}.
Together, these findings suggest that La$_3$Ni$_2$O$_7$ under pressure may provide a novel platform for exploring the interplay between orbital selectivity, strong correlations, and Hund's coupling.

Presently, the pairing mechanism in the pressurized La$_3$Ni$_2$O$_7$ remains an open question due to its complex electronic nature ~\cite{WangQH2023,Werner2023,oh2023type2,sui2023rno,YaoDX2023,Dagotto2023,cao2023flat,zhang2023structural,huang2023impurity,geisler2023structural,rhodes2023structural,zhang2023la3ni2o6,geisler2024optical,ZhangGM2023DMRG,Yi_Feng2023,qin2023high,tian2023correlation,luo2023high,kaneko2023pair,yang2024decom,Lu2024interplay,Ouyang2024absence,YangF2023,lechermann2023,Kuroki2023,HuJP2023,lu2023bilayertJ,liao2023electron,qu2023bilayer,jiang2023high,zhang2023trends,jiang2023pressure,lu2023sc,kitamine2023,zhang2023strong,pan2023rno,sakakibara2023La4Ni3O10,lange2023mixedtj,yang2023strong,lange2023feshbach,fan2023sc,Talantsev2024analysis,wu2024deconfined,ryee2024quenched,kaneko2025tj}.
A key issue among theoretical proposals is determining which orbitals are most relevant for the SC.
Some perspectives suggest that the SC of pressurized La$_3$Ni$_2$O$_7$ is significantly related to hybridization between the nearest-neighbor (NN) $3d_{z^2}$ and $3d_{x^2-y^2}$ orbitals~\cite{ZhangGM2023DMRG,Yi_Feng2023,qin2023high,tian2023correlation,luo2023high,kaneko2023pair,yang2024decom,Lu2024interplay,Ouyang2024absence}, 
while others emphasize the critical role of Hund's coupling in driving the superconducting behavior ~\cite{lu2023bilayertJ,oh2023type2,qu2023bilayer,tian2023correlation,zhang2023strong,pan2023rno,lange2023mixedtj,yang2023strong,lange2023feshbach,kaneko2023pair,wu2024deconfined,Lu2024interplay,Ouyang2024absence,oh2024high}.
There is also a possibility that both factors are involved.

In this work, we propose and study a mixed spin-$1$ and spin-$\frac{1}{2}$ bilayer $E_g$-orbital $t$-$J$ model, 
incorporating the spin coupling among and between the two orbitals. 
This model establishes a strong-coupling framework for identifying the dominant pairing channel in the bilayer system, 
while explicitly linking Hund's rule to superconducting strength.
We solve the ground state properties using slave-particle mean-field (SPMF) ~\cite{kotliar1988,lee2006doping} and the density matrix renormalization group (DMRG) ~\cite{white1993dmrg,weng1999dmrg} methods. 
The numerical results reveal a phase diagram where the dominant pairing occurs in the interlayer $3d_{x^2-y^2}$ orbitals. 
The Hund's coupling $J_H$ promotes SC pairing and attracts a few $3d_{z^2}$-orbital electrons to the $3d_{x^2-y^2}$-orbital. 
The effect of doping is further explored, where electron-doping will enhance the SC pairing.

The remainder of this paper is organized as follows. 
Sect.~\ref{eq:sect:tJmodel} introduces the effective strong-coupling $t$-$J$ model Hamiltonian. 
Sect.~\ref{sect:SPMFT} details the SPMF analysis and the pairing nature. 
Sect.~\ref{sect:DMRG} presents the DMRG calculations. 
Finally, Sect.~\ref{sect:Con} summarizes our findings and provides concluding remarks.


\section{Effective Bilayer two-orbital Model}
\label{eq:sect:tJmodel}
The electronic characteristics of the bilayer La$_3$Ni$_2$O$_7$ under pressure are predominantly influenced by the $d_{x^2 - y^2}$ and $d_{z^2}$ orbitals within the NiO$_2$ planes, which are naively quarter-filling and half-filling, respectively.
The $d_{x^2-y^2}$ orbital is primarily responsible for in-plane conduction and dominates the electronic states near the Fermi level, contributing to the metallic behavior observed in the normal state. 
On the other hand, the $d_{z^2}$ orbital is more localized, with its lobes extending out of the NiO$_2$ planes, 
thereby mediating the interlayer coupling between adjacent NiO$_2$ planes and leads to the formation of bonding and anti-bonding states \cite{Wang2023LNO}.

\begin{figure}[!t]
    \centering
    \includegraphics[width=1\linewidth]{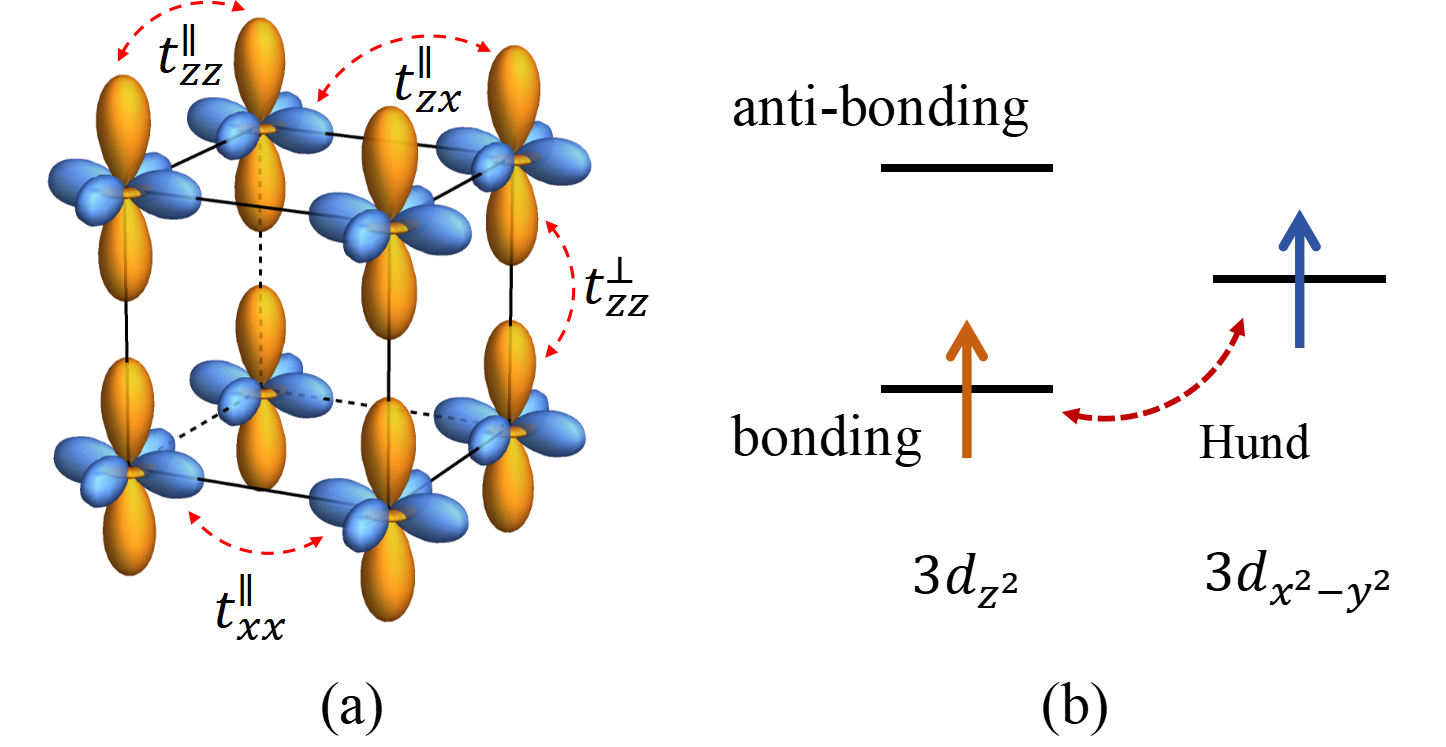}
    \caption{(a) Schematic illustration of the $d_{z^2}$ and $d_{x^2-y^2}$ orbitals within the bilayer structure of the nickelate La$_3$Ni$_2$O$_7$,
    including the relevant hopping integrals. 
    (b) Strong interlayer hybridization splits the $d_{z^2}$-band into lower-energy bonding and higher-energy anti-bonding bands, respectively. 
    The bonding band is energetically favored. 
    Due to strong Hund's coupling, electrons occupying the $d_{z^2}$ and $d_{x^2-y^2}$ orbitals tend to form a spin-triplet state.}
    \label{TB}
\end{figure}

The electronic properties of the bilayer La$_3$Ni$_2$O$_7$ are descried by a two-orbital Hubbard model on the bilayer square lattice given as $H_{\text{Hubbard}}=H_t+H_\text{Int}$.
Here, the kinetic part is given by
\begin{equation}
\begin{aligned}
\label{eq:kinetic}
H_t = &\sum_{i\mu\alpha\sigma}\epsilon_{\alpha}n_{i\mu\alpha\sigma} 
- \sum_{\langle ij\rangle\mu\sigma} t_{xx}^{\parallel} \left(c^{\dagger}_{i\mu x\sigma}c_{j\mu x\sigma} + \text{h.c.}\right)\\
&- \sum_{\langle ij\rangle\mu\sigma} t_{zx}^{\parallel} \left( c^{\dagger}_{i\mu z\sigma}c_{j\mu x\sigma} + (z\leftrightarrow x) + \text{h.c.}\right)\\
&- \sum_{\langle ij\rangle\mu\sigma} t_{zz}^{\parallel} \left(c^{\dagger}_{i\mu z\sigma}c_{j\mu z\sigma} + \text{h.c.}\right)\\
&- \sum_{i\sigma} t_{zz}^{\perp} \left(c^{\dagger}_{itz\sigma}c_{ibz\sigma} + \text{h.c.}\right),
\end{aligned}
\end{equation}
where $c_{i\mu\alpha\sigma}^\dag$ creates an $\alpha = \left\{d_{z^2}(z),d_{x^2-y^2}(x)\right\}$-orbital electron with spin $\sigma = \left\{\uparrow,\downarrow\right\}$ at the lattice site $i$ in the layer $\mu=\left\{\text{top}(t),\text{bottom}(b)\right\}$;
$\langle ij\rangle$ represents the intralayer NN bonds;
$t_{xx}^{\parallel}, t_{zx}^{\parallel}$, $t_{zz}^{\parallel}$ and $t_{zz}^\perp$ are hopping integrals as shown in Fig.~\ref{TB}~(a), 
of which $t_{zx}^{\parallel}$ represents the non-zero NN hybridization between the two $E_g$ orbitals, exhibiting opposite signs along $x$- and $y$-directions due to the symmetry constraint, $t_{zx,x}^\parallel = -t_{zx,y}^\parallel = t_{zx}^\parallel$;
the interlayer hopping of the $d_{x^2-y^2}$ orbital is negligible small due to the wavefunction distribution of it;
$\epsilon_{\alpha}$ denotes the onsite energy for each orbital.

The interacting part for the two-orbital system is given by 
\begin{equation}
\begin{aligned}
\label{eq:Int}
H_{\text{Int}} = &U\sum_{i\mu\alpha} n_{i\mu\alpha\uparrow} n_{i\mu\alpha\downarrow} + V\sum_{i\mu\sigma\sigma^{\prime}}n_{i\mu z\sigma}n_{i\mu x\sigma^{\prime}} \\
&- 2J_H\sum_{i\mu}\left(\bm{S}_{i\mu z}\cdot\bm{S}_{i\mu x} + \frac{1}{4}n_{i\mu z}n_{i\mu x}\right)\\
&+ J_H\sum_{i\mu}\left(c^{\dagger}_{i\mu z\uparrow} c^{\dagger}_{i\mu z\downarrow} c_{i\mu x\downarrow} c_{i\mu x\uparrow} + \mathrm{h.c.}\right).
\end{aligned}
\end{equation}
Here, $\bm{S}_{i\mu\alpha} = \frac{1}{2} c^\dagger_{i\mu\alpha\sigma} \left[\bm{\sigma}\right]_{\sigma\sigma^\prime} c_{i\mu\alpha\sigma^\prime}$ is the spin operator with Pauli matrices $\bm{\sigma}=\left(\sigma_x,\sigma_y,\sigma_z\right)$. 
$n_{i\mu\alpha\sigma} = c_{i\mu\alpha\sigma}^{\dag} c_{i\mu\alpha\sigma}$ is the particle occupancy number operator. 
$U$ and $V$ represent the onsite intra- and inter-orbital Coulomb repulsions, respectively. 
The Hund's coupling $J_H$ consists of the spin exchange (third term in the second line) and pair-hopping (fourth terms in the third lines) terms.
The pair-hopping term can be omitted when the on-site intraorbital no-double occupancy condition is imposed.
The spin-exchange part favors a spin-triplet formation between the $d_{z^2}$ and $d_{x^2-y^2}$ electrons at the same site, as shown in Fig.~\ref{TB}~(b).
The condition of orbital rotational symmetry induces $U=V+2J_H$ \cite{castellani1978}.

\begin{figure*}[!t]
    \centering
    \includegraphics[width=0.8\textwidth]{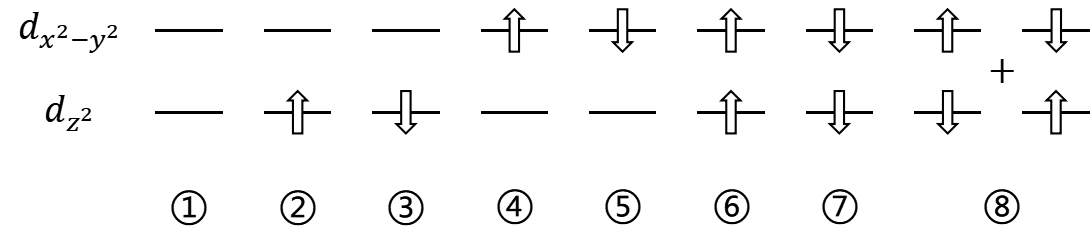}
    \caption{The low-energy local Hilbert space within a Ni site for the two $E_g$ orbitals ($d_{z^2}$ and $d_{x^2-y^2}$) in the strong coupling limit upon doping Ni $3d^8$ states, relevant for the derived $t$-$J$ model.
    It comprises the double-hole state, spin-$\frac{1}{2}$ singlons, and spin-$1$ triplet doublons.
    States with double occupancy within the same orbital are projected out due to strong Coulomb replusion, leaving these eight configurations as the effective basis.
    }
    \label{configurations}
\end{figure*}

In the strong coupling limit, the large onsite Hubbard repulsion $U$ forbids the formation of double occupancy for each orbital.
When both the $d_{z^2}$ and $d_{x^2-y^2}$-orbitals are single occupied on the same site, Hund's rule energetically favors the formation of an interorbital spin-triplet ($S=1$) state. 
Considering these constraints, the relevant low-energy local Hilbert space at each site comprises eight possible configurations, depicted in Fig.~\ref{configurations}.

Treating the kinetic hopping terms as perturbations to the dominant interaction terms, an effective low-energy Hamiltonian can be derived via the superexchange mechanism using the standard second-order perturbation theory. 
This procedure yields an effective bilayer $t$-$J$ type model that describes the dynamics of the allowed local stats, including spin-$\frac{1}{2}$ singlons (single occupancy in either $d_{z^2}$ or $d_{x^2-y^2}$) and spin-$1$ triplet doublons (single occupancy in both orbitals).
The resulting Hamiltonian is given by:
\begin{equation}
\begin{aligned}
H = &\ H_t + H_J^\parallel + H_J^\perp + V\sum_{i\mu\sigma\sigma^{\prime}}n_{i\mu z\sigma} n_{i\mu x\sigma^{\prime}} \\
&- 2J_H\sum_{i\mu}\left(\bm{S}_{i\mu z}\cdot\bm{S}_{i\mu x} + \frac{1}{4}n_{i\mu z} n_{i\mu x}\right)
\end{aligned}\label{eq:Ham}
\end{equation}
where $H_t$ represents the effective hopping terms (projected onto the restricted Hilbert space);
$H_J^\parallel$ and $H_J^\perp$ contain the intralayer and interlayer superexchange interactions, respectively.
The $V$- and $J_H$-terms account for the interorbital Coulomb repulsion and the Hund's coupling, respectively, which influence the local energy of each configuration in Fig.~\ref{configurations}.

The intralayer superexchange Hamiltonian takes the form:
\begin{equation}
\begin{aligned}
&H_J^\parallel=
\sum_{\langle ij\rangle\mu}J^{\parallel}_{zz}\left(\bm{S}_{i\mu z}\cdot\bm{S}_{j\mu z}-\frac{1}{4}n_{i\mu z}n_{j\mu z}\right)\\
&+\sum_{\langle ij\rangle\mu}J^{\parallel}_{xx}\left(\bm{S}_{i\mu x}\cdot\bm{S}_{j\mu x}-\frac{1}{4}n_{i\mu x}n_{j\mu x}\right)\\
&+\sum_{\langle ij\rangle\mu}J^{\parallel}_{dd}\left(\bm{S}_{i\mu d}\cdot\bm{S}_{j\mu d}-\frac{1}{4}n_{i\mu d}n_{j\mu d}\right)\\
&+\sum_{\langle ij\rangle\mu}J^{\parallel}_{zx}\left(\bm{S}_{i\mu z}\cdot\bm{S}_{j\mu x}-\frac{1}{4}n_{i\mu z}n_{j\mu x}+\left( i\leftrightarrow j\right)\right)\\
&+\sum_{\langle ij\rangle\mu}J^{\parallel}_{zd}\left(\bm{S}_{i\mu z}\cdot\bm{S}_{j\mu d}-\frac{1}{4}n_{i\mu z}n_{j\mu d}+\left( i\leftrightarrow j\right)\right)\\
&+\sum_{\langle ij\rangle\mu}J^{\parallel}_{xd}\left(\bm{S}_{i\mu x}\cdot\bm{S}_{j\mu d}-\frac{1}{4}n_{i\mu x}n_{j\mu d}+\left( i\leftrightarrow j\right)\right).
\end{aligned}\label{H_intraS}
\end{equation}
Here, $\bm{S}_{i\mu\alpha}$ (with $\alpha=z$ or $x$) represents the spin-$\frac{1}{2}$ operator when there is only one electron occupying the $d_{z^2}$ or $d_{x^2-y^2}$ orbital, respectively. 
$n_{i\mu\alpha} = \sum_{\sigma} n_{i\mu\alpha\sigma}$ represents the total particle number of the $\alpha$ orbital at the site $i$ in the layer $\mu$. 
$\bm{S}_{i\mu d}$ represents the spin-$1$ operator under Hund's rule when both the $d_{z^2}$ and $d_{x^2-y^2}$ orbitals are singly occupied, and $n_{i\mu d}= \sum_{\alpha} n_{i\mu\alpha}$ is the total particle number of the site $i$ in the layer $\mu$ under this occupation. 
The various superexchange parameters $J$ quantify the effective antiferromagnetic spin couplings between different types of local states across NN bonds;
for example, $J_{zd}^\parallel$ describes the intralayer spin coupling between a $d_{z^2}$ singlon and a triplet doublon. 
Explicit expressions relating these spin-exchange $J$ parameters to the original Hubbard model parameters are provided in Appendix~\ref{app_A}.
Similarly, we can also derive the interlayer superexchange Hamiltonian:
\begin{equation}
\begin{aligned}
&H_J^\perp =
\sum_{i}J^{\perp}_{zz}\left(\bm{S}_{itz}\cdot\bm{S}_{ibz}-\frac{1}{4}n_{itz}n_{ibz}\right)\\
&+\sum_{i}J^{\perp}_{dd}\left(\bm{S}_{itd}\cdot\bm{S}_{ibd}-\frac{1}{4}n_{itd}n_{ibd}\right)\\
&+\sum_{i}J^{\perp}_{zx}\left(\bm{S}_{itz}\cdot\bm{S}_{ibx}-\frac{1}{4}n_{itz}n_{ibx}+\left( t\leftrightarrow b\right)\right)\\
&+\sum_{i}J^{\perp}_{zd}\left(\bm{S}_{itz}\cdot\bm{S}_{ibd}-\frac{1}{4}n_{itz}n_{ibd}+\left( t\leftrightarrow b\right)\right)\\
&+\sum_{i}J^{\perp}_{xd}\left(\bm{S}_{itx}\cdot\bm{S}_{ibd}-\frac{1}{4}n_{itx}n_{ibd}+\left( t\leftrightarrow b\right)\right).
\end{aligned}\label{H_interS}
\end{equation}
Various interlayer superexchange parameters $J$ are defined in a similar way to those of the intralayer ones, and their expressions are also derived in Appendix~\ref{app_A}. 
The interlayer superexchange within a rung for the singly-occupied $d_{x^2-y^2}$-orbital spin-$\frac{1}{2}$ approximate to $0$ due to the negligible interlayer hopping between them.
To demonstrate the validity of the perturbation theory, we also solve the two-orbital Hubbard model combining Eqs. (\ref{eq:kinetic}) and (\ref{eq:Int}) on lattices of small sizes using exact diagonalization (ED) and DMRG methods. The results are provided in Appendix~\ref{app_C}.

We adopt the data obtained from the DFT calculations ~\cite{YaoDX2023} as input physical parameters for $H_t$. 
The intralayer hopping parameters are $t^\parallel_{zz}=0.110$ eV, $t^\parallel_{zx}=0.239$ eV, and $t^\parallel_{xx}=0.483$ eV. 
The interlayer hopping for the $d_{z^2}$ orbital is $t_{zz}^\perp=0.635$ eV, while the ones involving the $d_{x^2-y^2}$ orbital nearly vanish. 
The onsite energies are set to $\epsilon_z = 0.409$ eV and $\epsilon_x = 0.776$ eV. 
With the relationship of interaction strengths $U=V+2J_H$ \cite{castellani1978}, the onsite Coulomb repulsion is chosen as $U=5$ eV, and the Hund's coupling $J_H$ ranges from $\frac{U}{8}$ to $\frac{U}{4}$ approximately.

\section{Slave-Particle Mean-Field Study}
\label{sect:SPMFT}
In this section, we present the construction of the physical Hilbert space and apply the slave-particle mean-field theory to study the superconducting state.

\subsection{Constructions of physical states}
Within the framework of the slave-particle method, the restriction to the physical Hilbert space can be systematically imposed. 
The fully empty $3d^{6}$ state is defined by acting with the holon creation operator $h_{i\mu}^{\dagger}$ on a fictitious vacuum state,
\begin{equation}
|\text{empty}\rangle_{i\mu}
=h_{i\mu}^{\dagger} |\text{vac}\rangle_{i\mu},
\end{equation}
where $|\text{vac}\rangle$ represents the vacuum of slave particles and lies outside the physical space.
For simplicity, the onsite energy of the empty state
$|\text{empty}\rangle_{i\mu}$
in the atomic limit is set to zero, and its occupancy number is denoted by $\tilde{n}_h$.

The four singly-occupied $3d^7$ states, corresponding to the spin-$\frac{1}{2}$ singlon configurations shown in Fig.~\ref{configurations}, are represented using bosonic spinon creation operators, 
\begin{equation}
\begin{aligned}
|\uparrow_z \rangle_{i\mu} 
=&b_{i\mu z\uparrow}^{\dagger} |\text{vac}\rangle_{i\mu},\quad
|\uparrow_x \rangle_{i\mu} 
=b_{i\mu x\uparrow}^{\dagger} |\text{vac}\rangle_{i\mu},\\
|\downarrow_z \rangle_{i\mu} 
=&b_{i\mu z\downarrow}^{\dagger} |\text{vac}\rangle_{i\mu},\quad
|\downarrow_x \rangle_{i\mu} 
=b_{i\mu x\downarrow}^{\dagger} |\text{vac}\rangle_{i\mu},
\end{aligned}
\label{eq:spinHalfslave}
\end{equation}
where $b^\dag_{i\mu\alpha\sigma}$ is a creation operator of bosonic spinon, acting on the vacuum state to generate four spin-$\frac{1}{2}$ configurations. 
The occupancy numbers for these singly occupied state are $\tilde{n}_x$ and $\tilde{n}_z$ for the $3d_{x^2-y^2}$ and $3d_{z^2}$ orbitals, respectively.
The corresponding spin-$\frac{1}{2}$ operators are defined in the conventional manner. 
For instance, the spin operator for a state with only the $d_{x^2-y^2}$ orbital occupied is given by $\bm{S}_{i\mu x} = \frac{1}{2} b_{i\mu x\sigma}^{\dagger} \left[\bm{\sigma}\right]_{\sigma\sigma^\prime} b_{i\mu x\sigma^\prime}$, where $\bm{\sigma} = (\sigma_x, \sigma_y, \sigma_z)$ are the Pauli matrices. 
The onsite energy of the single-occupied state in the $d_{z^2}$ orbital is set to zero, while for the $d_{x^2-y^2}$ orbital, a finite energy difference $\Delta_g=\epsilon_x-\epsilon_z$ is considered.

For the three $3d^8$ spin-$1$ triplet doublons, we employ a three-component Schwinger fermion representation. 
The Schwinger fermion $f_{i\mu} = (f_{i\mu,+1}, f_{i\mu,0}, f_{i\mu,-1})^T$ labels each spin projection:
\begin{equation}
\begin{aligned}
|+1\rangle_{i\mu}=& f_{i\mu,+1}^{\dagger} |\text{vac}\rangle_{i\mu}, \\
|0\rangle_{i\mu}=& f_{i\mu,0}^{\dagger} |\text{vac}\rangle_{i\mu}, \\
|-1\rangle_{i\mu}=& f_{i\mu,-1}^{\dagger} |\text{vac}\rangle_{i\mu}.
\end{aligned}
\label{eq:spin1slave}
\end{equation}
The corresponding spin-$1$ operators are expressed as
\begin{equation}
\begin{aligned}
\hat{S}_{i\mu d}^+
=& f_{i\mu}^{\dagger} S_+ f_{i\mu}
=\sqrt{2} \big( f_{i\mu,+1}^{\dagger} f_{i\mu,0} +f_{i\mu,0}^{\dagger} f_{i\mu,-1} \big),	\\
\hat{S}_{i\mu d}^-
=& f_{i\mu}^{\dagger} S_- f_{i\mu}
=\sqrt{2} \big( f_{i\mu,0}^{\dagger} f_{i\mu,+1} +f_{i\mu,-1}^{\dagger} f_{i\mu,0} \big),	\\
\hat{S}_{i\mu d}^z=& f_{i\mu}^{\dagger} S_z f_{i\mu}
=f_{i\mu,+1}^{\dagger} f_{i\mu,+1}
-f_{i\mu,-1}^{\dagger} f_{i\mu,-1},
\end{aligned}
\end{equation}
where the ladder operators are $\hat{S}_{i\mu}^{\pm} = \hat{S}_{i\mu}^x \pm i \hat{S}_{i\mu}^y$, and the matrices $\{S_+, S_-, S_z\}$ form the spin-$1$ irreducible representation of the $\mathrm{SU(2)}$ generators. 
The occupancy number of these spin-triplet state is denoted as $\tilde{n}_d$.
The onsite energy of these spin-$1$ states is set to $V - J_H + \Delta_g$.

The original electron operators can be expressed in terms of these slave particle operators. 
For example, the creation operator for a $d_{x^2-y^2}$-orbital electron at site $i$ in layer $\mu$ with spin $\uparrow$ is
\begin{equation}
c_{i\mu x\uparrow}^{\dagger}
=f_{i\mu,+1}^{\dagger} b_{i\mu z\uparrow}
+\frac{1}{\sqrt{2}} f_{i\mu,0}^{\dagger} b_{i\mu z\downarrow}
+ b_{i\mu x\uparrow}^{\dagger} h_{i\mu},
\end{equation}
with similar expressions for the spin-$\downarrow$ state and the $d_{z^2}$ orbital.
To preserve fermionic statistics, we assign fermionic statistics to both $f$- and $h$-operators, while the $b$-operators are bosonic.
The physical Hilbert space is further constrained by the following local condition
\begin{equation}
\begin{aligned}
&f_{i\mu,+1}^{\dagger} f_{i\mu,+1}
+f_{i\mu,0}^{\dagger} f_{i\mu,0}
+f_{i\mu,-1}^{\dagger} f_{i\mu,-1}\\
+&h_{i\mu}^{\dagger} h_{i\mu}
+b_{i\mu x\uparrow}^{\dagger} b_{i\mu x\uparrow}
+b_{i\mu x\downarrow}^{\dagger} b_{i\mu x\downarrow}\\
+&b_{i\mu z\uparrow}^{\dagger} b_{i\mu z\uparrow}
+b_{i\mu z\downarrow}^{\dagger} b_{i\mu z\downarrow}
=1,
\end{aligned}
\end{equation}
which corresponds to a local $\mathrm{U(1)}$ gauge symmetry associated with charge conservation. 
For averaged $3d^{7.5}$ configuration relevant to La$_3$Ni$_2$O$_{7}$, 
the occupancy number are further related by $\tilde{n}_d-\tilde{n}_h=0.5$.


\subsection{Correlated pairing — mixed spin-1 and spin-$\frac{1}{2}$ valence bonds}
\begin{figure}[t!]
    \centering
    \includegraphics[width=1\linewidth]{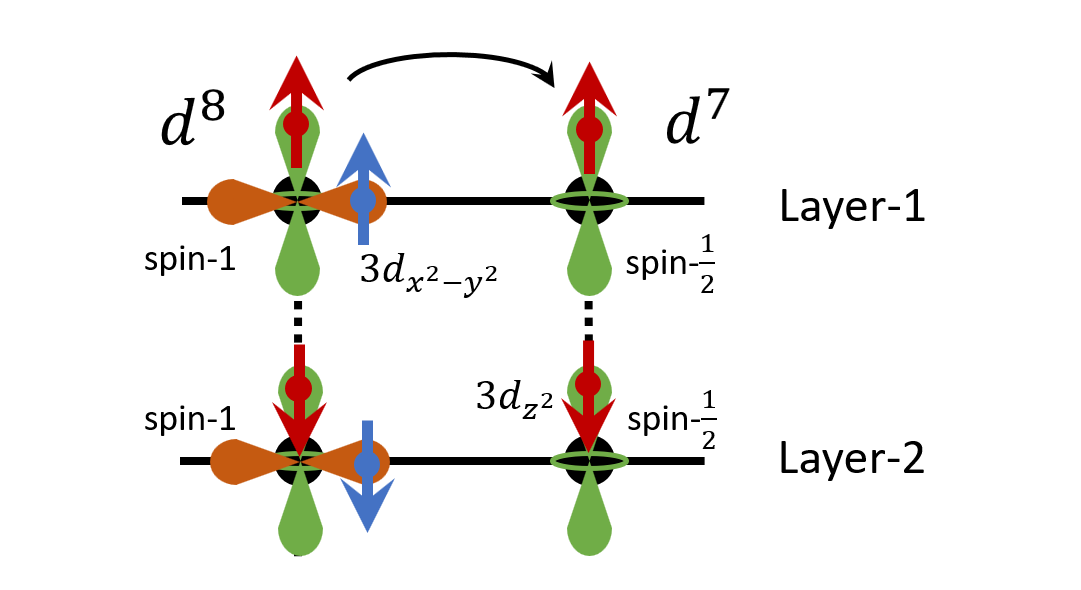}
    \caption{Schematic illustration of the two types of interlayer singlet bonds formed on a rung (vertical pair of sites) driven by strong interlayer superexchange. 
    Left: A spin-$1$ singlet bond formed between two local spin-triplets (characteristic of the $d^8$ electronic configuration). 
    Right: A spin-$1$ singlet bond formed between two localized spins (predominantly $d_{z^2}$ electrons, characteristic of the $d^7$ configuration emerging upon doping).} 
    \label{fig:Spin1singlet}
\end{figure}

When interlayer superexchange dominates, the two localized spin-triplet states along a rung tend to form a spin-$1$ singlet bond, as illustrated in Fig.~\ref{fig:Spin1singlet}.
The corresponding rung spin-singlet pairing operator at lattice site $j$ is given by
\begin{equation}
F_{j}^{\dagger} = \frac{1}{\sqrt{3}} \Big( f_{jt,1}^{\dagger} f_{jb,-1}^{\dagger} - f_{jt,0}^{\dagger} f_{jb,0}^{\dagger} + f_{jt,-1}^{\dagger} f_{jb,1}^{\dagger} \Big).
\end{equation}
This describes the formation of a spin-$1$ singlet bond between neighboring sites $jt$ and $jb$ on a rung. 
In the mother $3d^8$ configuration, the system forms a spin-$1$ interlayer valence-bond solid (VBS) state, where the four electrons residing in the two $E_g$ orbitals are strongly entangled, forming an interlayer spin singlet on each rung.

In the case of La$_3$Ni$_2$O$_7$, however, the actual electronic configuration is closer to $3d^{7.5}$, where additional holes prefer to occupy the $d_{x^2-y^2}$ orbitals. 
This emergence of SC in La$_3$Ni$_2$O$_7$ under pressure can be considered as doping the $d^8$ spin-$1$ VBS state. 
Upon doping, the spin-triplet configuration, characteristic of the $3d^8$ VBS, largely reduces to a $3d^7$ spin-$\frac{1}{2}$ configuration, predominantly involving the $d_{z^2}$ orbital electron. 
As a result, the occupancy number of the spin-$1$ states, denoted $n_1$, is approximately $0.5$, and the occupancy number of the spin-$\frac{1}{2}$ states in the $d_{z^2}$ orbital, $n_z$, is similarly around $0.5$. 
Consequently, the strong interlayer exchange $J_{zz}^{\perp}$ drives the formation of a spin-$\frac{1}{2}$ singlet bond along the rungs:
\begin{align}
B_{j}^{\dagger} =\frac{1}{\sqrt{2}}
\Big( b_{jtz\uparrow}^{\dagger} b_{jbz\downarrow}^{\dagger}
- b_{jtz\downarrow}^{\dagger} b_{jbz\uparrow}^{\dagger} \Big),
\end{align}
where the $d_{x^2-y^2}$ orbital remains essentially empty on the rungs. 
This description captures the fundamental change in the electronic structure as the system transitions from a spin-$1$ to a spin-$\frac{1}{2}$-dominated regime.

\begin{figure}[t!]
    \centering
    \includegraphics[width=1\linewidth]{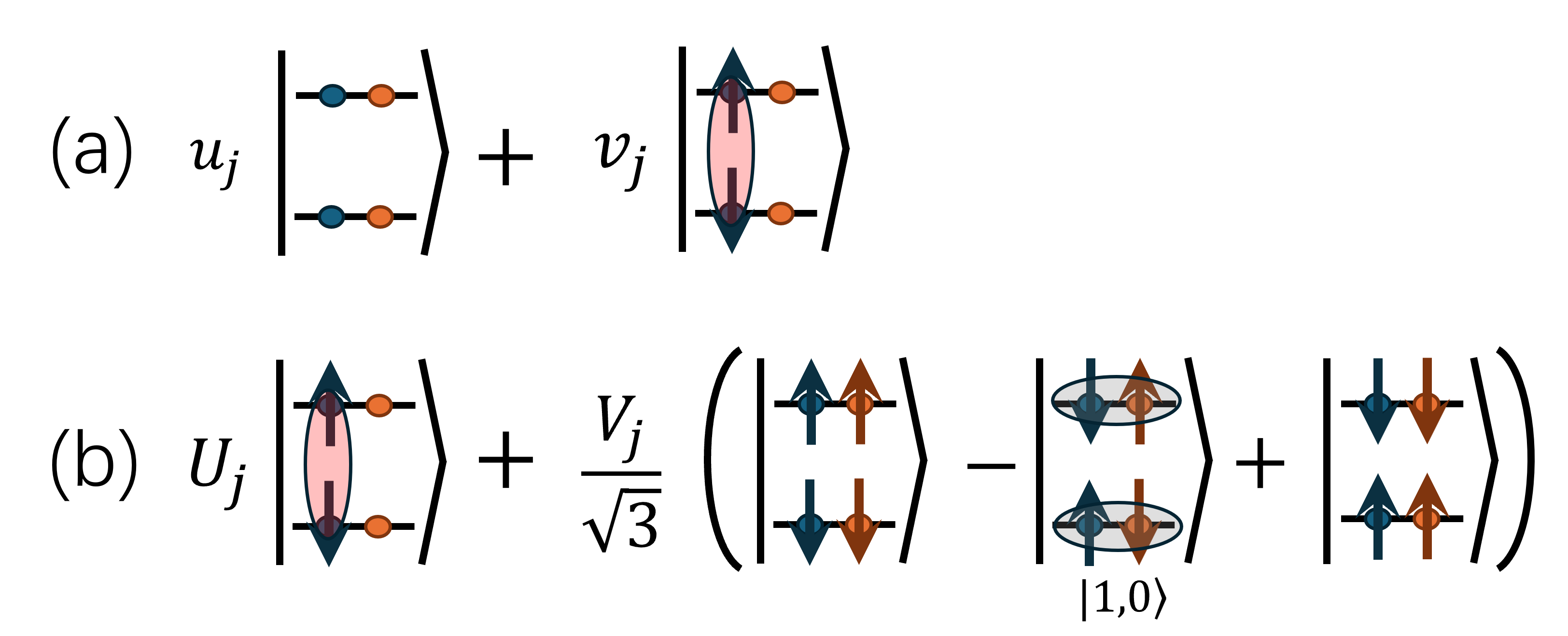}
    \caption{Schematic diagrams for the conventional BCS pairing (a) and the correlated pairing given by Eq.~(\ref{eq:co-pair}) (b).
    The former is represented as a coherent superposition of a fully paired electron state and the vacuum state. 
    The latter is a superposition of two valence bond singlets composed by a pair of spin-$1$ and a pair of spin-$\frac{1}{2}$ states, respectively. 
    The blue circles represent the $d_{z^2}$ orbitals, while the orange ones represent the $d_{x^2-y^2}$ orbitals. 
    $|1,0\rangle$ means a triplet with total spin $S=1$ and total $z$-component spin $S^z$.
    } 
    \label{fig:Spin1BCS}
\end{figure}

The nature of superconducting pairing in this doped spin-$1$ VBS state is fundamentally different from that of Cooper pairs in a Bardeen-Cooper-Schrieffer (BCS) superconductor \cite{tinkham2004introduction}, as depicted in Fig.~\ref{fig:Spin1BCS}. 
In a conventional superconductor, the BCS wavefunction is a coherent superposition of the fully paired electron state and the vacuum (empty) state. 
In contrast, the exotic spinon pairing mechanism considered here involves a combination of two distinct singlets: 
one associated with spin-$\frac{1}{2}$ singlet states and the other with spin-$1$ singlet states. 
In the slave-particle formalism, this pairing can be represented as
\begin{align}
|\text{spinon pair}\rangle 
=\prod_{j} \Big( U_j B_j^{\dagger} +V_j F_j^{\dagger} \Big) |\text{vac}\rangle,
\label{eq:co-pair}
\end{align}
where $U_j$, $V_j$ are coefficients that reflect the relative weights of the two components. 
This mixed pairing structure reflects the complex interplay between spin-$1$ and spin-$\frac{1}{2}$ physics in the doped system, and the resulting physical electron pairing emerges as a composite of these slave-particle pairings.

\subsection{Slave-particle Mean-field Results}
Based on the preceding model analysis, we perform a slave-particle mean-field calculation using the Hamiltonian Eq.~\ref{eq:Ham} on the $2\times200\times200$ bilayer lattice with the temperature $10^{-4}$ to further investigate the physical picture described earlier. 
In this mean-field approach, we concentrate on the SC pairing channels.
The superexchange terms considered most relevant and consequently decoupled into spin-singlet pairing channels are: 
the interlayer spin-$1$ interaction $J_{dd}^{\perp} \bm{S}_{itd}\cdot\bm{S}_{ibd}$, 
the interlayer $3d_{z^2}$ spin-$\frac{1}{2}$ interaction $J_{zz}^{\perp} \bm{S}_{itz}\cdot\bm{S}_{ibz}$, 
and the intralayer $3d_{x^2-y^2}$ spin-$\frac{1}{2}$ interaction $J_{xx}^{\parallel} \bm{S}_{i\mu x}\cdot\bm{S}_{j\mu x}$.
The decoupling of the spin-$\frac{1}{2}$ exchange interactions follows the conventional slave-boson mean-field theory \cite{kotliar1988}.
For the spin-$1$ interactions, the pairing and hopping order parameters for the total spin-singlet channels are introduced as:
\begin{equation*}
\begin{aligned}
\Delta_{0}^{\dagger} &= \frac{1}{\sqrt{3}}
\Big( f_{it,+1}^{\dagger} f_{id,-1}^{\dagger}
- f_{it,0}^{\dagger} f_{id,0}^{\dagger}
+ f_{it,-1}^{\dagger} f_{id,+1}^{\dagger} \Big), \\
\chi_{0}^{\dagger} &= \frac{1}{\sqrt{3}}
\Big( f_{it,+1}^{\dagger} f_{id,+1}
+ f_{it,0}^{\dagger} f_{id,0}
+ f_{it,-1}^{\dagger} f_{id,-1} \Big).
\end{aligned}
\end{equation*}
Here, $\Delta_{0}^{\dagger}$ corresponds to the rung spin-singlet pairing operator $F_i^{\dagger}$.
The spin-$1$ exchange interaction can then be decoupled into these singlet channels as:
\begin{align}
{\bm{S}}_{it} \cdot {\bm{S}}_{id}
= -2 \Delta_{0}^{\dagger} \Delta_{0}
-2 \chi_{0}^{\dagger} \chi_{0}.
\end{align}
In this mean-field decoupling scheme, contributions from higher-energy non-singlet channels (e.g., total spin-$1$ and spin-$2$ configurations) are omitted.
The resulting mean-field hopping and pairing order parameters are then determined by solving the equations self-consistently.
See Appendix~ \ref{app_SPMF} for the details.

Several other terms present in the full Hamiltonian (Eqs.~\ref{H_intraS} and \ref{H_interS}) are neglected within this mean-field ansatz.
Terms coupling the spin-$1$ doublon state ($f$-fermion) with spin-$\frac{1}{2}$ singlon states ($b$-boson), 
such as $\bm{S}_{d}\cdot\bm{S}_{x}$ and $\bm{S}_{d}\cdot\bm{S}_{z}$, do not lead to pairing and are therefore omitted.
Additionally, other superexchange terms like the intralayer $\bm{S}_{z}\cdot\bm{S}_{z}$ and $\bm{S}_{z}\cdot\bm{S}_{x}$, as well as the interlayer $\bm{S}_{x}\cdot\bm{S}_{x}$ and $\bm{S}_{z}\cdot\bm{S}_{x}$, 
are assumed to have significantly weaker coupling strengths and thus negligible contributions to the primary pairing mechanism; these are also omitted for simplicity.
These approximations allows us to focus more clearly on the behavior of the onsite occupancy numbers and dominant pairing amplitudes as functions of the Hund's coupling $J_H$, and doping levels. 
Importantly, our mean-field approach does not decouple the onsite interorbital Coulomb repulsion $V$ or the Hund's coupling $J_H$ terms. 
A naive mean-field decomposition of Hund's coupling, for instance, could generate unphysical intra-ion triplet pairing. 
The effect of these crucial onsite terms ($V$ and $J_H$) are instead treated exactly by modifying the local onsite energies of the physical slave-particle states, as detailed in previous sections.

\begin{figure}[t!]
    \centering
    \includegraphics[width=1\linewidth]{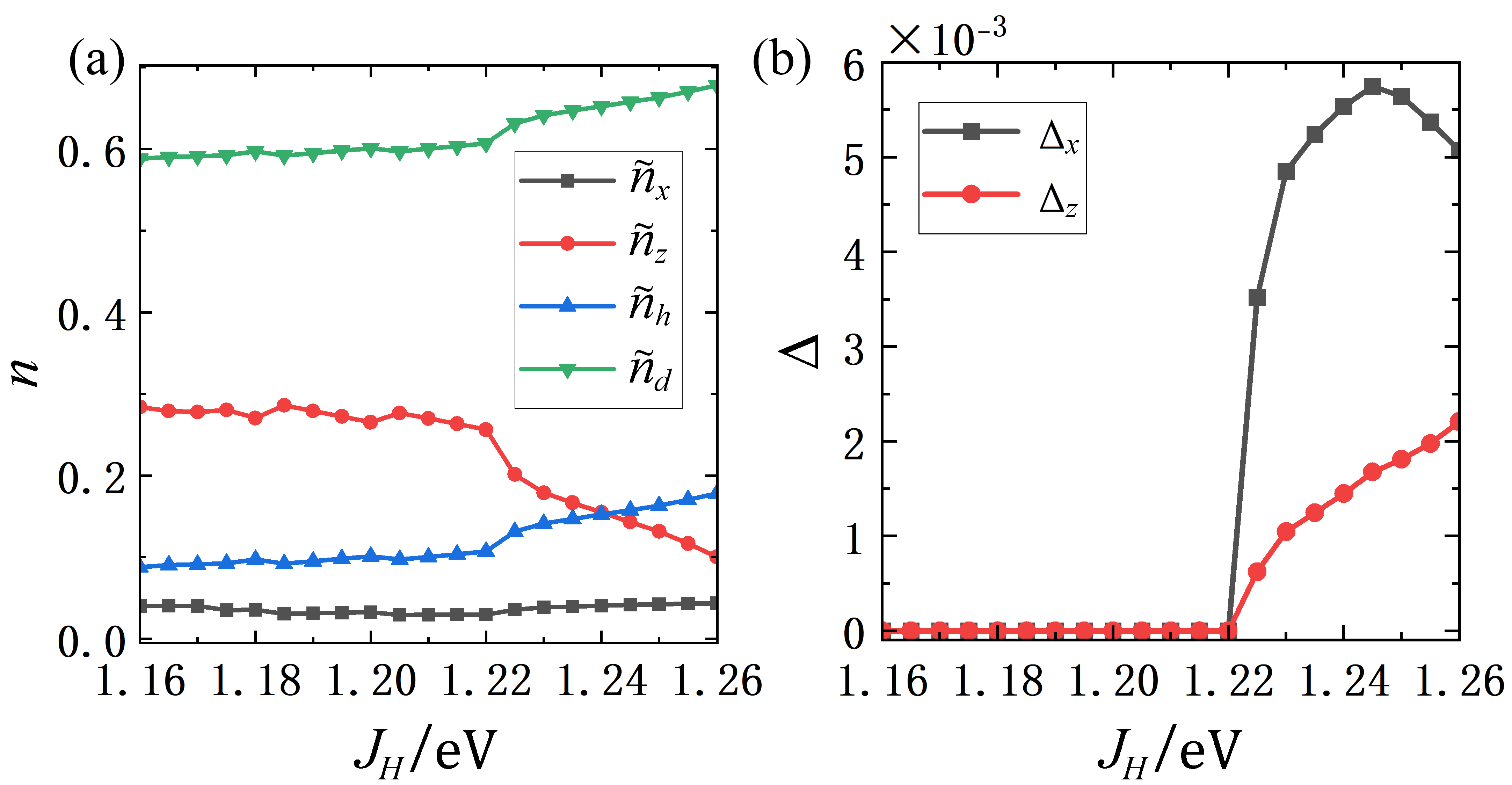}
    \caption{Results based on the SPMF method results as a function of Hund's coupling $J_H$.
    Plotted are (a) the onsite occupancy numbers ($\tilde{n}_x,\tilde{n}_z,\tilde{n}_h,\tilde{n}_d$) of different local electronic configurations
    and (b) the interlayer singlet pairing amplitudes $(\Delta_x,\Delta_z)$.
    The results indicate that superconducting paring is triggered by increasing $J_H$, and the contribution is dominated by the interlayer pairing $\Delta_x$ of the $d_{x^2-y^2}$ orbital.}
    \label{fig:NumResultsJH}
\end{figure}

\begin{figure}[tp]
    \centering
    \includegraphics[width=0.98\linewidth]{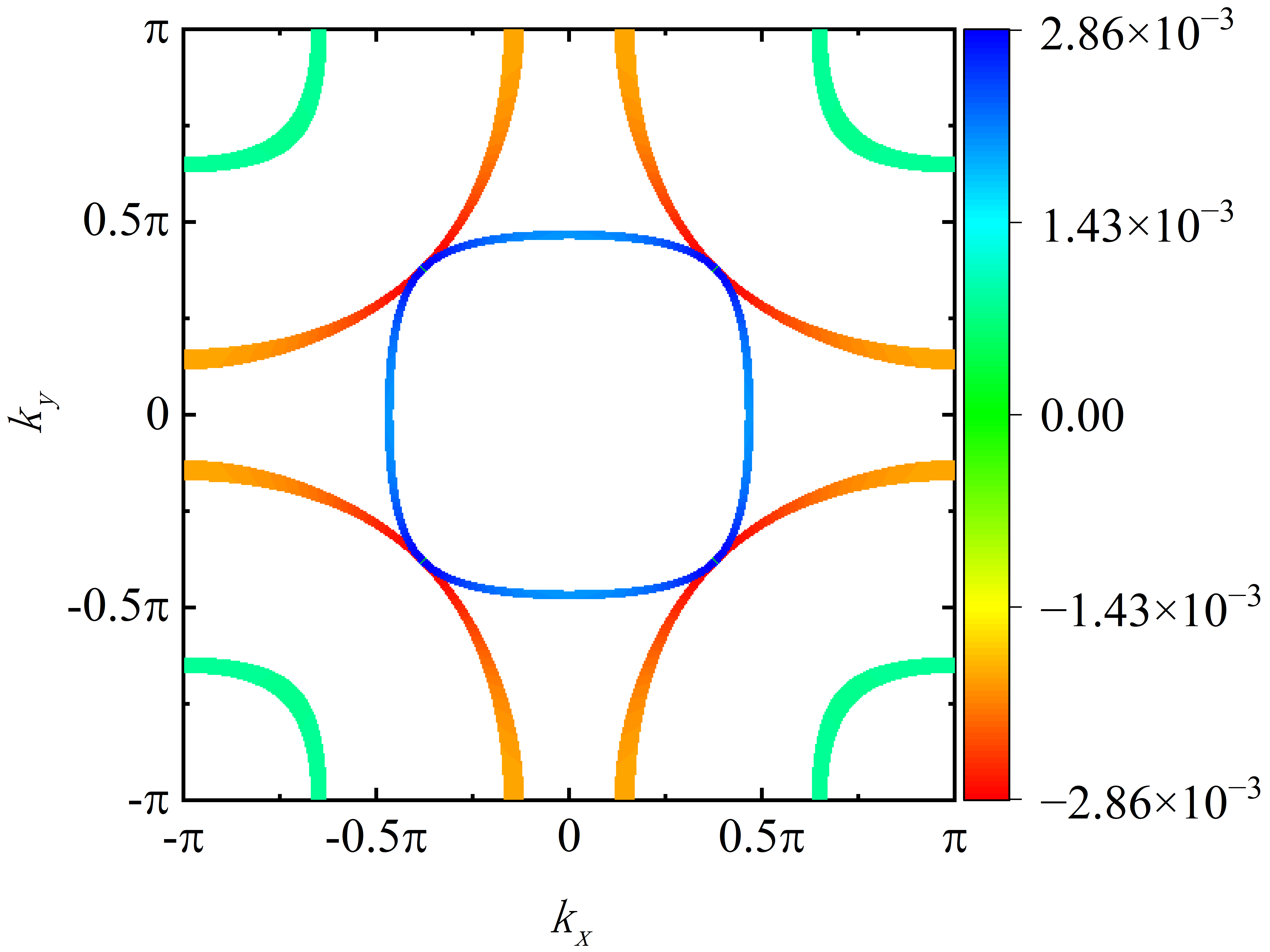}
    \caption{The projected superconducting gap distribution on the Fermi surface from the SPMF calculations. 
    }
    \label{fig:gapfunction}
\end{figure}

The calculated onsite occupancy numbers as a function of $J_H$ are shown in Fig.~\ref{fig:NumResultsJH} (a). 
Here, $\tilde{n}_d$ represents the onsite occupancy number for spin-$1$ state, while $\tilde{n}_z$ and $\tilde{n}_x$ correspond to the onsite occupancy numbers for the spin-$\frac{1}{2}$ states in the $d_{z^2}$ and $d_{x^2-y^2}$ orbitals, respectively.
Additionally, $\tilde{n}_h$ denotes the occupancy of the empty state. 

Fig.~\ref{fig:NumResultsJH} (a) reveals the physical figure that the Hund's coupling $J_H$ promotes the combination of a $d_{z^2}$ singlon and a $d_{x^2-y^2}$ singlon into a spin-$1$ triplet doublon and a double-hole state, while the hybridization induces the transformation of the $d_{z^2}$ singlons into the $d_{x^2-y^2}$ singlons, resulting in the decreased $\tilde{n}_z$, increased $\tilde{n}_h$ and $\tilde{n}_d$ and the almost unchanged $\tilde{n}_x$. 
This result is consistent with our analysis of low-energy subspace, which points out that the transformation of two spin-$\frac{1}{2}$ singlons into a spin-$1$ doublon and a double-hole state lifts the energy of $V-J_H$ (See Appendix \ref{app_A} for more details), a process that will be more likely to occur when $J_H$ increases. 
Furthermore, the noise behavior appearing in Fig.~\ref{fig:NumResultsJH} (a) results from the finite-size effect, which does not affect the critical physical figure captured.

In the atomic limit, there is one electron in the $d_{z^2}$-orbital, and half electron in the $d_{x^2-y^2}$-orbital. 
In the slave-particle picture outlined in Fig.~\ref{configurations}, the occupation numbers for the spin-1 state, the spin-$\frac{1}{2}$ state in the $d_{z^2}$-orbital are both 0.5. 
Then 
\begin{eqnarray}
\tilde{n}_d=\tilde{n}_z=0.5, \ \ \,
\tilde{n}_x=\tilde{n}_h=0.
\end{eqnarray}
However, the presence of intraorbital hopping and interorbital hybridization leads to significant deviations from these idealized atomic orbital occupancies.
A key factor influencing these deviations is Hund's coupling, $J_H$.
As $J_H$ increases, the onsite energy of the doublon states is systematically lowered, rendering these configuration energetically more favorable
Consequently, there is an enhanced tendency for spinons originating from the $3d_{x^2}$ and $3d_{z^2-y^2}$ orbitals to combine and form doublon-holon pairs.
This process manifests as an increase in the calculated average doublon ($\tilde{n}_d$) and ($\tilde{n}_h$) occupancies. 
It is noteworthy that while both $\tilde{n}_d$ and $\tilde{n}_h$ increase, their difference ($\tilde{n}_d-\tilde{n}_h$, related to the net doping) remains constant.

Simultaneously, the occupancy $\tilde{n}_z$ of singly occupied $3d_{z^2}$ orbitals decreases,
whereas the $3d_{x^2-y^2}$ orbital occupancy $\tilde{n}_x$ shows minimal variation. 
These observations indicate a shift in the balance between the spin-$1$ and spin-$\frac{1}{2}$ states as Hund's coupling strengthens. 
The sensitivity of the onsite occupancy numbers to $J_H$ reflects the delicate balance between intraorbital and interorbital interactions in determining the electronic ground state.

\begin{figure}[tp]
    \centering
    \includegraphics[width=1\linewidth]{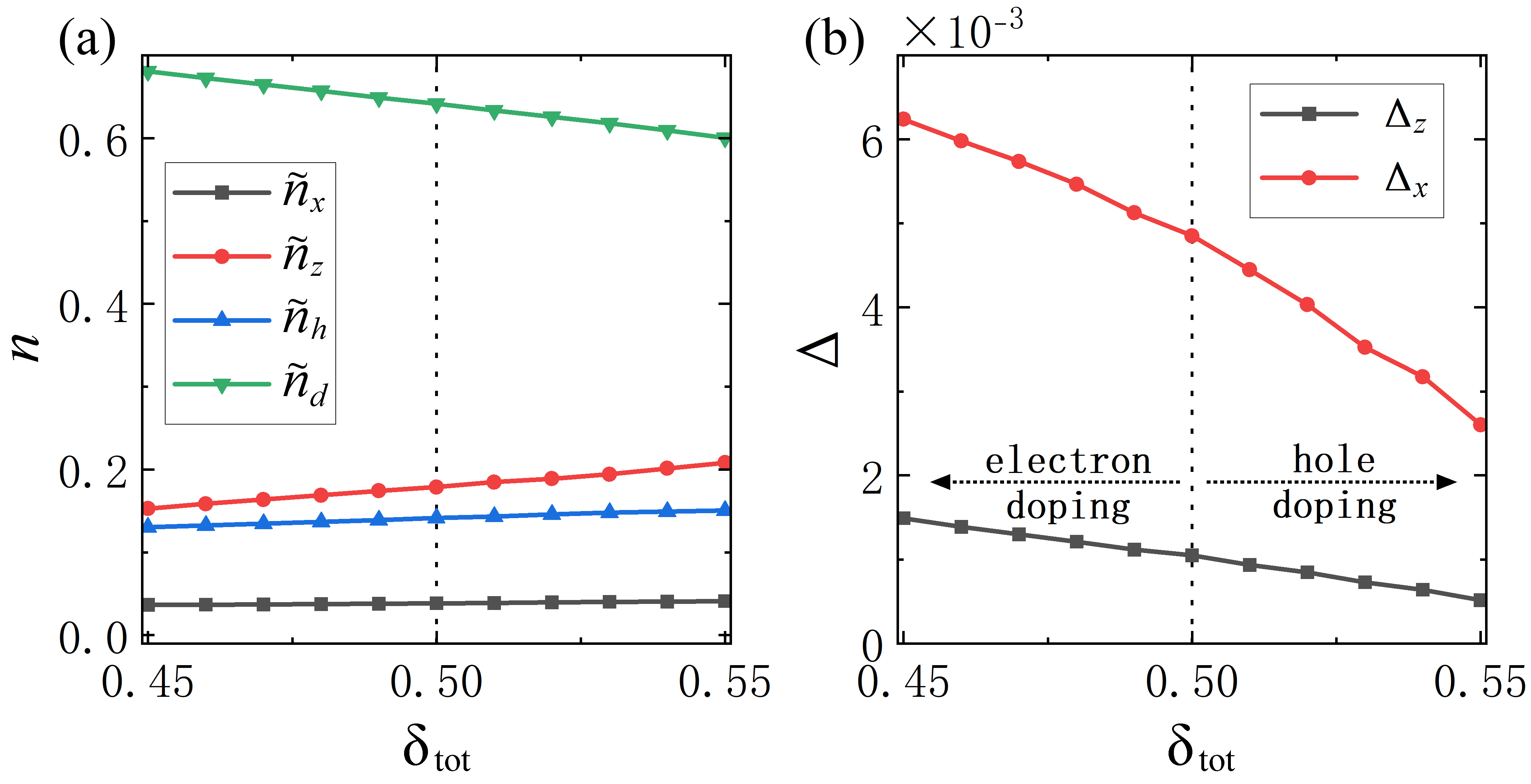}
    \caption{SPMF results as functions of the total hole density $\delta_{\text{tot}}$ (where $\delta_{\text{tot}}=0.5$ corresponds to the nominal $d^{7.5}$ configuration in La$_3$Ni$_2$O$_7$).
    (a) Onsite occupancy numbers for the different local states.
    (b) Interlayer singlet pairing amplitudes.
    A clear trend is observed where electron doping ($\delta_{\text{tot}}<0.5$, increased electron density) enhances the superconducting pairing amplitude (primarily the dominant $\Delta_x$), while hole doping ($\delta_{\text{tot}}>0.5$) has the opposite effect.} 
    \label{fig:NumResultsDelta}
\end{figure}

In real materials, SC is ultimately characterized by physical electron Cooper pairs.
The interlayer spin-$1$ singlet pairing introduced above constitutes both interlayer spin-$\frac{1}{2}$ singlet and triplet pairings for the $3d_{x^2-y^2}$ and $3d_{z^2}$ orbitals, while the singlet pairing channels are dominated.
The physical interlayer singlet pairing amplitudes for electrons in the $3d_{x^2-y^2}$ and $3d_{z^2}$ orbitals,
$\Delta_x$ and $\Delta_z$, can be understood as arising from the projection of the fundamental slave-particle pairing $\Delta_{0}$ onto these specific orbital channels. 
Crucially, the establishment of phase coherence for these physical electron pairs depends on the condensation of bosonic singlons. 
Consequently, the effective pairing amplitudes can be schematically expressed as $\Delta_x\sim \tilde{n}_z\langle\Delta_{0}\rangle$ and $\Delta_z\sim \tilde{n}_x\langle\Delta_{0}\rangle$.

An intriguing outcome of this formalism is that the phase coherence of the $3d_{x^2-y^2}$ interlayer pairing $\Delta_x$ is thus intrinsically linked to, and effectively controlled by, the occupancy and coherence of $3d_{z^2}$ singlons (and vice-versa for $\Delta_z)$.
This relationship is consistent with the underlying slave-particle construction: a local spin-$1$ doublon consists of one electron in the $3d_{x^2-y^2}$ orbital and and one in the $3d_{z^2}$ orbital.
If a hole is introduced into the $3d_{x^2-y^2}$ component of this doublon, namely, the $d_{x^2-y^2}$ electron is removed, the remnant on that site is a $3d_{z^2}$ singlon.
The dynamics of these singlons are therefore essential for the emergence of the physical superconducting state.

Fig.~\ref{fig:NumResultsJH} (b) shows the $J_H$-dependence of the pairing amplitude for interlayer singlet pairing in the $d_{z^2}$ and $d_{x^2-y^2}$ orbitals. 
As expected, pairing emerges rapidly with increasing $J_H$, with the interlayer pairing amplitude $\Delta_{x}$ dominating across the entire parameter range. 
This dominance of interlayer pairing originates from the robust interlayer spin-$1$ superexchange, which drives the formation of the rung pairs in this bilayer structure.
Meanwhile the $d_{z^2}$ pairing is suppressed by the small $d_{x^2-y^2}$ singlon number, leading to the dominance of the $d_{x^2-y^2}$ pairing.
As the main contributor to the superconducting pairing is the spin-$1$ doublon, when $J_H$ increases within a certain range, the increasing $\tilde{n}_d$ enhances $\left\langle\Delta_0\right\rangle$ promptly, and thus enhances $\Delta_z$ and $\Delta_x$ promptly.

To further analyze the superconducting pairing symmetry, we calculate the projected superconducting gap distribution on the Fermi surface in Fig.~\ref{fig:gapfunction}. The results indicates that the local interlayer pairing results in an extended $s$-wave order parameter, with a fully gapped Fermi surface.

In addition, the effect of doping on the electronic structure and pairing behavior is investigated, as illustrated in Fig.~\ref{fig:NumResultsDelta}.
The total hole density $\delta_{\text{tot}}=0.5$ corresponds to the nominal $3d^{7.5}$ electronic configuration of Ni in the La$_3$Ni$_2$O$_7$.
Hole doping ($\delta_{\text{tot}}>0.5$) and electron doping ($<0.5$) alter the occupancy numbers in distinct ways.
As shown in Fig.~\ref{fig:NumResultsDelta} (a), hole doping tends to reduce $\tilde{n}_d$, while increasing both $\tilde{n}_x$ and $\tilde{n}_z$.
Conversely, electron doping increases $\tilde{n}_d$, while decreasing $\tilde{n}_x$ and $\tilde{n}_z$.
Also, the difference between the doublon and holon are fixed, $\tilde{n}_d -\tilde{n}_h=\delta_{\text{tot}}$.

The superconducting pairing amplitude, shown in Fig.~\ref{fig:NumResultsDelta} (b), exhibits a clear dependence on doping, with the interlayer $d_{x^2-y^2}$ pairing amplitude $\Delta_{x}$ remaining dominant across the full doping range. 
As hole doping increases, the pairing amplitudes decrease, suggesting that hole doping weakens the overall superconducting state. 
In contrast, electron doping enhances the pairing amplitudes.

\section{Density Matrix Renormalization Group Study}
\label{sect:DMRG}
To further corroborate the physical picture derived above, we employ the state-of-art DMRG method \cite{white1993dmrg,weng1999dmrg} at zero temperature.

\begin{figure}[t!]
    \centering
    \includegraphics[width=0.6\linewidth]{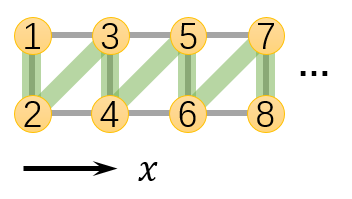}
    \caption{Illustration of the zigzag path used in the DMRG calculations for the $2\times 1\times L_x$ ladder.} 
    \label{ladder_zigzag}
\end{figure}

The DMRG calculations are performed using tensor libraries TensorKit \cite{jutho2024} and FiniteMPS \cite{li2024mps}, which provide an implementation of the required $\mathrm{U(1)}_\text{charge}\times \mathrm{SU(2)}_\text{spin}$ symmetries \cite{weichselbaum2012,weichselbaum2020}. 
Since the bond dimension $D$ is constrained by computational complexity, the DMRG simulation struggles to accurately solve two-dimensional (2D) systems.
In our work, we investigate the model on one-dimensional (1D) 
geometries of sizes $2\times 1\times 96$ and $2\times2\times 48$ with open boundary conditions in all directions. 
Although this setup deviates significantly from genuine bilayer 2D systems, we can still predict the existence and strength of order parameters in 2D systems by analyzing the decay behavior of the corresponding correlation functions in the 1D systems.
The matrix product state (MPS) representation is constructed using a zigzag path along the ladder, as illustrated in Fig.~\ref{ladder_zigzag}. 
We retain up to $D=10000$ $\mathrm{U(1)}_\text{charge}\times \mathrm{SU(2)}_\text{spin}$ multiplets in the DMRG simulations and a convergence threshold of $10^{-8}$ for the ground state energy.

To characterize the ground state properties, we numerically simulate the particle-number distribution and various correlation functions. 
The particle-number operator for orbital $\alpha$ at site $i$ (averaged over the two layers) is defined as:
\begin{equation}
n_{i\alpha}
= \frac{1}{2} \sum_{\mu\sigma} c^\dagger_{i\mu\alpha\sigma}c_{i\mu\alpha\sigma},
\end{equation}
where $c^\dagger_{i\mu\alpha\sigma}$ creates an electron in orbital $\alpha$ with spin-$\sigma$ at lattice site $i$ in layer $\mu$.
Its expectation value yields the particle-number distribution.
The single-particle correlation function for the combination of $d_{z^2}$ and $d_{x^2-y^2}$ orbitals is given by:
\begin{equation}
G(r) = \frac{1}{4} \sum_{\mu\sigma} \left\langle \left(c_{i\mu z\sigma}^\dagger+c^\dagger_{i\mu x\sigma}\right) \left(c_{j\mu z\sigma}+c_{j\mu x\sigma}\right) + \text{h.c.} \right\rangle,
\end{equation}
where $r=|i-j|$ is the distance between the sites $i$ and $j$. 
The charge density correlation function for the $\alpha$-orbital is defined as:
\begin{equation}
D_{\alpha}(r) = \left\langle n_{i\alpha} n_{j\alpha} \right\rangle - \left\langle n_{i\alpha} \right\rangle \left\langle n_{j\alpha} \right\rangle.
\end{equation}
The spin-spin correlation function averaged over layers is given by:
\begin{equation}
F(r) = \frac{1}{2} \sum_{\mu} \left\langle \bm{S}_{i\mu} \cdot \bm{S}_{j\mu} \right\rangle,
\end{equation}
where $\bm{S}_{i\mu}=\bm{S}_{i\mu z}+\bm{S}_{i\mu x}+\bm{S}_{i\mu d}$ is the total spin operator at site $i$ in layer $\mu$, 
containing contributions from $d_{z^2}$ singlons, $d_{x^2-y^2}$ singlons, and doublon states.

To characterize the superconducting tendencies, we investigate various pairing channels.
The interlayer ($\perp$) and intralayer ($\parallel$) pairing operators are given by:
\begin{equation}
\begin{aligned}
\Delta^{\perp\dag}_{i\alpha} 
=& \frac{1}{\sqrt{2}}\left(c^{\dag}_{it\alpha\uparrow}c^{\dag}_{ib\alpha\downarrow}-c^{\dag}_{it\alpha\downarrow}c^{\dag}_{ib\alpha\uparrow}\right), \\
\Delta^{\parallel\dag}_{i\mu\alpha} 
=& \frac{1}{\sqrt{2}}\left(c^{\dag}_{i\mu\alpha\uparrow}c^{\dag}_{i+1,\mu\alpha\downarrow}-c^{\dag}_{i\mu\alpha\downarrow}c^{\dag}_{i+1,\mu\alpha\uparrow}\right).
\end{aligned}
\end{equation}
The long-distance behavior of these pairings is characterized by their correlation functions:
\begin{equation}
\begin{aligned}
\Phi^{\perp}_{\alpha}(r) &= \left\langle \Delta^{\perp\dag}_{i\alpha} \Delta^{\perp}_{j\alpha} \right\rangle,\\
\Phi^{\parallel}_{\alpha}(r) &= \frac{1}{2}\sum_\mu \left\langle\Delta_{i\mu\alpha}^{\parallel\dag} \Delta^{\parallel}_{j\mu\alpha} \right\rangle.
\end{aligned}
\end{equation}
In a 1D system, pairing correlations in each channel typically decays algebraically as $r^{-K_\text{SC}}$, leading to quasi-long-ranged order.
Here, the decay power exponent $K_\text{SC}$ is related to the Luttinger parameter of the corresponding channel, and a value $K_{SC}<2$ generally indicates a divergence of superconductivity susceptibility in each channel.
The channel with the smallest value of $K_{SC}$ is the dominant one. 
Furthermore, due to the limitations of the $\mathrm{U(1)_\text{charge}\times SU(2)_\text{spin}}$ symmetry, we define the square root of the structure factors as the singlet pairing order parameters, 
\begin{equation}
\begin{aligned}
\langle\Delta^{\perp}_\alpha\rangle &= \sqrt{\frac{1}{N_b}\sum_{i,j}\left\langle \Delta^{\perp\dag}_{i\alpha} \Delta^{\perp}_{j\alpha} \right\rangle},\\
\langle\Delta^{\parallel}_\alpha\rangle &= \sqrt{\frac{1}{2N_b}\sum_{i,j,\mu}\left\langle \Delta^{\parallel\dag}_{i\mu\alpha} \Delta^{\parallel}_{j\mu\alpha} \right\rangle},
\end{aligned}
\end{equation}
where the sums over $i$, $j$ are restricted to $\frac{L_x}{4}\leq i,j\leq \frac{3L_x}{4}$, and $N_b$ is the number of contributing pairs in the sum.

Similarly, the triplet pairing operators (denoted by superscript $t$) are formulated as
\begin{equation}
\begin{aligned}
\Delta^{t,\perp\dag}_{i\alpha} 
=& c^{\dag}_{it\alpha\uparrow}c^{\dag}_{ib\alpha\uparrow} + c^{\dag}_{it\alpha\downarrow}c^{\dag}_{ib\alpha\downarrow} \\
&+ \frac{1}{\sqrt{2}}\left(c^{\dag}_{it\alpha\uparrow}c^{\dag}_{ib\alpha\downarrow} + c^{\dag}_{it\alpha\downarrow}c^{\dag}_{ib\alpha\uparrow}\right), \\
\Delta_{i\mu\alpha}^{t,\parallel\dag} 
=& c^{\dag}_{i\mu\alpha\uparrow}c^{\dag}_{i+1,\mu\alpha\uparrow} + c^{\dag}_{i\mu\alpha\downarrow}c^{\dag}_{i+1,\mu\alpha\downarrow} \\
&+ \frac{1}{\sqrt{2}}\left(c^{\dag}_{i\mu\alpha\uparrow}c^{\dag}_{i+1,\mu\alpha\downarrow} + c^{\dag}_{i\mu\alpha\downarrow}c^{\dag}_{i+1,\mu\alpha\uparrow}\right).
\end{aligned}
\end{equation}
The corresponding triplet pairing correlation functions are
\begin{equation}
\begin{aligned}
\Phi^{t,\perp}_{\alpha}(r) &= \left\langle \Delta^{t,\perp\dag}_{i\alpha} \Delta^{t,\perp}_{j\alpha} \right\rangle,\\
\Phi^{t,\parallel}_{\alpha}(r) &= \frac{1}{2}\sum_\mu 
\left\langle\Delta_{i\mu\alpha}^{t,\parallel\dag} \Delta^{t,\parallel}_{j\mu\alpha} \right\rangle.
\end{aligned}
\end{equation}

\begin{figure}[t!]
    \centering
    \includegraphics[width=1\linewidth]{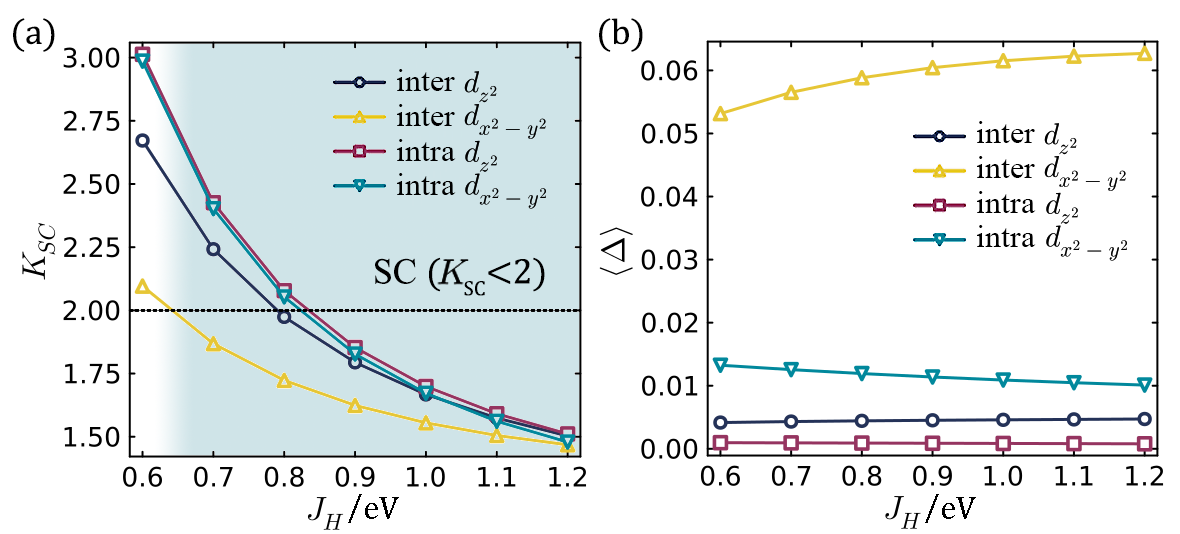}
    \caption{DMRG results as functions of Hund's coupling $J_H$. 
    (a) The decay power $K_\text{SC}$ of singlet pairing correlations $\Phi^{\perp}_{\alpha}(r)$ and $\Phi^{\parallel}_{\alpha}(r)$ in the $d_{z^2}$ and $d_{x^2-y^2}$ orbitals, respectively. 
    The legend \textit{inter} and \textit{intra} correspond to $\Phi^{\perp}_{\alpha}(r)$ and $\Phi^{\parallel}_{\alpha}(r)$ respectively. 
    Superconductivity ($K_{\text{SC}}<2$) is indicated for $J_H>0.7\text{ eV}$,
    dominated by interlayer $d_{x^2-y^2}$ pairing with the lowest $K_{\text{SC}}$.
    (b) The singlet pairing order parameters $\langle\Delta^{\perp}_\alpha\rangle$ and $\langle\Delta^{\parallel}_\alpha\rangle$.} 
    \label{JH}
\end{figure}

The impact of Hund's coupling on SC and particle distribution, as determined by the DMRG calculations, is presented in Fig.~\ref{JH}.
In Fig.~\ref{JH} (a), the emergence of the superconducting ($K_\text{SC}<2$) corresponds to the region where $J_H\gtrsim 0.7$ eV. 
Across the investigated range of $J_H$ from $0.6\text{ eV}$ to $1.2\text{ eV}$, 
interlayer singlet pairing in the $d_{x^2-y^2}$ orbital is consistently dominant, and the strength of this superconducting tendency increases monotonically with $J_H$. 
For $J_H \gtrsim 1$, the differences in $K_\text{SC}$ values among the various strong pairing channels become less significant, 
potentially due to proximity effects between competing or coexisting orders. 
To further substantiate the dominance of the interlayer $d_{x^2-y^2}$ orbital pairing,
the pairing order parameters $\langle\Delta^{\perp}_\alpha\rangle$ and $\langle\Delta^{\parallel}_\alpha\rangle$ are also calculated, as shown in Fig.~\ref{JH} (b).

\begin{figure}[t!]
    \centering
    \includegraphics[width=1\linewidth]{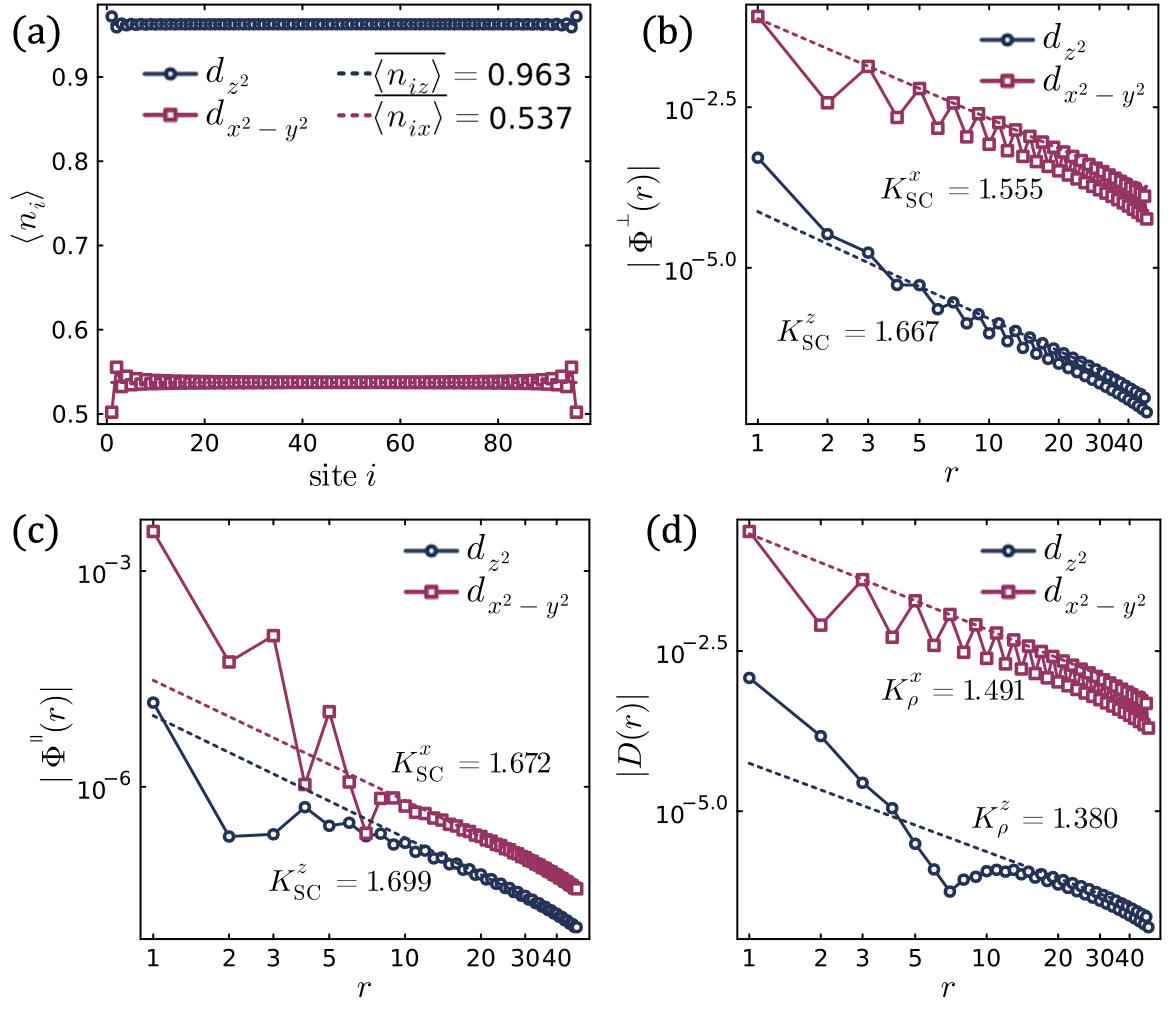}
    \caption{DMRG simulated results of the order parameters at $J_H=1$ eV. 
    (a) The electron densities in the two $E_g$ orbitals. 
    (b) The interlayer singlet pairing correlation functions of each orbital. 
    (c) The intralayer singlet pairing correlation functions. All of the interlayer and intralayer singlet pairing correlation functions follow an algebraic decay. 
    (d) The charge density correlation functions for each orbital, exhibiting algebraic decay.}
    \label{JH1_main}
\end{figure}

\begin{figure}[htbp]
    \centering 
    \includegraphics[width=1\linewidth]{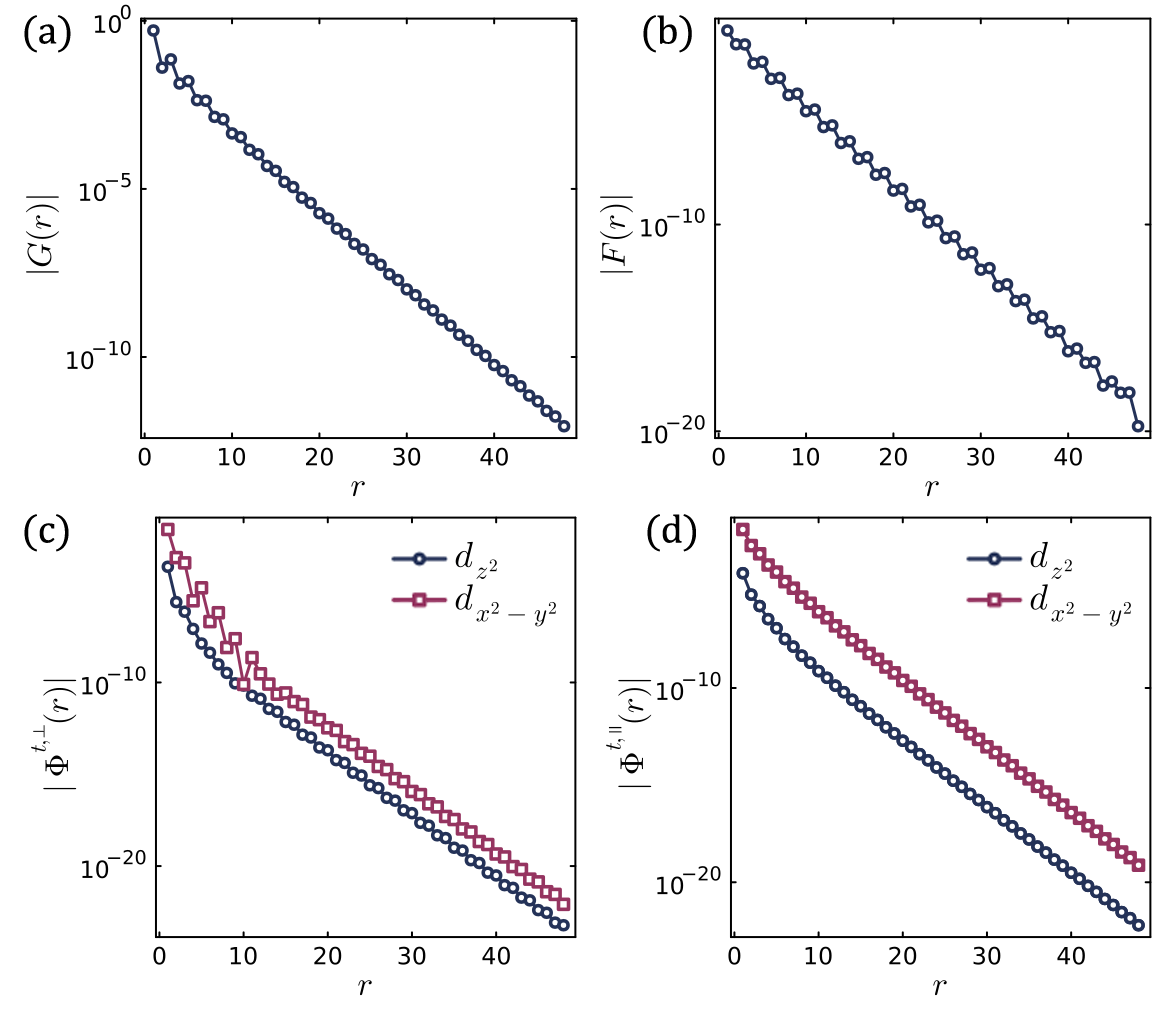}
    \caption{DMRG results of the order parameters exhibiting exponential decay behaviors, simulated at $J_H=1$ eV. (a) The single-particle correlation function $G(r)$. 
    (b) The spin-spin correlation function $F(r)$. 
    The interlayer (c) and intralayer (d) triplet pairing correlation functions across each orbital. }
    \label{JH1_minor}
\end{figure}

Focusing on the representative case of $J_H=1\mathrm{eV}$, additional detailed ground state properties are presented in Fig.~\ref{JH1_main} and \ref{JH1_minor}.
Fig.~\ref{JH1_main} (a) displays the particle number distribution in the two $E_g$ orbitals. 
The interlayer and intralayer singlet pairing correlation functions, shown in Fig.~\ref{JH1_main}(b) and (c) respectively, both exhibit clear algebraic decay. 
This behavior is characteristic of a Luther-Emery liquid, where spin excitations are gapped, allowing for dominant superconducting correlations \cite{luther1974}; 
also see below for the behavior of spin excitation. 
Furthermore, the charge density correlation function, depicted in Fig.~\ref{JH1_main} (d), also decays algebraically.
This latter finding suggests that the full two-dimensional system might exhibit a complex interplay or coexistence of SC and charge density wave ordering, with the interlayer $d_{x^2-y^2}$ orbital as the primary source of SC.

In addition, a typical 1D single-orbital Luther-Emery liquid has the property: $K_\text{SC}\times K_\rho=1$, while our simulation results show that $K^\alpha_\text{SC}\times K^\alpha_\rho$ range from $2.3$ to $2.6$. The deviation results from the combination of the strong inter-orbital interactions and the geometrical structure of the ladder, which allows the $2 \times 1 \times L_x$ two-orbital system to be physically mapped to a four-orbital system with complex inter-orbital interactions, deviating from a strict 1D single-orbital system.
 
As shown in Fig.~\ref{JH1_minor} (a), the single-particle correlation function $G(r)$ exhibits exponential decay, indicating the gapped charge degree of freedom.
The spin-spin correlation function $F(r)$, presented in Fig.~\ref{JH1_minor} (b), also exhibit exponential decay, confirming the presence of a spin gap, consistent with the Luther-Emery liquid picture. 
Of course, the spin gap is a result on the ladder, and the existence of magnetic order in the bulk bilayer square lattice remains an open question.
We also computed the interlayer and intralayer triplet pairing correlation functions, shown in Fig.~\ref{JH1_minor} (c) and (d). 
Both of these decay exponentially with distance. 
These results suggest the absence of the corresponding order parameters in the pressurized La$_3$Ni$_2$O$_7$ system, leaving singlet interlayer $3d_{x^2-y^2}$-orbital SC pairing as the dominant low-energy phenomena.

More DMRG results on the convergence of the pairing correlation functions and on the lattice of size $2\times2\times48$ are shown in Appendix \ref{app_D}.

\section{Conclusion and Discussion}
\label{sect:Con}
In this work, we propose and study a bilayer two-orbital model for the superconducting La$_3$Ni$_2$O$_7$ under high pressure, which fully accounts for the effects of the onsite triplet doublons. 
The results presented here provide significant insight into the interplay between Hund's coupling, electron correlation, and SC in the bilayer nickelate material. 
The underlying mother state is formed within a strong-coupling regime, characterized by robust Hund's coupling that leads to spin-triplet configurations for the two $E_g$ orbitals in the Ni $3d^8$ electronic state. 
The strong interlayer superexchange interactions promote the formation of interlayer spin-$1$ singlet states, resulting in an interlayer VBS structure. 
When the system approaches the physically relevant Ni $d^{7.5}$ valence in pressurized La$_3$Ni$_2$O$_7$, these spin-$1$ singlets gradually evolve into spin-$\frac{1}{2}$ singlets primarily involving the $d_{z^2}$ orbitals. 
Doping introduces additional holes, which mediate phase coherence among these singlet bonds, eventually giving rise to SC.

The central finding of our work is the dominance of interlayer spinon singlet pairing across both the Hund's coupling ($J_H$) and doping regimes, suggesting that interlayer superexchange plays a pivotal role in stabilizing the superconducting state. 
This distinction points to the novelty of the superconducting mechanism in La$_3$Ni$_2$O$_7$, where the bilayer architecture, combined with the strong interlayer superexchange, offers a platform to explore unconventional SC. 
This structure fosters unique pairing mechanisms that rely heavily on the coupling between layers, distinguishing this system from the in-plane-dominated SC typically seen in cuprates and other layered systems.

Furthermore, our findings reveal that the coexistence of spin-$1$ and spin-$\frac{1}{2}$ singlets creates an unusual quantum entangled state that diverges from traditional models of SC. 
This entanglement between different spin channels suggests that both spin-$1$ and spin-$\frac{1}{2}$ singlets contribute to the overall pairing mechanism, raising fundamental questions about the precise nature of the superconducting phase in bilayer nickelates. 
The transition from a spin-$1$ valence bond solid to a mixed spin-$1$/spin-$\frac{1}{2}$ state, driven by doping, introduces the possibility of a rich and complex phase diagram. 
This phase diagram may exhibit a variety of superconducting properties, 
which is left for future works and may be experimentally clarified.

\textbf{Note added}: 
When finalizing the manuscript, we become aware of a related preprint \cite{kaneko2025tj} that also considers an effective bilayer two-orbital $t$-$J$ model incorporating strong Hund's coupling, similar to the model presented here. 
Our study builds upon such a model framework by performing detailed theoretical and numerical analyses (SPMF and DMRG) to investigate the nature of the SC pairing mechanism and symmetry.

\section*{Data availability}
All data are displayed in the main text and Appendix. 
Data within this paper are available from the corresponding author upon request. 

\section*{Acknowledgments}
We are grateful to the stimulating discussions with Wei Li, Xing-Zhou Qu and Jialin Chen. 
C.W. is supported by the National Natural Science Foundation of China under the Grants No. 12234016 and No. 12174317. 
F.Y. is supported by the National Natural Science Foundation of China under the Grants No. 12074031. 
C.L. is supported by the National Natural Science Foundation of China under the Grants No. 12304180. 
This work has been supported by the New Cornerstone Science Foundation.

\appendix
\renewcommand{\theequation}{S\arabic{equation}}
\renewcommand{\thefigure}{A\arabic{figure}}
\setcounter{equation}{0}
\setcounter{figure}{0}


\section{Details on the bilayer $t-J-J_H$ model}\label{app_A}
\begin{figure*}[t!]
    \centering
    \includegraphics[width=1\linewidth]{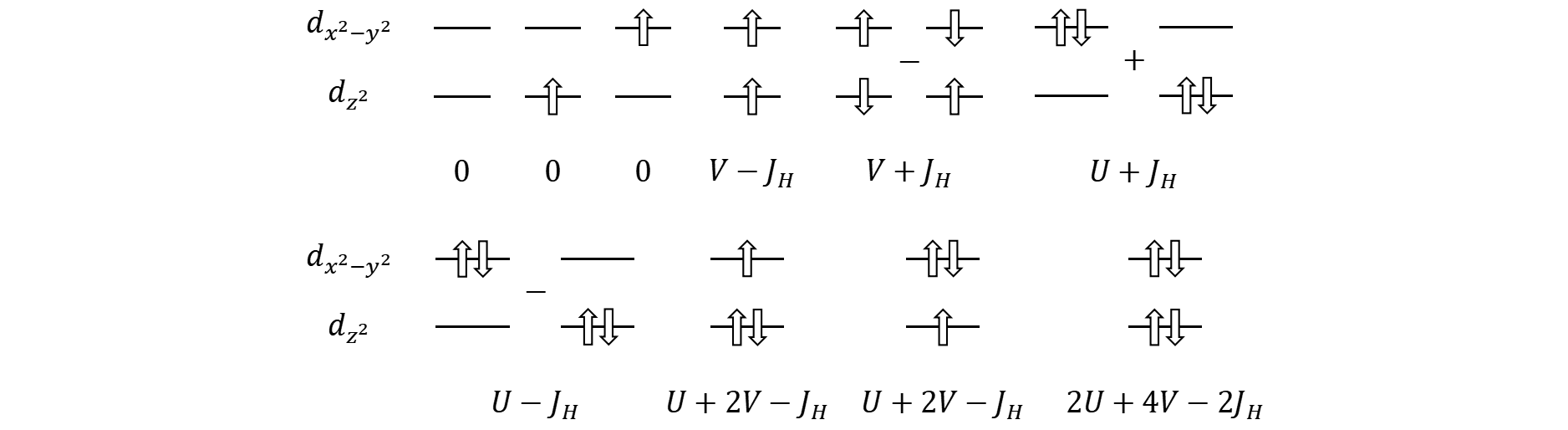}
    \caption{The zero-order energies and the corresponding typical two-orbital configurations.}
    \label{conf}
\end{figure*}

In the second-order perturbation theory, the electron interactions are treated as zero-order terms, while the kinetic terms are considered as first-order perturbations. The zero-order energies of some typical two-orbital configurations are shown in Fig.~\ref{conf}.
Near the natural filling of La$_3$Ni$_2$O$_7$, the local Hilbert space at each site is constrained to the eight configurations shown in Fig.~\ref{configurations} in the strong correlation limit. 
The hole configuration (neither orbital is occupied) is important with two main considerations in mind: (i) It avoids unreasonable ferromagnetic superexchange terms, which takes the form $-\frac{t^2}{V-J_H+\Delta\epsilon}$ in certain components of $J$; (ii) In the mean-field approach, the reservation of the hole configuration avoids momentum locking at zero, as the two orbitals are independent of each other without it.
In addition, its reservation resulted in the zero-order terms in Hamiltonian (\ref{eq:Int}) being partially retained in the resulting Hamiltonian (\ref{eq:Ham}).

The effects of the discarded high-energy subspace are captured through the superexchange terms, whose strength is determined by second-order perturbation theory. 
The intralayer superexchange parameters of the bilayer $t-J-J_H$ model are related to the parameters of the original two-band Hubbard model through the following equations
\begin{widetext}
\begin{equation}
\begin{aligned}
J^{\parallel}_{zz} = &\ \frac{4t^{\parallel2}_{zz}}{U}+\frac{2t^{\parallel2}_{zx}}{V+J_H+\epsilon_x-\epsilon_z},\qquad 
J^{\parallel}_{xx} = \frac{4t^{\parallel2}_{xx}}{U}+\frac{2t^{\parallel2}_{zx}}{V+J_H+\epsilon_z-\epsilon_x},\\
J^{\parallel}_{dd} = &\ \frac{t^{\parallel2}_{zz}+t^{\parallel2}_{xx}}{U+J_H} + t^{\parallel2}_{zx}\left[\frac{1}{U+J_H+\epsilon_z-\epsilon_x}+\frac{1}{U+J_H+\epsilon_x-\epsilon_z}\right],\\
J^{\parallel}_{zx} = &\ \frac{t^{\parallel2}_{zz}+t^{\parallel2}_{xx}}{V+J_H} + 2t^{\parallel2}_{zx}\left[\frac{1}{U+\epsilon_z-\epsilon_x}+\frac{1}{U+\epsilon_x-\epsilon_z}\right],\\
J^{\parallel}_{zd} = &\ t^{\parallel2}_{zz}\left[\frac{1}{U+V}+\frac{1}{U-V+J_H}\right] + \frac{t^{\parallel2}_{xx}}{4J_H} + \frac{t^{\parallel2}_{zx}}{2}\frac{1}{2J_H+\epsilon_x-\epsilon_z} \\
&+ t^{\parallel2}_{zx}\left[\frac{1}{U+V+\epsilon_x-\epsilon_z}+\frac{1}{U-V+J_H+\epsilon_z-\epsilon_x}\right],\\
J^{\parallel}_{xd} = &\ t^{\parallel2}_{xx}\left[\frac{1}{U+V}+\frac{1}{U-V+J_H}\right] + \frac{t^{\parallel2}_{zz}}{4J_H} + \frac{t^{\parallel2}_{zx}}{2}\frac{1}{2J_H+\epsilon_z-\epsilon_x} \\
&+ t^{\parallel2}_{zx}\left[\frac{1}{U+V+\epsilon_z-\epsilon_x}+\frac{1}{U-V+J_H+\epsilon_x-\epsilon_z}\right],
\end{aligned}
\end{equation}
\end{widetext}
and the interlayer parameters are as follows
\begin{equation}
\begin{aligned}
J^{\perp}_{zz} &= \frac{4t^{\perp2}_{zz}}{U},\qquad 
J^{\perp}_{dd} = \frac{t^{\perp2}_{zz}}{U+J_H},\\
J^{\perp}_{zx} &= \frac{t^{\perp2}_{zz}}{V+J_H},\qquad
J^{\perp}_{xd} = \frac{t^{\perp2}_{zz}}{4J_H},\\
J^{\perp}_{zd} &= t^{\perp2}_{zz}\left[\frac{1}{U+V}+\frac{1}{U-V+J_H}\right].
\end{aligned}
\end{equation}

\begin{figure}[htbp]
    \centering
    \includegraphics[width=1\linewidth]{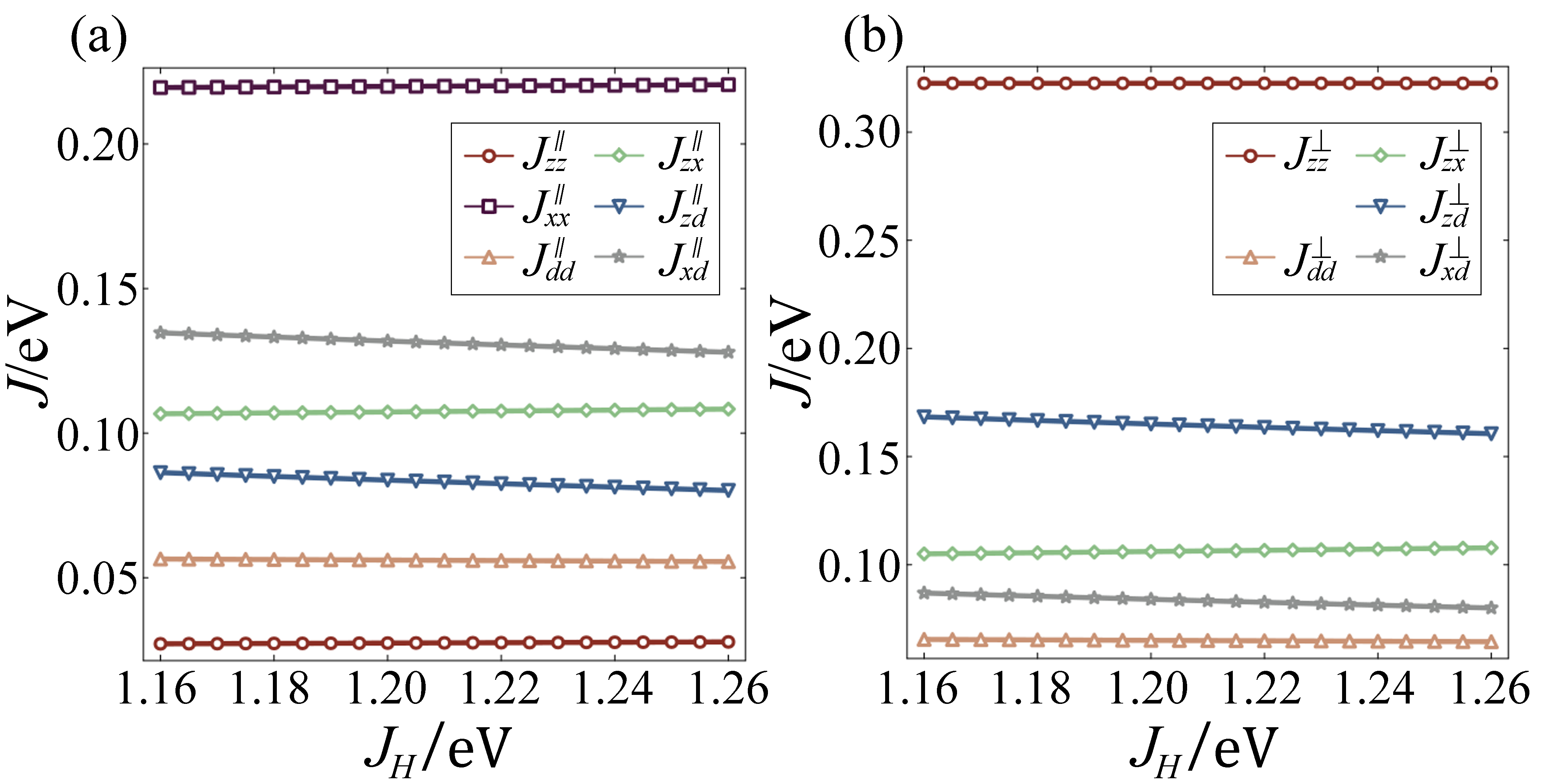}
    \caption{(a) The intralayer superexchange parameters $J^\parallel$ and (b) the interlayer superexchange parameters $J^\perp$ as functions of Hund's coupling $J_H$. }
    \label{fig:J_JH}
\end{figure}

To visually show the trend of these parameters with $J_H$, we show the changes in these parameters within the range set in the SPMF method in Fig.\ref{fig:J_JH}.

Fig.~\ref{fig:J_JH} (a) shows how the intralayer superexchange parameters $J^\parallel$ vary with $J_H$, exhibiting that $J^\parallel_{zz}$, $J^\parallel_{zx}$ and the dominant parameter $J^\parallel_{xx}$ are positively correlated with $J_H$, while $J^\parallel_{zd}$, $J^\parallel_{xd}$ and $J^\parallel_{dd}$ are opposite. 
Similarly, the trend of the interlayer superexchange parameters $J^\perp$ with $J_H$ is shown in Fig.~\ref{fig:J_JH} (b), indicating that $J^\perp_{zx}$ is positively correlated with $J_H$, while $J^\parallel_{zd}$, $J^\parallel_{xd}$ and $J^\parallel_{dd}$ are opposite and the dominant parameter $J^\perp_{zz}$ is independent of $J_H$.

\section{Details of the slave-particle theory} \label{app_SPMF}
In the slave-particle mean-field approach, the spin-$\frac{1}{2}$ exchange interaction between two nearest-neighbor sites, 
labeled $1$ and $2$, can be conventionally decoupled into singlet hopping and pairing channels:
\begin{align*}
\bm{S}_1\cdot \bm{S}_2
&=-\frac{3}{8} \chi_{12}^{\dagger} \chi_{12}
-\frac{3}{8} \Delta_{12}^{\dagger} \Delta_{12},
\end{align*}
where hopping $\chi$ and pairing $\Delta$ are given as,
\begin{align*}
\chi_{12}
=b_{1\uparrow}^{\dagger} b_{2\uparrow}
+b_{1\downarrow}^{\dagger} b_{2\downarrow},\quad
\Delta_{12}
=b_{1\downarrow} b_{2\uparrow }
-b_{1\uparrow} b_{2\downarrow}.
\end{align*}
Here, the labels $1$ and $2$ are composite indices. 
For the interlayer $d_{z^2}$ spin-$\frac{1}{2}$ interaction $J_{zz}^{\perp} \bm{S}_{i t z}\cdot\bm{S}_{i b z}$, these indices represent $(i, t, z)$ and $(i, b, z)$ respectively, denoting site $i$, top ($t$) or bottom ($b$) layer, and the $d_{z^2}$ orbital (labeled $z$). 
For the intralayer $d_{x^2-y^2}$ spin-$\frac{1}{2}$ interaction $J_{xx}^{\parallel} \bm{S}_{i\mu x}\cdot\bm{S}_{j\mu x}$, they represent $(i, \mu, x)$ and $(j, \mu, x)$, indicating sites $i$ and $j$ within the same layer $\mu$ and involving the $d_{x^2-y^2}$ orbital (labeled $x$). 
The operators $b_{s\sigma}$ ($b_{s\sigma}^\dagger$) annihilate (create) a slave boson for spin $\sigma$ at composite site $s$.

Similarly, the interlayer spin-$1$ interaction $J_{dd}^{\perp} \bm{S}_{itd}\cdot\bm{S}_{ibd}$, involving $E_g$ spin-$1$ moments on the top ($t$) and bottom ($b$) layers at site $i$, can also be decoupled into singlet channels.
For this spin-$1$ exchange interaction, the situation is more involved. 
The spin state of two individual spin $1$ can be denoted by $|m_1, m_2\rangle$, where $m_1, m_2 \in \{+1, 0, -1\}$ are the $z$-components of their respective spins. 
The combined two-spin states, denoted by $|J,M\rangle$ (where $J$ is the total spin and $M$ is its $z$-component), can be decomposed using Clebsch-Gordan coefficients into total spin-$0$ (singlet), spin-$1$ channel and spin-$2$ channels.
The total spin singlet $|J=0,M=0\rangle$ for two spin-$1$ entities, analogous to the spin-$\frac{1}{2}$ singlet state, is given by,
\begin{align*}
|\underset{J}{0},\underset{M}{0}\rangle =&\frac{1}{\sqrt{3}}\Big(|1,-1\rangle
-|0,0\rangle +|-1,1\rangle\Big).	
\end{align*}
We define the total spin-singlet hopping and pairing operators using fermionic $f$-particles,
\begin{align*}
\Delta_{0}^{\dagger} &= \frac{1}{\sqrt{3}}
\Big( f_{it,+1}^{\dagger} f_{id,-1}^{\dagger}
- f_{it,0}^{\dagger} f_{id,0}^{\dagger}
+ f_{it,-1}^{\dagger} f_{id,+1}^{\dagger} \Big), \\
\chi_{0}^{\dagger} &= \frac{1}{\sqrt{3}}
\Big( f_{it,+1}^{\dagger} f_{id,+1}
+ f_{it,0}^{\dagger} f_{id,0}
+ f_{it,-1}^{\dagger} f_{id,-1} \Big).
\end{align*}
The mean-field decoupling of the spin-$1$ exchange interaction, when projected onto these total spin-singlet channels, is then given by:
\begin{align*}
{\bm{S}}_{it} \cdot {\bm{S}}_{id}
= -2 \Delta_{0}^{\dagger} \Delta_{0}
-2 \chi_{0}^{\dagger} \chi_{0}.
\end{align*}
In this decomposition, contributions from non-singlet channels (i.e., $J=1$ and $J=2$) have been omitted, focusing on the formation of local singlets.

As highlighted in the main text, SC pairing can be conceptually understood as a quantum superposition of an ``empty" electronic state and a ``paired" electronic state. 
Focusing specifically on the ``paired" state where four electrons occupy the $E_g$ orbitals across a bilayer rung, a distinct pairing mechanism emerges. 
Initially, within each Ni ion, the intra-atomic Hund's rule dictates that the two $E_g$ electrons (one in the $d_{x^2-y^2}$ orbital and one in the $d_{z^2}$ orbital) will align their spins, forming a local spin-$1$ triplet state on each layer. 
Subsequently, these two local spin-$1$ moments, residing on the two layers of the rung, couple AFM via superexchange interaction, resulting in an interlayer spin-$0$ singlet state, as shown in the above $|\text{spin-1 singlet}\rangle\equiv |\underset{J}{0},\underset{M}{0}\rangle$. 
Critically, this spin-$1$ mediated singlet configuration cannot be simply factorized into a direct product of two independent spin-$\frac{1}{2}$ singlet configurations, signifying a notable deviation from the conventional BCS theory of Cooper pairing.

In contrast, conventional electron singlet pairing in such a system would typically involve a spin configuration composed of two independent interlayer spin singlets, one formed by the $3d_{x^2-y^2}$ orbitals on the two layers and the other by the $3d_{z^2}$ orbitals. 
It's important to recognize that these two total spin-$0$ configurations – the spin-$1$ mediated singlet and the direct product of two spin-$\frac{1}{2}$ singlets – are not orthogonal. 
Mathematically, these configurations can be rearranged into two orthogonal total spin-$0$ states, reflecting the two distinct pathways to form a total spin-$0$ singlet from four spin-$\frac{1}{2}$ particles, as illustrated by the Clebsch-Gordan series decomposition $\frac{1}{2}\otimes\frac{1}{2}\otimes\frac{1}{2}\otimes\frac{1}{2}=(0\oplus 1) \otimes (0\oplus 1)= 0\oplus 0 \oplus 1 \oplus 1 \oplus 1 \oplus 2$. 
To further elucidate the interlayer $3d_{x^2-y^2}$ singlet component originating from this spin-$1$ configuration, we acknowledge that the spin-$1$ state on each layer is a triplet formed by the combination of the $d_{x^2-y^2}$ and $d_{z^2}$ electron spins, and the interlayer total spin singlet arises from the coupling of these two triplets.
Projecting on the interlayer $3d_{x^2-y^2}$ spin-$\frac{1}{2}$ singlet channel $|3d_{x^2-y^2}\text{ singlet}\rangle$, we can have $\langle 3d_{x^2-y^2}\text{ singlet}|\text{spin-1 singlet}\rangle=\frac{\sqrt{3}}{2}$.
In this sense, the physical $3d_{x^2-y^2}$ electron interlayer singlet pairing can be roughly given by,
\begin{align*}
\Delta_x \propto \tilde{n}_z \langle \Delta_{0}\rangle,
\end{align*}
where $\tilde{n}_z$ arising from the holon condensation, leading to the phase coherence of the pairing.

\begin{figure}[htbp]
    \centering
    \includegraphics[width=1\linewidth]{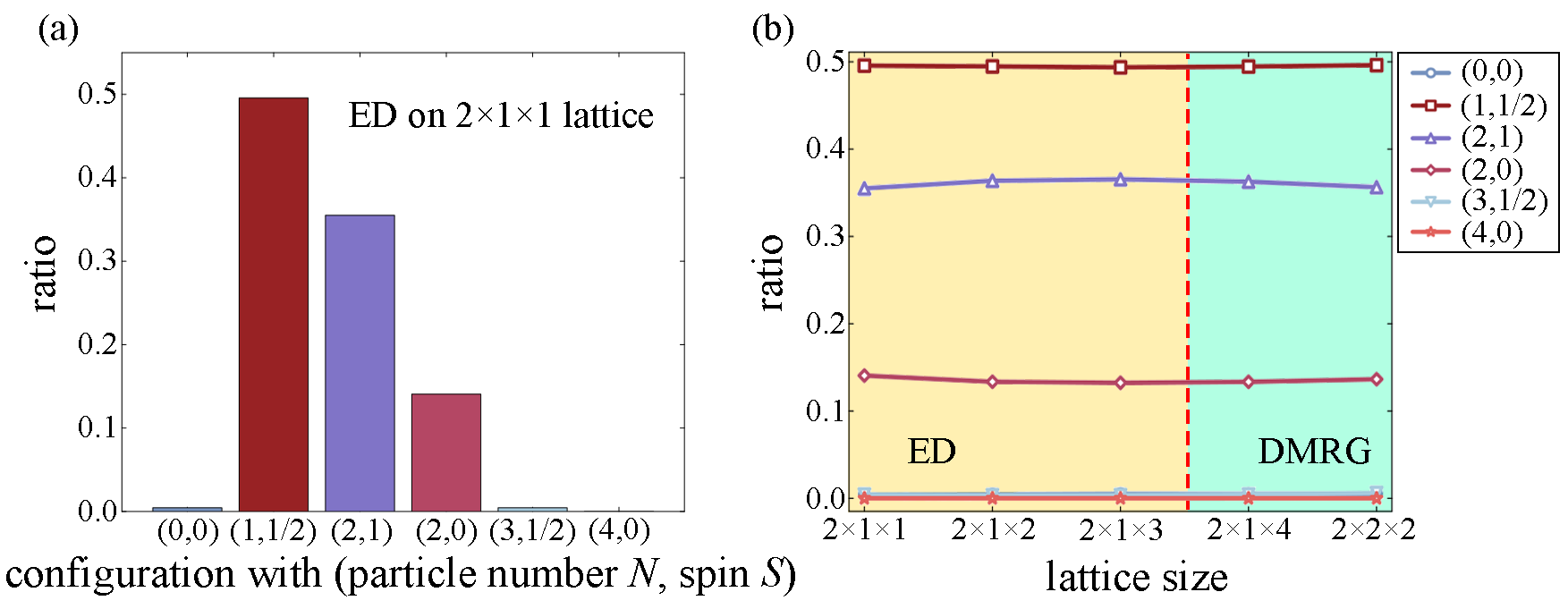}
    \caption{(a) The ratios of configurations with total particle number $N$ and total spin $S$ in the ground state of the two-orbital Hubbard model on a $2\times1\times1$ lattice. (b) The ratios on lattices of different sizes.}
    \label{fig:project}
\end{figure}

\begin{figure*}[t!]
    \centering
    \includegraphics[width=1\linewidth]{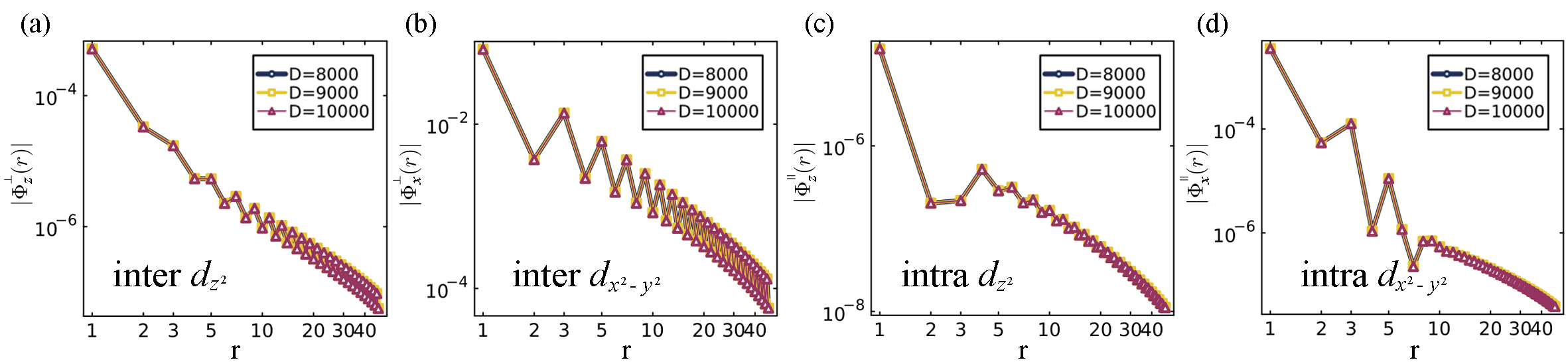}
    \caption{The singlet pairing correlation functions calculated by DMRG with different bond dimensions $D$ for (a) interlayer pairing in $d_{z^2}$ orbitals, (b) interlayer pairing in $d_{x^2-y^2}$ orbitals, (c) intralayer pairing in $d_{z^2}$ orbitals and (d) intralayer pairing in $d_{x^2-y^2}$ orbitals.}
    \label{fig:conv}
\end{figure*}

\section{Results on the bilayer two-orbital Hubbard model}\label{app_C}

Fig. \ref{configurations} in the main text shows the eight configurations constituting the low-energy effective local Hilbert space of our bilayer $t-J-J_H$ model. 
As analyzed in the main text and in Appendix \ref{app_A}, the four spin-$\frac{1}{2}$ singlons with total particle number $N=1$ and total spin $S=\frac{1}{2}$ (i.e. the configurations $\textcircled{2}\sim\textcircled{5}$ in Fig. \ref{configurations}) and the three spin-$1$ triplet doublons  with $N=2$ and $S=1$ (i.e. the configurations $\textcircled{6}\sim\textcircled{8}$ in Fig. \ref{configurations}) are the true low-energy configurations with $U$, $V$ and $J_H$ as zero-order terms. 
The additional hole configuration with $N=0$ and $S=0$ (i.e. the configuration $\textcircled{1}$ in Fig. \ref{configurations}) is retained due to the anomalous behavior of partial superexchange parameters and numerical necessity, which introduces the residual zero-order terms in Eq. (\ref{eq:Ham}) of the main text.

In this section, we solve the bilayer two-orbital Hubbard model combining Eqs. (\ref{eq:kinetic}) and (\ref{eq:Int}) using exact diagonalization and density matrix renormalization group methods, to demonstrate the validity of the perturbation theory by quantifying the ratios of the configurations $\textcircled{2}\sim\textcircled{8}$ in Fig. \ref{configurations} in the ground state.

The results are shown in Fig. \ref{fig:project}.
Fig. \ref{fig:project} (a) shows that the ratios of the configurations with different $(N, S)$ in the ground state of the bilayer two-orbital Hubbard model with the same parameter set of our bilayer $t$-$J$-$J_H$ model ($J_H = 1$) on a $2\times1\times1$ lattice. 
Fig. \ref{fig:project} (b) exhibits the results on the lattices of different sizes, where we apply the exact diagonalization method on $2\times1\times L_x$ lattices where $L_x=1,2,3$, and the density matrix renormalization group method is employed on $2\times1\times4$ and $2\times2\times2$ lattices. 
The results indicate that the spin-$\frac{1}{2}$ singlon and spin-$1$ triplet doublon are significantly dominant with the total ratio of about $0.87$, exactly consistent with our analysis based on the perturbation theory.

\section{More DMRG results on the bilayer $t-J-J_H$ model}\label{app_D}

In this section, we provide more DMRG results, in particular on convergence verification and on the lattice of size $2\times2\times48$.

Fig. \ref{fig:conv} exhibits the singlet pairing correlation functions calculated by DMRG with different bond dimensions $D$ for (a) interlayer pairing in $d_{z^2}$ orbitals, (b) interlayer pairing in $d_{x^2-y^2}$ orbitals, (c) intralayer pairing in $d_{z^2}$ orbitals and (d) intralayer pairing in $d_{x^2-y^2}$ orbitals. The results show that these pairing correlation functions highly overlap in the bond dimensions $D=8000$, $9000$ and $10000$, indicating that our DMRG results have converged with $D=10000$.

\begin{figure}[htbp]
    \centering
    \includegraphics[width=1\linewidth]{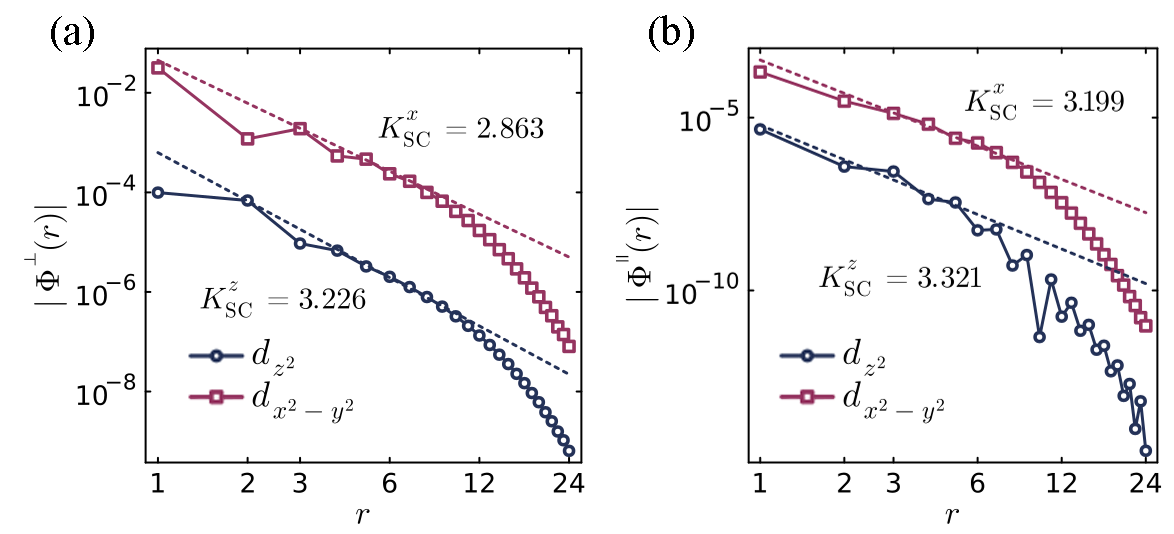}
    \caption{The DMRG results on a $2\times2\times48$ lattice. (a) The interlayer singlet pairing correlation functions. (b) The intralayer singlet pairing correlation functions.}
    \label{fig:DMRGonTube}
\end{figure}

Furthermore, we also solve the bilayer $t-J-J_H$ model by DMRG on a $2\times2\times48$ lattice with $J_H=1$ and the same set of other parameters on the $2\times1\times96$ ladder, retaining up to $D=10000$ $\mathrm{U(1)}_\text{charge}\times \mathrm{SU(2)}_\text{spin}$ multiplets as well. 
The new results are shown in Fig. \ref{fig:DMRGonTube}. 
Fig. \ref{fig:DMRGonTube} (a) exhibits the interlayer singlet pairing correlation functions, while Fig. \ref{fig:DMRGonTube}  (b) shows the intralayer one.
Although arithmetic constraints prevent us from extending $D$ enough to observe the algebraic decay of the pairing correlation functions over the full distance we examine, the dominance of the interlayer $d_{x^2-y^2}$ pairing can still be determined from the dominant amplitude and the slowest algebraic decay rate fitted in a small range, which is qualitatively consistent with our SPMF results on a two-dimension bilayer square lattice and the DMRG results on the $2\times1\times96$ ladder.
In addition, the result in $2\times2\times64$ lattice that the SC exponents are larger than 2, different from the ones in $2\times1\times128$, is likely due to the finite size effect. We predict that our two-orbital model may also support SC exponents smaller than 2 in a lattice with width equal to or larger than 3, which implies the existence of SC in exact 2D systems, as our SPMF results show.

\bibliography{reference}

@article{lu2023bilayertJ,
  title = {Interlayer-Coupling-Driven High-Temperature Superconductivity in {L}a$_3${N}i$_2${O}$_7$ under Pressure},
  author = {Lu, Chen and Pan, Zhiming and Yang, Fan and Wu, Congjun},
  journal = {Phys. Rev. Lett.},
  volume = {132},
  issue = {14},
  pages = {146002},
  numpages = {6},
  year = {2024},
  month = {Apr},
  publisher = {American Physical Society},
  doi = {10.1103/PhysRevLett.132.146002},
  url = {https://link.aps.org/doi/10.1103/PhysRevLett.132.146002}
}

@article{kotliar1988,
  title = {Superexchange mechanism and $d$-wave superconductivity},
  author = {Kotliar, Gabriel and Liu, Jialin},
  journal = {Phys. Rev. B},
  volume = {38},
  issue = {7},
  pages = {5142--5145},
  numpages = {0},
  year = {1988},
  month = {Sep},
  publisher = {American Physical Society},
  doi = {10.1103/PhysRevB.38.5142},
  url = {https://link.aps.org/doi/10.1103/PhysRevB.38.5142}
}

@article{su2024strongly,
  title={Strongly Anisotropic Charge Dynamics in {L}a$_3${N}i$_2${O}$_7$ with Coherent-to-Incoherent Crossover of Interlayer Charge Dynamics},
  author={Su, Bo and Huang, Chaoxin and Zhao, Jianzhou and Huo, Mengwu and Luo, Jianlin and Wang, Meng and Chen, Zhi-Guo},
  journal={arXiv:2411.10786},
  year={2024},
  url={https://arxiv.org/abs/2411.10786}
}

@article{shi2025prerequisite,
  title={Prerequisite of superconductivity: SDW rather than tetragonal structure in double-layer {L}a$_3${N}i$_2${O}$_{7-\delta}$},
  author={Shi, Mengzhu and Peng, Di and Li, Yikang and Xing, Zhenfang and Wang, Yuzhu and Fan, Kaibao and Li, Houpu and Wu, Rongqi and Zeng, Zhidan and Zeng, Qiaoshi and Ying, Jianjun and Chen, Xianhui},
  journal={arXiv:2501.14202},
  year={2025},
  url={https://arxiv.org/abs/2501.14202}
}

@article{li2025ambient,
  title={Ambient pressure growth of bilayer nickelate single crystals with superconductivity over 90 {K} under high pressure},
  author={Li, Feiyu and Xing, Zhenfang and Peng, Di and Dou, Jie and Guo, Ning and Ma, Liang and Zhang, Yulin and Wang, Lingzhen and Luo, Jun and Yang, Jie and Zhang, Jian and Chang, Tieyan and Chen, Yu-Sheng and Cai, Weizhao and Cheng, Jinguang and Wang, Yuzhu and Zeng, Zhidan and Zheng, Qiang and Zhou, Rui and Zeng, Qiaoshi and Tao, Xutang and Zhang, Junjie},
  journal={arXiv:2501.14584},
  year={2025},
  url={https://arxiv.org/abs/2501.14584}
}

@article{huo2025low,
  title={Low volume fraction of high-${T}_c$ superconductivity in {L}a$_3${N}i$_2${O}$_7$ at 80 {K} and ambient pressure},
  author={Huo, Mengwu and Ma, Peiyue and Huang, Chaoxin and Huang, Xing and Sun, Hualei and Wang, Meng},
  journal={arXiv:2501.15929},
  year={2025},
  url={https://arxiv.org/abs/2501.15929}
}

@article{zhang2025damage,
  title={Damage of bilayer structure in {L}a$_3${N}i$_2${O}$_{7-\delta}$ induced by high p{O}$_2$ annealing},
  author={Zhang, Yulin and Pei, Cuiying and Guo, Ning and Fan, Longlong and Zhang, Mingxin and Wang, Lingzhen and Zhang, Gongting and Li, Feiyu and Wang, Yunong and Ma, Chao and Cheng, Wenyong and Wang, Shanpeng and Zheng, Qiang and Qi, Yanpeng and Zhang, Junjie},
  journal={arXiv:2502.01501},
  year={2025},
  url={https://arxiv.org/abs/2502.01501}
}

@article{zhang2024doping,
  title={Doping evolution of the normal state magnetic excitations in pressurized {L}a$_3${N}i$_2${O}$_7$},
  author={Zhang, Hai-Yang and Bai, Yu-Jie and Kong, Fan-Jie and Wu, Xiu-Qiang and Xing, Yu-Heng and Xu, Ning},
  journal={New J. Phys.},
  volume={26},
  number={12},
  pages={123027},
  year={2024},
  publisher={IOP Publishing},
  url={https://iopscience.iop.org/article/10.1088/1367-2630/ada0d4/}
}

@article{khasanov2024pressure,
  author = {Khasanov, Rustem and Hicken, Thomas J. and Gawryluk, Dariusz J. and Sazgari, Vahid and Plokhikh, Igor and Sorel, Lo{\"\i}c Pierre and Bartkowiak, Marek and B{\"o}tzel, Steffen and Lechermann, Frank and Eremin, Ilya M. and Luetkens, Hubertus and Guguchia, Zurab},
  title = {Pressure-enhanced splitting of density wave transitions in {L}a$_3${N}i$_2${O}$_{7-\delta}$},
  journal = {Nature Physics},
  volume = {21},
  number = {3},
  pages = {430--436},
  year = {2025},
  doi = {10.1038/s41567-024-02754-z},
  url = {https://doi.org/10.1038/s41567-024-02754-z},
  date = {2025-03-01}
}

@article{chen2024evidence,
  title={Evidence of Spin Density Waves in {L}a$_3${N}i$_2${O}$_{7-\delta}$},
  author={Chen, Kaiwen and Liu, Xiangqi and Jiao, Jiachen and Zou, Muyuan and Jiang, Chengyu and Li, Xin and Luo, Yixuan and Wu, Qiong and Zhang, Ningyuan and Guo, Yanfeng and Shu, Lei},
  journal={Phys. Rev. Lett.},
  volume={132},
  number={25},
  pages={256503},
  year={2024},
  publisher={APS},
  url={https://journals.aps.org/prl/abstract/10.1103/PhysRevLett.132.256503}
}

@article{Kakoi2024,
author = {Kakoi ,Masataka and Oi ,Takashi and Ohshita ,Yujiro and Yashima ,Mitsuharu and Kuroki ,Kazuhiko and Kato ,Takeru and Takahashi ,Hidefumi and Ishiwata ,Shintaro and Adachi ,Yoshinobu and Hatada ,Naoyuki and Uda ,Tetsuya and Mukuda ,Hidekazu},
title = {Multiband Metallic Ground State in Multilayered Nickelates {L}a$_3${N}i$_2${O}$_7$ and {L}a$_4${N}i$_3${O}$_{10}$ Probed by $^{139}${L}a-{NMR} at Ambient Pressure},
journal = {J. Phys. Soc. Jpn.},
volume = {93},
number = {5},
pages = {053702},
year = {2024},
doi = {10.7566/JPSJ.93.053702},
URL = {https://doi.org/10.7566/JPSJ.93.053702}
}

@article{xu2024pressure,
      title={Pressure-Induced Phase Transitions in Bilayer {L}a$_3${N}i$_2${O}$_7$}, 
      author={Mingyu Xu and Greeshma C. Jose and Aya Rutherford and Haozhe Wang and Stephen Zhang and Robert J. Cava and Haidong Zhou and Wenli Bi and Weiwei Xie},
      year={2024},
      journal={arXiv:2410.18840},
      url={https://arxiv.org/abs/2410.18840}, 
}

@article{kaneko2025tj,
      title={$t$-${J}$ model for strongly correlated two-orbital systems: Application to bilayer nickelate superconductors}, 
      author={Tatsuya Kaneko and Masataka Kakoi and Kazuhiko Kuroki},
      year={2025},
      journal={arxiv:2504.10114},
      url={https://arxiv.org/abs/2504.10114}, 
}

@article{zhang2024electronic,
  title = {Electronic structure, self-doping, and superconducting instability in the alternating single-layer trilayer stacking nickelates {L}a$_{3}${N}i$_{2}${O}$_{7}$},
  author = {Zhang, Yang and Lin, Ling-Fang and Moreo, Adriana and Maier, Thomas A. and Dagotto, Elbio},
  journal = {Phys. Rev. B},
  volume = {110},
  issue = {6},
  pages = {L060510},
  numpages = {7},
  year = {2024},
  month = {Aug},
  publisher = {American Physical Society},
  doi = {10.1103/PhysRevB.110.L060510},
  url = {https://link.aps.org/doi/10.1103/PhysRevB.110.L060510}
}

@article{zhao2024spin,
  title = {Pressure-enhanced spin-density-wave transition in double-layer nickelate {L}a$_3${N}i$_2${O}$_{7-\delta}$},
  author = {Dan Zhao and Yanbing Zhou and Mengwu Huo and Yu Wang and Linpeng Nie and Ye Yang and Jianjun Ying and Meng Wang and Tao Wu and Xianhui Chen},
  journal = {Science Bulletin},
  volume = {70},
  number = {8},
  pages = {1239-1245},
  year = {2025},
  issn = {2095-9273},
  doi = {https://doi.org/10.1016/j.scib.2025.02.019},
  url = {https://www.sciencedirect.com/science/article/pii/S2095927325001811}
}

@article{chen2024electronic,
  title={Electronic and magnetic excitations in {L}a$_3${N}i$_2${O}$_7$}, 
  author={Xiaoyang Chen and Jaewon Choi and Zhicheng Jiang and Jiong Mei and Kun Jiang and Jie Li and Stefano Agrestini and Mirian Garcia-Fernandez and Xing Huang and Hualei Sun and Dawei Shen and Meng Wang and Jiangping Hu and Yi Lu and Ke-Jin Zhou and Donglai Feng},
  journal={Nat. Commun.},
  volume = {15},
  number = {1},
  pages = {9597},
  year = {2024},
  month = {nov},
  doi = {10.1038/s41467-024-53863-5},
  url = {https://doi.org/10.1038/s41467-024-53863-5}
}

@article{Fukamachi2001,
title = {$^{139}${L}a {NMR} studies of layered perovskite systems {L}a$_3${N}i$_2${O}$_{7-\delta}$ and {L}a$_4${N}i$_3${O}$_{10}$},
journal = {J. Phys. Chem. Solids},
volume = {62},
number = {1},
pages = {195-198},
year = {2001},
issn = {0022-3697},
doi = {https://doi.org/10.1016/S0022-3697(00)00127-X},
url = {https://www.sciencedirect.com/science/article/pii/S002236970000127X},
author = {T Fukamachi and Y Kobayashi and T Miyashita and M Sato}
}

@article{wang2024self,
  title={Self-doped Molecular {M}ott Insulator for Bilayer High-Temperature Superconducting {L}a$_3${N}i$_2${O}$_7$},
  author={Wang, Zhan and Jiang, Kun and Zhang, Fu-Chun},
  journal={arXiv:2412.18469},
  year={2024},
  url={https://arxiv.org/abs/2412.18469}
}

@article{xu2025competition,
  title={Incommensurate spin-fluctuations and competing pairing symmetries in {L}a$_3${N}i$_2${O}$_7$},
  author={Han-Xiang Xu and Daniel Guterding},
  journal={arXiv:2501.05254},
  year={2025},
  url={https://arxiv.org/abs/2501.05254}
}

@article{bhatta2025structural,
  title={Structural and Electronic Evolution of Bilayer Nickelates Under Biaxial Strain},
  author={Bhatta, HC and Zhang, Xiaoliang and Zhong, Yong and Jia, Chunjing},
  journal={arXiv:2502.01624},
  year={2025},
  url={https://arxiv.org/abs/2502.01624}
}

@article{shi2025effect,
  title={The effect of Carrier Doping and Thickness on the Electronic Structures of {L}a$_3${N}i$_2${O}$_7$ Thin Films},
  author={Shi, Haoliang and Huo, Zihao and Li, Guanlin and Ma, Hao and Cui, Tian and Yao, Dao-Xin and Duan, Defang},
  journal={arXiv:2502.04255},
  year={2025},
  url={https://arxiv.org/abs/2502.04255}
}

@article{liu2024growth,
  title = {Growth and characterization of the {L}a$_{3}${N}i$_{2}${O}$_{7-\delta}$ thin films: Dominant contribution of the $d_{x^{2}-y^{2}}$ orbital at ambient pressure},
  author = {Liu, Yuecong and Ou, Mengjun and Chu, Haifeng and Yang, Huan and Li, Qing and Zhang, Ying-Jie and Wen, Hai-Hu},
  journal = {Phys. Rev. Mater.},
  volume = {8},
  issue = {12},
  pages = {124801},
  numpages = {8},
  year = {2024},
  month = {Dec},
  publisher = {American Physical Society},
  doi = {10.1103/PhysRevMaterials.8.124801},
  url = {https://link.aps.org/doi/10.1103/PhysRevMaterials.8.124801}
}

@article{Wang2022LNO,
  title={Evidence for charge and spin density waves in single crystals of {L}a$_3${N}i$_2${O}$_7$ and {L}a$_3${N}i$_2${O}$_6$},
  author={Liu, Zengjia and Sun, Hualei and Huo, Mengwu and Ma, Xiaoyan and Ji, Yi and Yi, Enkui and Li, Lisi and Liu, Hui and Yu, Jia and Zhang, Ziyou and Chen, Zhiqiang and Liang, Feixiang and Dong, Hongliang and Guo, Hanjie and Zhong, Dingyong and Shen, Bing and Li, Shiliang and Wang, Meng},
  journal={Sci. China-Phys. Mech. Astron.},
  volume={66},
  number={1},
  pages={217411},
  year={2023},
  publisher={Springer},
  url = {https://link.springer.com/article/10.1007/s11433-022-1962-4}
}

@article{Wang2023LNO,
   author = {Sun, Hualei and Huo, Mengwu and Hu, Xunwu and Li, Jingyuan and Liu, Zengjia and Han, Yifeng and Tang, Lingyun and Mao, Zhongquan and Yang, Pengtao and Wang, Bosen and Cheng, Jinguang and Yao, Dao-Xin and Zhang, Guang-Ming and Wang, Meng},
   title = {Signatures of superconductivity near 80{K} in a nickelate under high pressure},
journal={Nature},
year={2023},
month={Sep},
day={01},
volume={621},
number={7979},
pages={493-498},
issn={1476-4687},
doi={10.1038/s41586-023-06408-7},
url={https://doi.org/10.1038/s41586-023-06408-7}
}

@article{liu2024electronic,
author={Liu, Zhe
and Huo, Mengwu
and Li, Jie
and Li, Qing
and Liu, Yuecong
and Dai, Yaomin
and Zhou, Xiaoxiang
and Hao, Jiahao
and Lu, Yi
and Wang, Meng
and Wen, Hai-Hu},
title={Electronic correlations and partial gap in the bilayer nickelate {L}a$_3${N}i$_2${O}$_7$},
journal={Nat. Commun.},
year={2024},
month={Aug},
day={31},
volume={15},
number={1},
pages={7570},
issn={2041-1723},
doi={10.1038/s41467-024-52001-5}
}

@article{Wang2023LNOb,
   author = {Jun Hou and Peng-Tao Yang and Zi-Yi Liu and Jing-Yuan Li and Peng-Fei Shan and Liang Ma and Gang Wang and Ning-Ning Wang and Hai-Zhong Guo and Jian-Ping Sun and Yoshiya Uwatoko and Meng Wang and Guang-Ming Zhang and Bo-Sen Wang and Jin-Guang Cheng},
   title = {Emergence of High-Temperature Superconducting Phase in Pressurized {L}a$_{3}${N}i$_{2}${O}$_7$ Crystals},
   publisher = {Chin. Phys. Lett.},
   year = {2023},
   journal = {Chin. Phys. Lett.},
   volume = {40},
   number = {11},
   eid = {117302},
   pages = {117302},
   url = {https://cpl.iphy.ac.cn/EN/abstract/article_116425.shtml},
   doi = {10.1088/0256-307X/40/11/117302}
}

@article{wang2024review,
doi = {10.1088/0256-307X/41/7/077402},
url = {https://dx.doi.org/10.1088/0256-307X/41/7/077402},
year = {2024},
month = {jul},
publisher = {Chinese Physical Society and IOP Publishing Ltd},
volume = {41},
number = {7},
pages = {077402},
author = {Wang, Meng and Wen, Hai-Hu and Wu, Tao and Yao, Dao-Xin and Xiang, Tao},
title = {Normal and Superconducting Properties of {L}a$_3${N}i$_2${O}$_7$},
journal = {Chin. Phys. Lett.},
}

@article{YuanHQ2023LNO,
author={Zhang, Yanan
and Su, Dajun
and Huang, Yanen
and Shan, Zhaoyang
and Sun, Hualei
and Huo, Mengwu
and Ye, Kaixin
and Zhang, Jiawen
and Yang, Zihan
and Xu, Yongkang
and Su, Yi
and Li, Rui
and Smidman, Michael
and Wang, Meng
and Jiao, Lin
and Yuan, Huiqiu},
title={High-temperature superconductivity with zero resistance and strange-metal behaviour in {L}a$_3${N}i$_2${O}$_{7-\delta}$},
journal={Nat. Phys.},
year={2024},
month={Aug},
day={01},
volume={20},
number={8},
pages={1269-1273},
issn={1745-2481},
doi={10.1038/s41567-024-02515-y},
}

@article{zhou2025investigations,
    author = {Zhou, Yazhou and Guo, Jing and Cai, Shu and Sun, Hualei and Li, Chengyu and Zhao, Jinyu and Wang, Pengyu and Han, Jinyu and Chen, Xintian and Chen, Yongjin and Wu, Qi and Ding, Yang and Xiang, Tao and Mao, Ho-kwang and Sun, Liling},
    title = {Investigations of key issues on the reproducibility of high-{$T_c$} superconductivity emerging from compressed {L}a$_3${N}i$_2${O}$_7$},
    journal = {Matter and Radiation at Extremes},
    volume = {10},
    number = {2},
    pages = {027801},
    year = {2025},
    month = {01},
    issn = {2468-2047},
    doi = {10.1063/5.0247684},
    url = {https://pubs.aip.org/aip/mre/article/10/2/027801/3331819/Investigations-of-key-issues-on-the}
}

@article{yang2024orbital,
  title={Orbital-dependent electron correlation in double-layer nickelate {L}a$_3${N}i$_2${O}$_7$},
  author={Yang, Jiangang and Sun, Hualei and Hu, Xunwu and Xie, Yuyang and Miao, Taimin and Luo, Hailan and Chen, Hao and Liang, Bo and Zhu, Wenpei and Qu, Gexing and Chen, Cui-Qun and Huo, Mengwu and Huang, Yaobo and Zhang, Shenjin and Zhang, Fengfeng and Yang, Feng and Wang, Zhimin and Peng, Qinjun and Mao, Hanqing and Liu, Guodong and Xu, Zuyan and Qian, Tian and Yao, Dao-Xin and Wang, Meng and Zhao, Lin and Zhou, X. J.},
  journal={Nat. Commun.},
  volume={15},
  number={1},
  pages={4373},
  year={2024},
  publisher={Nature Publishing Group UK London},
  url={https://www.nature.com/articles/s41467-024-48701-7}
}

@article{zhang2023pressure,
  title={Effects of pressure and doping on {R}uddlesden-{P}opper phases {L}a$_{n+1}${N}i$_n${O}$_{3n+1}$},
  author={Zhang, Mingxin and Pei, Cuiying and Wang, Qi and Zhao, Yi and Li, Changhua and Cao, Weizheng and Zhu, Shihao and Wu, Juefei and Qi, Yanpeng},
  journal={J. Mater. Sci. Technol.},
  volume={185},
  pages={147--154},
  year={2024},
  publisher={Elsevier},
  url={https://www.sciencedirect.com/science/article/pii/S1005030223009829}
}

@article{wang2023LNOpoly,
  title = {Pressure-Induced Superconductivity In Polycrystalline {L}a$_3${N}i$_2${O}$_7$},
  author = {Wang, G. and Wang, N. N. and Shen, X. L. and Hou, J. and Ma, L. and Shi, L. F. and Ren, Z. A. and Gu, Y. D. and Ma, H. M. and Yang, P. T. and Liu, Z. Y. and Guo, H. Z. and Sun, J. P. and Zhang, G. M. and Calder, S. and Yan, J.-Q. and Wang, B. S. and Uwatoko, Y. and Cheng, J.-G.},
  journal = {Phys. Rev. X},
  volume = {14},
  issue = {1},
  pages = {011040},
  numpages = {8},
  year = {2024},
  month = {Mar},
  publisher = {American Physical Society},
  doi = {10.1103/PhysRevX.14.011040},
  url = {https://link.aps.org/doi/10.1103/PhysRevX.14.011040}
}

@article{wang2023la2prnio7,
  title={Observation of high-temperature superconductivity in the high-pressure tetragonal phase of {L}a$_2${P}r{N}i$_2${O}$_{7-\delta}$}, 
  author={Gang Wang and Ningning Wang and Yuxin Wang and Lifen Shi and Xiaoling Shen and Jun Hou and Hanming Ma and Pengtao Yang and Ziyi Liu and Hua Zhang and Xiaoli Dong and Jianping Sun and Bosen Wang and Kun Jiang and Jiangping Hu and Yoshiya Uwatoko and Jinguang Cheng},
  journal={arXiv:2311.08212},
  url = {https://arxiv.org/abs/2311.08212},
  year={2023}
}

@article{mijit2024local,
  title={Local electronic properties of {L}a$_3${N}i$_2${O}$_7$ under pressure},
  author={Mijit, Emin and Ma, Peiyue and Sahle, Christoph J and Rosa, Angelika D and Hu, Zhiwei and De Angelis, Francesco and Lopez, Alberto and Amatori, Simone and Tchoudinov, Georghii and Joly, Yves and Irifune, Tetsuo and Rodrigues, Joao Elias F. S. and Garbarino, Gaston and Parra, Samuel Gallego and Wang, Meng and Yu, Runze and Mathon, Olivier},
  journal={arXiv:2412.08269},
  year={2024},
  url={https://arxiv.org/abs/2412.08269}
}

@article{wang2023structure,
  title={Structure responsible for the superconducting state in {L}a$_3${N}i$_2${O}$_7$ at low temperature and high pressure conditions}, 
  author={Luhong Wang and Yan Li and Shengyi Xie and Fuyang Liu and Hualei Sun and Caoxin Huang and Yang Gao and Takeshi Nakagawa and Boyang Fu and Bo Dong and Zhenhui Cao and Runze Yu and Saori I. Kawaguchi and Hirokazu Kadobayashi and Meng Wang and Changqing Jin and Ho-kwang Mao and Haozhe Liu},
  journal = {Journal of the American Chemical Society},
  year = {2024},
  month = {mar},
  volume = {146},
  number = {11},
  pages = {7506--7514},
  doi = {10.1021/jacs.3c13094},
  url = {https://doi.org/10.1021/jacs.3c13094}
}

@article{li2024pressure,
  title={Identification of the superconductivity in bilayer nickelate {L}a$_3${N}i$_2${O}$_7$ upon 100 {GP}a}, 
  author={Li, Jingyuan and Peng, Di and Ma, Peiyue and Zhang, Hengyuan and Xing, Zhenfang and Huang, Xing and Huang, Chaoxin and Huo, Mengwu and Hu, Deyuan and Dong, Zixian and Chen, Xiang and Xie, Tao and Dong, Hongliang and Sun, Hualei and Zeng, Qiaoshi and Mao, Ho-kwang and Wang, Meng},
  journal={arXiv:2404.11369},
  url={https://arxiv.org/abs/2404.11369},
  year={2024}
}

@article{Ueki2025pressure,
author = {Ueki ,Yuta and Sakurai ,Hiroya and Nagata ,Hibiki and Yamane ,Kazuki and Matsumoto ,Ryo and Terashima ,Kensei and Hirose ,Keisuke and Ohta ,Hiroto and Kato ,Masaki and Takano ,Yoshihiko},
title = {Phase Diagram of Pressure-Induced High Temperature Superconductor {L}a$_3${N}i$_2${O}$_{7+\delta}$},
journal = {J. Phys. Soc. Jpn.},
volume = {94},
number = {1},
pages = {013703},
year = {2025},
doi = {10.7566/JPSJ.94.013703},
URL = {https://doi.org/10.7566/JPSJ.94.013703}
}

@article{chen2025unveiling,
  title = {Unveiling the multiband metallic nature of the normal state in the nickelate {L}a$_3${N}i$_2${O}$_7$},
  author = {Chen, Bowen and Zhang, Hengyuan and Li, Jingyuan and Hu, Deyuan and Huo, Mengwu and Wang, Shuyang and Xi, Chuanying and Wang, Zhaosheng and Sun, Hualei and Wang, Meng and Shen, Bing},
  journal = {Phys. Rev. B},
  volume = {111},
  issue = {5},
  pages = {054519},
  numpages = {7},
  year = {2025},
  month = {Feb},
  publisher = {American Physical Society},
  doi = {10.1103/PhysRevB.111.054519},
  url = {https://link.aps.org/doi/10.1103/PhysRevB.111.054519}
}

@article{sui2023rno,
  title = {Electronic properties of the bilayer nickelates {R}$_3${N}i$_2${O}$_7$ with oxygen vacancies ({R}={L}a or {C}e)},
  author = {Sui, Xuelei and Han, Xiangru and Jin, Heng and Chen, Xiaojun and Qiao, Liang and Shao, Xiaohong and Huang, Bing},
  journal = {Phys. Rev. B},
  volume = {109},
  issue = {20},
  pages = {205156},
  numpages = {12},
  year = {2024},
  month = {May},
  publisher = {American Physical Society},
  doi = {10.1103/PhysRevB.109.205156},
  url = {https://link.aps.org/doi/10.1103/PhysRevB.109.205156}
}

@article{YaoDX2023,
  title = {Bilayer Two-Orbital Model of {L}a$_3${N}i$_2${O}$_7$ under Pressure},
  author = {Luo, Zhihui and Hu, Xunwu and Wang, Meng and W\'u, W\'ei and Yao, Dao-Xin},
  journal = {Phys. Rev. Lett.},
  volume = {131},
  issue = {12},
  pages = {126001},
  numpages = {6},
  year = {2023},
  month = {Sep},
  publisher = {American Physical Society},
  doi = {10.1103/PhysRevLett.131.126001},
  url = {https://link.aps.org/doi/10.1103/PhysRevLett.131.126001}
}

@article{Dagotto2023,
  title = {Electronic structure, dimer physics, orbital-selective behavior, and magnetic tendencies in the bilayer nickelate superconductor {L}a$_3${N}i$_2${O}$_7$ under pressure},
  author = {Zhang, Yang and Lin, Ling-Fang and Moreo, Adriana and Dagotto, Elbio},
  journal = {Phys. Rev. B},
  volume = {108},
  issue = {18},
  pages = {L180510},
  numpages = {5},
  year = {2023},
  month = {Nov},
  publisher = {American Physical Society},
  doi = {10.1103/PhysRevB.108.L180510},
  url = {https://link.aps.org/doi/10.1103/PhysRevB.108.L180510}
}

@article{WangQH2023,
  title = {Possible ${S}_{\pm}$-wave superconductivity in {L}a$_3${N}i$_2${O}$_7$},
  author = {Yang, Qing-Geng and Wang, Da and Wang, Qiang-Hua},
  journal = {Phys. Rev. B},
  volume = {108},
  issue = {14},
  pages = {L140505},
  numpages = {5},
  year = {2023},
  month = {Oct},
  publisher = {American Physical Society},
  doi = {10.1103/PhysRevB.108.L140505},
  url = {https://link.aps.org/doi/10.1103/PhysRevB.108.L140505}
}

@article{lechermann2023,
  title = {Electronic correlations and superconducting instability in {L}a$_3${N}i$_2${O}$_7$ under high pressure},
  author = {Lechermann, Frank and Gondolf, Jannik and B\"otzel, Steffen and Eremin, Ilya M.},
  journal = {Phys. Rev. B},
  volume = {108},
  issue = {20},
  pages = {L201121},
  numpages = {6},
  year = {2023},
  month = {Nov},
  publisher = {American Physical Society},
  doi = {10.1103/PhysRevB.108.L201121},
  url = {https://link.aps.org/doi/10.1103/PhysRevB.108.L201121}
}

@article{Kuroki2023,
  title = {Possible High ${T}_{c}$ Superconductivity in {L}a$_3${N}i$_2${O}$_7$ under High Pressure through Manifestation of a Nearly Half-Filled Bilayer {H}ubbard Model},
  author = {Sakakibara, Hirofumi and Kitamine, Naoya and Ochi, Masayuki and Kuroki, Kazuhiko},
  journal = {Phys. Rev. Lett.},
  volume = {132},
  issue = {10},
  pages = {106002},
  numpages = {6},
  year = {2024},
  month = {Mar},
  publisher = {American Physical Society},
  doi = {10.1103/PhysRevLett.132.106002},
  url = {https://link.aps.org/doi/10.1103/PhysRevLett.132.106002}
}

@article{HuJP2023,
  title = {Effective model and pairing tendency in the bilayer {N}i-based superconductor {L}a$_{3}${N}i$_{2}${O}$_{7}$},
  author = {Gu, Yuhao and Le, Congcong and Yang, Zhesen and Wu, Xianxin and Hu, Jiangping},
  journal = {Phys. Rev. B},
  volume = {111},
  issue = {17},
  pages = {174506},
  numpages = {7},
  year = {2025},
  month = {May},
  publisher = {American Physical Society},
  doi = {10.1103/PhysRevB.111.174506},
  url = {https://link.aps.org/doi/10.1103/PhysRevB.111.174506}
}

@article{ZhangGM2023DMRG,
  title={Effective Bi-Layer Model Hamiltonian and Density-Matrix Renormalization Group Study for the High-${T}_c$ Superconductivity {L}a$_3${N}i$_2${O}$_7$ under High Pressure},
  author={Shen, Yang and Qin, Mingpu and Zhang, Guang-Ming},
  journal={Chin. Phys. Lett.},
  volume={40},
  number={12},
  pages={127401},
  year={2023},
  publisher={Chinese Physical Society},
  url={https://iopscience.iop.org/article/10.1088/0256-307X/40/12/127401}
}

@article{Werner2023,
  title = {Correlated Electronic Structure of {L}a$_3${N}i$_2${O}$_7$ under Pressure},
  author = {Christiansson, Viktor and Petocchi, Francesco and Werner, Philipp},
  journal = {Phys. Rev. Lett.},
  volume = {131},
  issue = {20},
  pages = {206501},
  numpages = {6},
  year = {2023},
  month = {Nov},
  publisher = {American Physical Society},
  doi = {10.1103/PhysRevLett.131.206501},
  url = {https://link.aps.org/doi/10.1103/PhysRevLett.131.206501}
}

@article{shilenko2023correlated,
  title = {Correlated electronic structure, orbital-selective behavior, and magnetic correlations in double-layer {L}a$_3${N}i$_2${O}$_7$ under pressure},
  author = {Shilenko, D. A. and Leonov, I. V.},
  journal = {Phys. Rev. B},
  volume = {108},
  issue = {12},
  pages = {125105},
  numpages = {9},
  year = {2023},
  month = {Sep},
  publisher = {American Physical Society},
  doi = {10.1103/PhysRevB.108.125105},
  url = {https://link.aps.org/doi/10.1103/PhysRevB.108.125105}
}

@article{WuWei2023charge,
  title={Superexchange and charge transfer in the nickelate superconductor {L}a$_3${N}i$_2${O}$_7$ under pressure},
  author={W{\'u}, W{\'e}i and Luo, Zhihui and Yao, Dao-Xin and Wang, Meng},
  journal={Sci. China-Phys. Mech. Astron.},
  volume={67},
  number={11},
  pages={117402},
  year={2024},
  publisher={Springer},
  url={https://link.springer.com/article/10.1007/s11433-023-2300-4}
}

@article{cao2023flat,
  title = {Flat bands promoted by {H}und's rule coupling in the candidate double-layer high-temperature superconductor {L}a$_3${N}i$_2${O}$_7$ under high pressure},
  author = {Cao, Yingying and Yang, Yi-feng},
  journal = {Phys. Rev. B},
  volume = {109},
  issue = {8},
  pages = {L081105},
  numpages = {6},
  year = {2024},
  month = {Feb},
  publisher = {American Physical Society},
  doi = {10.1103/PhysRevB.109.L081105},
  url = {https://link.aps.org/doi/10.1103/PhysRevB.109.L081105}
}

@article{chen2023critical,
  title = {Charge and spin instabilities in superconducting {L}a$_3${N}i$_2${O}$_7$},
  author = {Chen, Xuejiao and Jiang, Peiheng and Li, Jie and Zhong, Zhicheng and Lu, Yi},
  journal = {Phys. Rev. B},
  volume = {111},
  issue = {1},
  pages = {014515},
  numpages = {8},
  year = {2025},
  month = {Jan},
  publisher = {American Physical Society},
  doi = {10.1103/PhysRevB.111.014515},
  url = {https://link.aps.org/doi/10.1103/PhysRevB.111.014515}
}

@article{YangF2023,
  title = {s$^{\pm}$-Wave Pairing and the Destructive Role of Apical-Oxygen Deficiencies in {L}a$_3${N}i$_2${O}$_7$ under Pressure},
  author = {Liu, Yu-Bo and Mei, Jia-Wei and Ye, Fei and Chen, Wei-Qiang and Yang, Fan},
  journal = {Phys. Rev. Lett.},
  volume = {131},
  issue = {23},
  pages = {236002},
  numpages = {6},
  year = {2023},
  month = {Dec},
  publisher = {American Physical Society},
  doi = {10.1103/PhysRevLett.131.236002},
  url = {https://link.aps.org/doi/10.1103/PhysRevLett.131.236002}
}

@article{oh2023type2,
  title = {Type-{II} $t$-${J}$ model and shared superexchange coupling from {H}und's rule in superconducting {L}a$_3${N}i$_2${O}$_7$},
  author = {Oh, Hanbit and Zhang, Ya-Hui},
  journal = {Phys. Rev. B},
  volume = {108},
  issue = {17},
  pages = {174511},
  numpages = {8},
  year = {2023},
  month = {Nov},
  publisher = {American Physical Society},
  doi = {10.1103/PhysRevB.108.174511},
  url = {https://link.aps.org/doi/10.1103/PhysRevB.108.174511}
}

@article{zhang2023structural,
  title={Structural phase transition, $s_{\pm}$-wave pairing, and magnetic stripe order in bilayered superconductor {L}a$_3${N}i$_2${O}$_7$ under pressure},
  author={Zhang, Yang and Lin, Ling-Fang and Moreo, Adriana and Maier, Thomas A and Dagotto, Elbio},
  journal={Nat. Commun.},
  volume={15},
  number={1},
  pages={2470},
  year={2024},
  publisher={Nature Publishing Group UK London},
  url={https://www.nature.com/articles/s41467-024-46622-z}
}

@article{liao2023electron,
  title = {Electron correlations and superconductivity in {L}a$_{3}${N}i$_{2}${O}$_{7}$ under pressure tuning},
  author = {Liao, Zhiguang and Chen, Lei and Duan, Guijing and Wang, Yiming and Liu, Changle and Yu, Rong and Si, Qimiao},
  journal = {Phys. Rev. B},
  volume = {108},
  issue = {21},
  pages = {214522},
  numpages = {9},
  year = {2023},
  month = {Dec},
  publisher = {American Physical Society},
  doi = {10.1103/PhysRevB.108.214522},
  url = {https://link.aps.org/doi/10.1103/PhysRevB.108.214522}
}

@article{qu2023bilayer,
  title = {Bilayer $t$-${J}$-${J}_{\perp}$ Model and Magnetically Mediated Pairing in the Pressurized Nickelate {L}a$_3${N}i$_2${O}$_7$},
  author = {Qu, Xing-Zhou and Qu, Dai-Wei and Chen, Jialin and Wu, Congjun and Yang, Fan and Li, Wei and Su, Gang},
  journal = {Phys. Rev. Lett.},
  volume = {132},
  issue = {3},
  pages = {036502},
  numpages = {6},
  year = {2024},
  month = {Jan},
  publisher = {American Physical Society},
  doi = {10.1103/PhysRevLett.132.036502},
  url = {https://link.aps.org/doi/10.1103/PhysRevLett.132.036502}
}

@article{Yi_Feng2023,
  title = {Interlayer valence bonds and two-component theory for high-${T}_{c}$ superconductivity of {L}a$_3${N}i$_2${O}$_7$ under pressure},
  author = {Yang, Yi-Feng and Zhang, Guang-Ming and Zhang, Fu-Chun},
  journal = {Phys. Rev. B},
  volume = {108},
  issue = {20},
  pages = {L201108},
  numpages = {6},
  year = {2023},
  month = {Nov},
  publisher = {American Physical Society},
  doi = {10.1103/PhysRevB.108.L201108},
  url = {https://link.aps.org/doi/10.1103/PhysRevB.108.L201108}
}

@article{jiang2023high,
  title={High temperature superconductivity in {L}a$_3${N}i$_2${O}$_7$},
  author={Jiang, Kun and Wang, Ziqiang and Zhang, Fu-Chun},
  journal={Chin. Phys. Lett.},
  year={2023},
  url={https://iopscience.iop.org/article/10.1088/0256-307X/41/1/017402}
}

@article{zhang2023trends,
  title = {Trends in electronic structures and $s_{\pm}$-wave pairing for the rare-earth series in bilayer nickelate superconductor ${R}_3${N}i$_2${O}$_7$},
  author = {Zhang, Yang and Lin, Ling-Fang and Moreo, Adriana and Maier, Thomas A. and Dagotto, Elbio},
  journal = {Phys. Rev. B},
  volume = {108},
  issue = {16},
  pages = {165141},
  numpages = {8},
  year = {2023},
  month = {Oct},
  publisher = {American Physical Society},
  doi = {10.1103/PhysRevB.108.165141},
  url = {https://link.aps.org/doi/10.1103/PhysRevB.108.165141}
}

@article{huang2023impurity,
  title = {Impurity and vortex states in the bilayer high-temperature superconductor {L}a$_3${N}i$_2${O}$_7$},
  author = {Huang, Junkang and Wang, Z. D. and Zhou, Tao},
  journal = {Phys. Rev. B},
  volume = {108},
  issue = {17},
  pages = {174501},
  numpages = {7},
  year = {2023},
  month = {Nov},
  publisher = {American Physical Society},
  doi = {10.1103/PhysRevB.108.174501},
  url = {https://link.aps.org/doi/10.1103/PhysRevB.108.174501}
}

@article{qin2023high,
  title = {High-${T}_{c}$ superconductivity by mobilizing local spin singlets and possible route to higher ${T}_{c}$ in pressurized {L}a$_3${N}i$_2${O}$_7$},
  author = {Qin, Qiong and Yang, Yi-Feng},
  journal = {Phys. Rev. B},
  volume = {108},
  issue = {14},
  pages = {L140504},
  numpages = {6},
  year = {2023},
  month = {Oct},
  publisher = {American Physical Society},
  doi = {10.1103/PhysRevB.108.L140504},
  url = {https://link.aps.org/doi/10.1103/PhysRevB.108.L140504}
}

@article{tian2023correlation,
  title = {Correlation effects and concomitant two-orbital ${s}_{\pm}$-wave superconductivity in {L}a$_3${N}i$_2${O}$_7$ under high pressure},
  author = {Tian, Yi-Heng and Chen, Yin and Wang, Jia-Ming and He, Rong-Qiang and Lu, Zhong-Yi},
  journal = {Phys. Rev. B},
  volume = {109},
  issue = {16},
  pages = {165154},
  numpages = {6},
  year = {2024},
  month = {Apr},
  publisher = {American Physical Society},
  doi = {10.1103/PhysRevB.109.165154},
  url = {https://link.aps.org/doi/10.1103/PhysRevB.109.165154}
}

@article{jiang2023pressure,
  title = {Pressure Driven Fractionalization of Ionic Spins Results in Cupratelike High-${T}_{c}$ Superconductivity in {L}a$_3${N}i$_2${O}$_7$},
  author = {Jiang, Ruoshi and Hou, Jinning and Fan, Zhiyu and Lang, Zi-Jian and Ku, Wei},
  journal = {Phys. Rev. Lett.},
  volume = {132},
  issue = {12},
  pages = {126503},
  numpages = {7},
  year = {2024},
  month = {Mar},
  publisher = {American Physical Society},
  doi = {10.1103/PhysRevLett.132.126503},
  url = {https://link.aps.org/doi/10.1103/PhysRevLett.132.126503}
}

@article{lu2023sc,
  title={Superconductivity from Doping Symmetric Mass Generation Insulators: Application to {L}a$_3${N}i$_2${O}$_7$ under Pressure},
  author={Lu, Da-Chuan and Li, Miao and Zeng, Zhao-Yi and Hou, Wanda and Wang, Juven and Yang, Fan and You, Yi-Zhuang},
  journal={arXiv:2308.11195},
  year={2023},
  url = {https://arxiv.org/abs/2308.11195}
}

@article{kitamine2023,
  title={Theoretical designing of multiband Nickelate and Palladate superconductors with $d^{8+\delta}$ configuration}, 
  author={Naoya Kitamine and Masayuki Ochi and Kazuhiko Kuroki},
  journal={arXiv:2308.12750},
  year={2023},
  url = {https://arxiv.org/abs/2308.12750}
}

@article{luo2023high,
  title={High-{T}$_c$ superconductivity in {L}a$_3${N}i$_2${O}$_7$ based on the bilayer two-orbital t-{J} model},
author={Luo, Zhihui
and Lv, Biao
and Wang, Meng
and W{\'u}, W{\'e}i
and Yao, Dao-Xin},
journal={npj Quantum Mater.},
year={2024},
month={Aug},
day={13},
volume={9},
number={1},
pages={61},
issn={2397-4648},
doi={10.1038/s41535-024-00668-w},
url={https://doi.org/10.1038/s41535-024-00668-w}
}

@article{zhang2023strong,
  title={Strong Pairing Originated from an Emergent $\mathbb{Z}_2$ Berry Phase in {L}a$_3${N}i$_2${O}$_7$}, 
  author = {Zhang, Jia-Xin and Zhang, Hao-Kai and You, Yi-Zhuang and Weng, Zheng-Yu},
  journal = {Phys. Rev. Lett.},
  volume = {133},
  issue = {12},
  pages = {126501},
  numpages = {7},
  year = {2024},
  month = {Sep},
  publisher = {American Physical Society},
  doi = {10.1103/PhysRevLett.133.126501},
  url = {https://link.aps.org/doi/10.1103/PhysRevLett.133.126501}
}

@article{pan2023rno,
author = {Zhiming Pan and Chen Lu and Fan Yang and Congjun Wu},
title = {Effect of Rare-Earth Element Substitution in Superconducting {R}$_3${N}i$_2${O}$_7$ under Pressure},
publisher = {Chin. Phys. Lett.},
year = {2024},
journal = {Chin. Phys. Lett.},
volume = {41},
number = {8},
eid = {087401},
pages = {087401},
url = {https://cpl.iphy.ac.cn/EN/abstract/article_116616.shtml},
doi = {10.1088/0256-307X/41/8/087401}
}

@article{sakakibara2023La4Ni3O10,
  title = {Theoretical analysis on the possibility of superconductivity in the trilayer {R}uddlesden-{P}opper nickelate {L}a$_4${N}i$_3${O}$_{10}$ under pressure and its experimental examination: Comparison with {L}a$_3${N}i$_2${O}$_7$},
  author = {Sakakibara, Hirofumi and Ochi, Masayuki and Nagata, Hibiki and Ueki, Yuta and Sakurai, Hiroya and Matsumoto, Ryo and Terashima, Kensei and Hirose, Keisuke and Ohta, Hiroto and Kato, Masaki and Takano, Yoshihiko and Kuroki, Kazuhiko},
  journal = {Phys. Rev. B},
  volume = {109},
  issue = {14},
  pages = {144511},
  numpages = {10},
  year = {2024},
  month = {Apr},
  publisher = {American Physical Society},
  doi = {10.1103/PhysRevB.109.144511},
  url = {https://link.aps.org/doi/10.1103/PhysRevB.109.144511}
}

@article{lange2023mixedtj,
  title={Pairing dome from an emergent Feshbach resonance in a strongly repulsive bilayer model}, 
  author = {Lange, Hannah and Homeier, Lukas and Demler, Eugene and Schollw\"ock, Ulrich and Bohrdt, Annabelle and Grusdt, Fabian},
  journal = {Phys. Rev. B},
  volume = {110},
  issue = {8},
  pages = {L081113},
  numpages = {7},
  year = {2024},
  month = {Aug},
  publisher = {American Physical Society},
  doi = {10.1103/PhysRevB.110.L081113},
  url = {https://link.aps.org/doi/10.1103/PhysRevB.110.L081113}
}

@article{geisler2023structural,
  title={Structural transitions, octahedral rotations, and electronic properties of ${A}_3${N}i$_2${O}$_7$ rare-earth nickelates under high pressure},
  author={Geisler, Benjamin and Hamlin, James J and Stewart, Gregory R and Hennig, Richard G and Hirschfeld, PJ},
  journal={npj Quantum Mater.},
  volume={9},
  number={1},
  pages={38},
  year={2024},
  publisher={Nature Publishing Group UK London},
  url={https://www.nature.com/articles/s41535-024-00648-0}
}

@article{yang2023strong,
  title = {Strong pairing from a small {F}ermi surface beyond weak coupling: Application to {L}a$_3${N}i$_2${O}$_7$},
  author = {Yang, Hui and Oh, Hanbit and Zhang, Ya-Hui},
  journal = {Phys. Rev. B},
  volume = {110},
  issue = {10},
  pages = {104517},
  numpages = {21},
  year = {2024},
  month = {Sep},
  publisher = {American Physical Society},
  doi = {10.1103/PhysRevB.110.104517},
  url = {https://link.aps.org/doi/10.1103/PhysRevB.110.104517}
}

@article{rhodes2023structural,
  title = {Structural routes to stabilize superconducting {L}a$_3${N}i$_2${O}$_7$ at ambient pressure},
  author = {Rhodes, Luke C. and Wahl, Peter},
  journal = {Phys. Rev. Mater.},
  volume = {8},
  issue = {4},
  pages = {044801},
  numpages = {9},
  year = {2024},
  month = {Apr},
  publisher = {American Physical Society},
  doi = {10.1103/PhysRevMaterials.8.044801},
  url = {https://link.aps.org/doi/10.1103/PhysRevMaterials.8.044801}
}

@article{lange2023feshbach,
  title = {Feshbach resonance in a strongly repulsive ladder of mixed dimensionality: A possible scenario for bilayer nickelate superconductors},
  author = {Lange, Hannah and Homeier, Lukas and Demler, Eugene and Schollw\"ock, Ulrich and Grusdt, Fabian and Bohrdt, Annabelle},
  journal = {Phys. Rev. B},
  volume = {109},
  issue = {4},
  pages = {045127},
  numpages = {16},
  year = {2024},
  month = {Jan},
  publisher = {American Physical Society},
  doi = {10.1103/PhysRevB.109.045127},
  url = {https://link.aps.org/doi/10.1103/PhysRevB.109.045127}
}

@article{labollita2023ele,
  title={Electronic structure and magnetic properties of {L}a$_{3}${N}i$_{2}${O}$_{7}$ under pressure: active role of the {N}i-$d_{x^2-y^2}$ orbitals}, 
  author={Harrison LaBollita and Victor Pardo and Michael R. Norman and Antia S. Botana},
  journal={arXiv:2309.17279},
  year={2023},
  url = {https://arxiv.org/abs/2309.17279}
}

@article{Li2024ele,
doi = {10.1088/0256-307X/41/8/087402},
url = {https://dx.doi.org/10.1088/0256-307X/41/8/087402},
year = {2024},
month = {jul},
publisher = {Chinese Physical Society and IOP Publishing Ltd},
volume = {41},
number = {8},
pages = {087402},
author = {Yidian Li and Xian Du and Yantao Cao and Cuiying Pei and Mingxin Zhang and Wenxuan Zhao and Kaiyi Zhai and Runzhe Xu and Zhongkai Liu and Zhiwei Li and Jinkui Zhao and Gang Li and Yanpeng Qi and Hanjie Guo and Yulin Chen and Lexian Yang},
title = {Electronic Correlation and Pseudogap-Like Behavior of High-Temperature Superconductor {L}a$_3${N}i$_2${O}$_7$},
journal = {Chin. Phys. Lett.}
}

@article{feng2024unaltered,
  title = {Unaltered density wave transition and pressure-induced signature of superconductivity in {N}d-doped {L}a$_{3}${N}i$_{2}${O}$_{7}$},
  author = {Feng, Jun-Jie and Han, Tao and Song, Jiang-Peng and Long, Ming-Sheng and Hou, Xing-Yuan and Zhang, Chang-Jin and Mu, Qing-Ge and Shan, Lei},
  journal = {Phys. Rev. B},
  volume = {110},
  issue = {10},
  pages = {L100507},
  numpages = {7},
  year = {2024},
  month = {Sep},
  publisher = {American Physical Society},
  doi = {10.1103/PhysRevB.110.L100507},
  url = {https://link.aps.org/doi/10.1103/PhysRevB.110.L100507}
}

@article{meng2024density,
  author = {Meng, Yanghao and Yang, Yi and Sun, Hualei and Zhang, Sasa and Luo, Jianlin and Chen, Liucheng and Ma, Xiaoli and Wang, Meng and Hong, Fang and Wang, Xinbo and Yu, Xiaohui},
  title = {Density-wave-like gap evolution in {L}a$_3${N}i$_2${O}$_7$ under high pressure revealed by ultrafast optical spectroscopy},
  journal = {Nat. Commun.},
  volume = {15},
  number = {1},
  pages = {10408},
  year = {2024},
  doi = {10.1038/s41467-024-54518-1},
  url = {https://doi.org/10.1038/s41467-024-54518-1},
  date = {2024-11-29}
}

@article{wang2024bulk,
title={Bulk high-temperature superconductivity in the high-pressure tetragonal phase of bilayer {L}a$_2${P}r{N}i$_2${O}$_7$}, 
author={Ningning Wang and Gang Wang and Xiaoling Shen and Jun Hou and Jun Luo and Xiaoping Ma and Huaixin Yang and Lifen Shi and Jie Dou and Jie Feng and Jie Yang and Yunqing Shi and Zhian Ren and Hanming Ma and Pengtao Yang and Ziyi Liu and Yue Liu and Hua Zhang and Xiaoli Dong and Yuxin Wang and Kun Jiang and Jiangping Hu and Stuart Calder and Jiaqiang Yan and Jianping Sun and Bosen Wang and Rui Zhou and Yoshiya Uwatoko and Jinguang Cheng},
journal={Nature},
year={2024},
month={Oct},
day={01},
volume={634},
number={8034},
pages={579-584},
issn={1476-4687},
doi={10.1038/s41586-024-07996-8},
url={https://doi.org/10.1038/s41586-024-07996-8}
}

@article{LI2024distinct,
title = {Distinct ultrafast dynamics of bilayer and trilayer nickelate superconductors regarding the density-wave-like transitions},
journal = {Sci. Bull.},
volume = {70},
pages={180},
year = {2024},
issn = {2095-9273},
url = {https://www.sciencedirect.com/science/article/pii/S2095927324007503},
author = {Yidian Li and Yantao Cao and Liangyang Liu and Pai Peng and Hao Lin and Cuiying Pei and Mingxin Zhang and Heng Wu and Xian Du and Wenxuan Zhao and Kaiyi Zhai and Xuefeng Zhang and Jinkui Zhao and Miaoling Lin and Pingheng Tan and Yanpeng Qi and Gang Li and Hanjie Guo and Luyi Yang and Lexian Yang}
}

@article{yang2024decom,
  title = {Decomposition of multilayer superconductivity with interlayer pairing},
  author = {Yang, Yi-Feng},
  journal = {Phys. Rev. B},
  volume = {110},
  issue = {10},
  pages = {104507},
  numpages = {6},
  year = {2024},
  month = {Sep},
  publisher = {American Physical Society},
  doi = {10.1103/PhysRevB.110.104507},
  url = {https://link.aps.org/doi/10.1103/PhysRevB.110.104507}
}

@article{ryee2024quenched,
  title = {Quenched Pair Breaking by Interlayer Correlations as a Key to Superconductivity in {L}a$_{3}${N}i$_{2}${O}$_{7}$},
  author = {Ryee, Siheon and Witt, Niklas and Wehling, Tim O.},
  journal = {Phys. Rev. Lett.},
  volume = {133},
  issue = {9},
  pages = {096002},
  numpages = {7},
  year = {2024},
  month = {Aug},
  publisher = {American Physical Society},
  doi = {10.1103/PhysRevLett.133.096002},
  url = {https://link.aps.org/doi/10.1103/PhysRevLett.133.096002}
}

@article{Lu2024interplay,
  title = {Interplay of two ${E}_{g}$ orbitals in superconducting {L}a$_{3}${N}i$_{2}${O}$_{7}$ under pressure},
  author = {Lu, Chen and Pan, Zhiming and Yang, Fan and Wu, Congjun},
  journal = {Phys. Rev. B},
  volume = {110},
  issue = {9},
  pages = {094509},
  numpages = {16},
  year = {2024},
  month = {Sep},
  publisher = {American Physical Society},
  doi = {10.1103/PhysRevB.110.094509},
  url = {https://link.aps.org/doi/10.1103/PhysRevB.110.094509}
}

@article{li2024distinguishing,
      title={Distinguishing Electronic Band Structure of Single-layer and Bilayer {R}uddlesden-{P}opper Nickelates Probed by in-situ High Pressure {X}-ray Absorption Near-edge Spectroscopy}, 
      author={Mingtao Li and Yiming Wang and Cuiying Pei and Mingxin Zhang and Nana Li and Jiayi Guan and Monica Amboage and N-Diaye Adama and Qingyu Kong and Yanpeng Qi and Wenge Yang},
      year={2024},
      journal ={arXiv:2410.04230},
      url={https://arxiv.org/abs/2410.04230}, 
}

@article{zhou2024revealing,
      title={Revealing nanoscale structural phase separation in {L}a$_{3}${N}i$_{2}${O}$_{7-\delta}$ single crystal via scanning near-field optical microscopy}, 
      author={Xiaoxiang Zhou and Weihong He and Zijian Zhou and Kaipeng Ni and Mengwu Huo and Deyuan Hu and Yinghao Zhu and Enkang Zhang and Zhicheng Jiang and Shuaikang Zhang and Shiwu Su and Juan Jiang and Yajun Yan and Yilin Wang and Dawei Shen and Xue Liu and Jun Zhao and Meng Wang and Mengkun Liu and Zengyi Du and Donglai Feng},
      year={2024},
      journal ={arXiv:2410.06602},
      url={https://arxiv.org/abs/2410.06602}, 
}

@article{fan2024tunn,
  title = {Tunneling spectra with gaplike features observed in nickelate {L}a$_{3}${N}i$_{2}${O}$_{7}$ at ambient pressure},
  author = {Fan, Shengtai and Luo, Zhihui and Huo, Mengwu and Wang, Zhaohui and Li, Han and Yang, Huan and Wang, Meng and Yao, Dao-Xin and Wen, Hai-Hu},
  journal = {Phys. Rev. B},
  volume = {110},
  issue = {13},
  pages = {134520},
  numpages = {9},
  year = {2024},
  month = {Oct},
  publisher = {American Physical Society},
  doi = {10.1103/PhysRevB.110.134520},
  url = {https://link.aps.org/doi/10.1103/PhysRevB.110.134520}
}

@article{kaneko2023pair,
  title = {Pair correlations in the two-orbital {H}ubbard ladder: Implications for superconductivity in the bilayer nickelate {L}a$_3${N}i$_2${O}$_7$},
  author = {Kaneko, Tatsuya and Sakakibara, Hirofumi and Ochi, Masayuki and Kuroki, Kazuhiko},
  journal = {Phys. Rev. B},
  volume = {109},
  issue = {4},
  pages = {045154},
  numpages = {5},
  year = {2024},
  month = {Jan},
  publisher = {American Physical Society},
  doi = {10.1103/PhysRevB.109.045154},
  url = {https://link.aps.org/doi/10.1103/PhysRevB.109.045154}
}

@article{Talantsev2024analysis,
  author="E.F. Talantsev and V.V. Chistyakov",
  title="Debye temperature, electron-phonon coupling constant, and three-dome shape of crystalline strain as a function of pressure in highly compressed {L}a$_3${N}i$_2${O}$_{7-\delta}$",
  publisher="Letters on Materials",
  year="2024",
  journal="Letters on Materials",
  volume="14",
  number="3",
  pages="262-268",
  url="https://lettersonmaterials.com/en/Readers/Article.aspx?aid=44771",
  doi="10.48612/letters/2024-3-262-268"}

@article{zhang2023la3ni2o6,
  title = {Electronic structure, magnetic correlations, and superconducting pairing in the reduced {R}uddlesden-{P}opper bilayer {L}a$_3${N}i$_2${O}$_6$ under pressure: Different role of $d_{3z^2-r^2}$ orbital compared with {L}a$_3${N}i$_2${O}$_7$},
  author = {Zhang, Yang and Lin, Ling-Fang and Moreo, Adriana and Maier, Thomas A. and Dagotto, Elbio},
  journal = {Phys. Rev. B},
  volume = {109},
  issue = {4},
  pages = {045151},
  numpages = {10},
  year = {2024},
  month = {Jan},
  publisher = {American Physical Society},
  doi = {10.1103/PhysRevB.109.045151},
  url = {https://link.aps.org/doi/10.1103/PhysRevB.109.045151}
}

@article{grusdt2023lno,
  title={Superconductivity in the pressurized nickelate {L}a$_3${N}i$_2${O}$_7$ in the vicinity of a {BEC-BCS} crossover}, 
  author={Henning Schlömer and Ulrich Schollwöck and Fabian Grusdt and Annabelle Bohrdt},
  journal = {Commun. Phys.},
  volume = {7},
  number = {1},
  pages = {366},
  year = {2024},
  month = {nov},
  doi = {10.1038/s42005-024-01854-9},
  url = {https://doi.org/10.1038/s42005-024-01854-9}
}

@article{chen2023iPEPS,
  title = {Orbital-selective superconductivity in the pressurized bilayer nickelate {L}a$_3${N}i$_2${O}$_7$: An infinite projected entangled-pair state study},
  author = {Chen, Jialin and Yang, Fan and Li, Wei},
  journal = {Phys. Rev. B},
  volume = {110},
  issue = {4},
  pages = {L041111},
  numpages = {7},
  year = {2024},
  month = {Jul},
  publisher = {American Physical Society},
  doi = {10.1103/PhysRevB.110.L041111},
  url = {https://link.aps.org/doi/10.1103/PhysRevB.110.L041111}
}

@article{liu2023dxy,
  author = {Xia, Chengliang and Liu, Hongquan and Zhou, Shengjie and Chen, Hanghui},
  title = {Sensitive dependence of pairing symmetry on {N}i-$e_g$ crystal field splitting in the nickelate superconductor {L}a$_3${N}i$_2${O}$_7$},
  journal = {Nat. Commun.},
  volume = {16},
  number = {1},
  pages = {1054},
  year = {2025},
  doi = {10.1038/s41467-025-56206-0},
  url = {https://doi.org/10.1038/s41467-025-56206-0}
}

@article{Ouyang2024absence,
author={Ouyang, Zhenfeng
and Gao, Miao
and Lu, Zhong-Yi},
title={Absence of electron-phonon coupling superconductivity in the bilayer phase of {L}a$_3${N}i$_2${O}$_7$ under pressure},
journal={npj Quantum Mater.},
year={2024},
month={Oct},
day={15},
volume={9},
number={1},
pages={80},
issn={2397-4648},
doi={10.1038/s41535-024-00689-5},
url={https://doi.org/10.1038/s41535-024-00689-5}
}

@article{qu2023roles,
  title={Roles of {H}und's Rule and Hybridization in the Two-orbital Model for High-${T}_c$ Superconductivity in the Bilayer Nickelate}, 
  author={Xing-Zhou Qu and Dai-Wei Qu and Wei Li and Gang Su},
  journal={arXiv:2311.12769},
  url = {https://arxiv.org/abs/2311.12769},
  year={2023}
}

@article{zheng2023twoorbital,
  title = {${s}_{\pm}$-wave superconductivity in the bilayer two-orbital {H}ubbard model},
  author = {Zheng, Yao-Yuan and W\'u, W\'ei},
  journal = {Phys. Rev. B},
  volume = {111},
  issue = {3},
  pages = {035108},
  numpages = {7},
  year = {2025},
  month = {Jan},
  publisher = {American Physical Society},
  doi = {10.1103/PhysRevB.111.035108},
  url = {https://link.aps.org/doi/10.1103/PhysRevB.111.035108}
}

@article{kakoi2023pair,
  title = {Pair correlations of the hybridized orbitals in a ladder model for the bilayer nickelate {L}a$_3${N}i$_2${O}$_7$},
  author = {Kakoi, Masataka and Kaneko, Tatsuya and Sakakibara, Hirofumi and Ochi, Masayuki and Kuroki, Kazuhiko},
  journal = {Phys. Rev. B},
  volume = {109},
  issue = {20},
  pages = {L201124},
  numpages = {6},
  year = {2024},
  month = {May},
  publisher = {American Physical Society},
  doi = {10.1103/PhysRevB.109.L201124},
  url = {https://link.aps.org/doi/10.1103/PhysRevB.109.L201124}
}

@article{heier2023competing,
  title = {Competing ${d}_{xy}$ and ${s}_{\pm}$ pairing symmetries in superconducting {L}a$_3${N}i$_2${O}$_7$: $\mathrm{LDA}+\mathrm{FLEX}$ calculations},
  author = {Heier, Griffin and Park, Kyungwha and Savrasov, Sergey Y.},
  journal = {Phys. Rev. B},
  volume = {109},
  issue = {10},
  pages = {104508},
  numpages = {9},
  year = {2024},
  month = {Mar},
  publisher = {American Physical Society},
  doi = {10.1103/PhysRevB.109.104508},
  url = {https://link.aps.org/doi/10.1103/PhysRevB.109.104508}
}

@article{fan2023sc,
  title={Superconductivity in nickelate and cuprate superconductors with strong bilayer coupling},
  author = {Fan, Zhen and Zhang, Jian-Feng and Zhan, Bo and Lv, Dingshun and Jiang, Xing-Yu and Normand, Bruce and Xiang, Tao},
  journal = {Phys. Rev. B},
  volume = {110},
  issue = {2},
  pages = {024514},
  numpages = {10},
  year = {2024},
  month = {Jul},
  publisher = {American Physical Society},
  doi = {10.1103/PhysRevB.110.024514},
  url = {https://link.aps.org/doi/10.1103/PhysRevB.110.024514}
}

@article{geisler2024optical,
  title = {Optical properties and electronic correlations in {L}a$_3${N}i$_2${O}$_{7-\delta}$ bilayer nickelates under high pressure},
  author = {Geisler, Benjamin and Fanfarillo, Laura and Hamlin, James J. and Stewart, Gregory R. and Hennig, Richard G. and Hirschfeld, P. J.},
  journal = {npj Quantum Mater.},
  volume = {9},
  number = {1},
  pages = {89},
  year = {2024},
  month = {nov},
  doi = {10.1038/s41535-024-00690-y},
  url = {https://doi.org/10.1038/s41535-024-00690-y}
}

@article{oh2024high,
  title={High-temperature superconductivity from kinetic energy},
  author={Oh, Hanbit and Yang, Hui and Zhang, Ya-Hui},
  journal={arXiv:2411.07292},
  year={2024},
  url={https://arxiv.org/abs/2411.07292}
}

@article{wu2024deconfined,
  title={Deconfined {F}ermi liquid to {F}ermi liquid transition and superconducting instability},
  author = {Wu, Xiaofan and Yang, Hui and Zhang, Ya-Hui},
  journal = {Phys. Rev. B},
  volume = {110},
  issue = {12},
  pages = {125122},
  numpages = {18},
  year = {2024},
  month = {Sep},
  publisher = {American Physical Society},
  doi = {10.1103/PhysRevB.110.125122},
  url = {https://link.aps.org/doi/10.1103/PhysRevB.110.125122}
}

@article{nomura2022sc,
  title={Superconductivity in infinite-layer nickelates},
  author={Nomura, Yusuke and Arita, Ryotaro},
  journal={Rep. Prog. Phys.},
  volume={85},
  number={5},
  pages={052501},
  year={2022},
  publisher={IOP Publishing},
  url={https://iopscience.iop.org/article/10.1088/1361-6633/ac5a60/meta}
}

@article{xie2024neutron,
title = {Strong interlayer magnetic exchange coupling in {L}a$_3${N}i$_2${O}$_{7-\delta}$ revealed by inelastic neutron scattering},
journal = {Sci. Bull.},
volume = {69},
number = {20},
pages = {3221-3227},
year = {2024},
issn = {2095-9273},
doi = {https://doi.org/10.1016/j.scib.2024.07.030},
url = {https://www.sciencedirect.com/science/article/pii/S2095927324005164},
author = {Tao Xie and Mengwu Huo and Xiaosheng Ni and Feiran Shen and Xing Huang and Hualei Sun and Helen C. Walker and Devashibhai Adroja and Dehong Yu and Bing Shen and Lunhua He and Kun Cao and Meng Wang}
}

@article{gupta2024anisotropic,
      title={Anisotropic Spin Stripe Domains in Bilayer {L}a$_3${N}i$_2${O}$_7$}, 
      author={N. K Gupta and R. Gong and Y. Wu and M. Kang and C. T. Parzyck and B. Z. Gregory and N. Costa and R. Sutarto and S. Sarker and A. Singer and D. G. Schlom and K. M. Shen and D. G. Hawthorn},
      year={2024},
      journal={arXiv:2409.03210},
      url={https://arxiv.org/abs/2409.03210}, 
}

@article{ren2025resolving,
  title={Resolving the electronic ground state of {L}a$_3${N}i$_2${O}$_{7-\delta}$ films},
  author={Ren, Xiaolin and Sutarto, Ronny and Wu, Xianxin and Zhang, Jianfeng and Huang, Hai and Xiang, Tao and Hu, Jiangping and Comin, Riccardo and Zhou, Xingjiang and Zhu, Zhihai},
  journal={Communications Physics},
  volume={8},
  number={1},
  pages={52},
  year={2025},
  publisher={Nature Publishing Group UK London},
  url={https://www.nature.com/articles/s42005-025-01971-z}
}

@article{li2019sc,
  title={Superconductivity in an infinite-layer nickelate},
  author={Li, Danfeng and Lee, Kyuho and Wang, Bai Yang and Osada, Motoki and Crossley, Samuel and Lee, Hye Ryoung and Cui, Yi and Hikita, Yasuyuki and Hwang, Harold Y},
  journal={Nature},
  volume={572},
  number={7771},
  pages={624--627},
  year={2019},
  publisher={Nature Publishing Group UK London}, 
  url={https://www.nature.com/articles/s41586-019-1496-5}
}

@article{botana2020similar,
  title = {Similarities and Differences between {L}a{N}i{O}$_2$ and {C}a{C}u{O}$_2$ and Implications for Superconductivity},
  author = {Botana, A. S. and Norman, M. R.},
  journal = {Phys. Rev. X},
  volume = {10},
  issue = {1},
  pages = {011024},
  numpages = {6},
  year = {2020},
  month = {Feb},
  publisher = {American Physical Society},
  doi = {10.1103/PhysRevX.10.011024},
  url = {https://link.aps.org/doi/10.1103/PhysRevX.10.011024}
}

@article{white1993dmrg,
  title = {Density-matrix algorithms for quantum renormalization groups},
  author = {White, Steven R.},
  journal = {Phys. Rev. B},
  volume = {48},
  issue = {14},
  pages = {10345--10356},
  numpages = {0},
  year = {1993},
  month = {Oct},
  publisher = {American Physical Society},
  doi = {10.1103/PhysRevB.48.10345},
  url = {https://link.aps.org/doi/10.1103/PhysRevB.48.10345}
}

@article{weng1999dmrg,
  title = {Two-leg {t}-{J} ladder: A mean-field description},
  author = {Lee, Y. L. and Lee, Y. W. and Mou, C.-Y. and Weng, Z. Y.},
  journal = {Phys. Rev. B},
  volume = {60},
  issue = {19},
  pages = {13418--13428},
  numpages = {0},
  year = {1999},
  month = {Nov},
  publisher = {American Physical Society},
  doi = {10.1103/PhysRevB.60.13418},
  url = {https://link.aps.org/doi/10.1103/PhysRevB.60.13418}
}

@software{jutho2024,
author = {Jutho and Lukas Devos and Markus Hauru and maartenvd and ho-oto and Gertian and Lander Burgelman and tangwei94 and Julia TagBot and Stefanos Carlstr{\"o}m and Victor Vanthilt and Xiaoyu and qmortier},
doi = {10.5281/zenodo.13950435},
month = oct,
publisher = {Zenodo},
title = {Jutho/TensorKit.jl: v0.12.7},
url = {https://doi.org/10.5281/zenodo.13950435},
version = {v0.12.7},
year = 2024}

@software{li2024mps,
author = {Li, Qiaoyi},
doi = {10.5281/zenodo.14615184},
month = jan,
title = {{FiniteMPS.jl}},
url = {https://github.com/Qiaoyi-Li/FiniteMPS.jl},
version = {1.6.1},
year = {2025}
}

@article{weichselbaum2012,
title = {Non-abelian symmetries in tensor networks: A quantum symmetry space approach},
journal = {Ann. Phys.},
volume = {327},
number = {12},
pages = {2972-3047},
year = {2012},
issn = {0003-4916},
doi = {https://doi.org/10.1016/j.aop.2012.07.009},
url = {https://www.sciencedirect.com/science/article/pii/S0003491612001121},
author = {Andreas Weichselbaum},
}

@article{weichselbaum2020,
  title = {X-symbols for non-Abelian symmetries in tensor networks},
  author = {Weichselbaum, Andreas},
  journal = {Phys. Rev. Res.},
  volume = {2},
  issue = {2},
  pages = {023385},
  numpages = {16},
  year = {2020},
  month = {Jun},
  publisher = {American Physical Society},
  doi = {10.1103/PhysRevResearch.2.023385},
  url = {https://link.aps.org/doi/10.1103/PhysRevResearch.2.023385}
}

@article{luther1974,
  title = {Backward Scattering in the One-Dimensional Electron Gas},
  author = {Luther, A. and Emery, V. J.},
  journal = {Phys. Rev. Lett.},
  volume = {33},
  issue = {10},
  pages = {589--592},
  numpages = {0},
  year = {1974},
  month = {Sep},
  publisher = {American Physical Society},
  doi = {10.1103/PhysRevLett.33.589},
  url = {https://link.aps.org/doi/10.1103/PhysRevLett.33.589}
}

@article{zhang1988effective,
  title={Effective Hamiltonian for the superconducting {C}u oxides},
  author={Zhang, FC and Rice, TM},
  journal={Phys. Rev. B},
  volume={37},
  number={7},
  pages={3759},
  year={1988},
  publisher={APS}, 
  url={https://journals.aps.org/prb/abstract/10.1103/PhysRevB.37.3759}
}

@article{tsuei2000pairing,
  title={Pairing symmetry in cuprate superconductors},
  author={Tsuei, CC and Kirtley, JR},
  journal={Rev. Mod. Phys.},
  volume={72},
  number={4},
  pages={969},
  year={2000},
  publisher={APS}, 
  url={https://journals.aps.org/rmp/abstract/10.1103/RevModPhys.72.969}
}

@article{lee2006doping,
  title={Doping a {M}ott insulator: Physics of high-temperature superconductivity},
  author={Lee, Patrick A and Nagaosa, Naoto and Wen, Xiao-Gang},
  journal={Rev. Mod. Phys.},
  volume={78},
  number={1},
  pages={17--85},
  year={2006},
  publisher={APS}, 
  url={https://journals.aps.org/rmp/abstract/10.1103/RevModPhys.78.17}
}

@article{taillefer2010scattering,
  title={Scattering and pairing in cuprate superconductors},
  author={Taillefer, Louis},
  journal={Annu. Rev. Condens. Matter Phys.},
  volume={1},
  number={1},
  pages={51--70},
  year={2010},
  publisher={Annual Reviews},
  url={https://www.annualreviews.org/content/journals/10.1146/annurev-conmatphys-070909-104117}
}

@article{proust2019remarkable,
  title={The remarkable underlying ground states of cuprate superconductors},
  author={Proust, Cyril and Taillefer, Louis},
  journal={Ann. Rev. Condens. Matter Phys.},
  volume={10},
  number={1},
  pages={409--429},
  year={2019},
  publisher={Annual Reviews},
  url={https://www.annualreviews.org/content/journals/10.1146/annurev-conmatphys-031218-013210}
}

@article{castellani1978,
  title = {Magnetic structure of {V}$_{2}${O}$_{3}$ in the insulating phase},
  author = {Castellani, C. and Natoli, C. R. and Ranninger, J.},
  journal = {Phys. Rev. B},
  volume = {18},
  issue = {9},
  pages = {4945--4966},
  numpages = {0},
  year = {1978},
  month = {Nov},
  publisher = {American Physical Society},
  doi = {10.1103/PhysRevB.18.4945},
  url = {https://link.aps.org/doi/10.1103/PhysRevB.18.4945}
}

@book{tinkham2004introduction,
  title={Introduction to superconductivity},
  author={Tinkham, Michael},
  year={2004},
  publisher={Courier Corporation}
}

@article{wang2024experimental,
  title={Experimental progress in superconducting nickelates},
  author={Wang, Bai Yang and Lee, Kyuho and Goodge, Berit H},
  journal={Ann. Rev. Condens. Matter Phys.},
  volume={15},
  year={2024},
  publisher={Annual Reviews},
  url={https://www.annualreviews.org/content/journals/10.1146/annurev-conmatphys-032922-093307}
}

@article{ko2024signatures,
  title={Signatures of ambient pressure superconductivity in thin film {L}a$_3${N}i$_2${O}$_7$},
  author={Ko, Eun Kyo and Yu, Yijun and Liu, Yidi and Bhatt, Lopa and Li, Jiarui and Thampy, Vivek and Kuo, Cheng-Tai and Wang, Bai Yang and Lee, Yonghun and Lee, Kyuho and Lee, Jun-Sik and Goodge, Berit H and Muller, David A and Hwang, Harold Y},
  journal={Nature},
  volume = {638},
  pages={935--940},
  year={2025},
  publisher={Nature Publishing Group UK London},
  url={https://www.nature.com/articles/s41586-024-08525-3}
}

@article{zhou2025ambient,
  title={Ambient-pressure superconductivity onset above 40 {K} in ({L}a,{P}r)$_3${N}i$_2${O}$_7$ films},
  author={Zhou, Guangdi and Lv, Wei and Wang, Heng and Nie, Zihao and Chen, Yaqi and Li, Yueying and Huang, Haoliang and Chen, Weiqiang and Sun, Yu-Jie and Xue, Qi-Kun and Chen, Zhuoyu},
  journal={Nature},
  volume={640},
  pages={641-646},
  year={2025},
  publisher={Nature Publishing Group UK London},
  url={https://www.nature.com/articles/s41586-025-08755-z}
}

@article{yue2025correlated,
  title={Correlated electronic structures and unconventional superconductivity in bilayer nickelate heterostructures},
  author={Yue, Changming and Miao, Jian-Jian and Huang, Haoliang and Hua, Yichen and Li, Peng and Li, Yueying and Zhou, Guangdi and Lv, Wei and Yang, Qishuo and Sun, Hongyi and Sun, Yu-Jie and Lin, Junhao and Xue, Qi-Kun and Chen, Zhuoyu and Chen, Wei-Qiang},
  journal={arXiv:2501.06875},
  year={2025},
  url={https://arxiv.org/abs/2501.06875}
}

@article{wang2025mottness,
  title={The {M}ottness and the {A}nderson localization in bilayer nickelate {L}a$_3${N}i$_2${O}$_{7-\delta}$},
  author={Wang, Yuxin and Chen, Ziyan and Zhang, Yi and Jiang, Kun and Hu, Jiangping},
  journal={arXiv:2501.08536},
  year={2025},
  url={https://arxiv.org/abs/2501.08536}
}

@article{shao2025band,
  title={Band Structure and Pairing Nature of {L}a$_3${N}i$_2${O}$_7$ Thin Film at Ambient Pressure},
  author={Shao, Zhi-Yan and Liu, Yu-Bo and Liu, Min and Yang, Fan},
  journal={arXiv:2501.10409},
  year={2025},
  url={https://arxiv.org/abs/2501.10409}
}

@article{bhatt2025resolving,
  title={Resolving Structural Origins for Superconductivity in Strain-Engineered {L}a$_3${N}i$_2${O}$_7$ Thin Films},
  author={Bhatt, Lopa and Jiang, Abigail Y and Ko, Eun Kyo and Schnitzer, Noah and Pan, Grace A and Segedin, Dan Ferenc and Liu, Yidi and Yu, Yijun and Zhao, Yi-Feng and Morales, Edgar Abarca and Brooks, Charles M and Botana, Antia S. and Hwang, Harold Y and Mundy, Julia A and Muller, David A and Goodge, Berit H},
  journal={arXiv:2501.08204},
  year={2025},
  url={https://arxiv.org/abs/2501.08204}
}

@article{liu2025superconductivity,
  title={Superconductivity and normal-state transport in compressively strained {L}a$_2${P}r{N}i$_2${O}$_7$ thin films},
  author={Liu, Yidi and Ko, Eun Kyo and Tarn, Yaoju and Bhatt, Lopa and Goodge, Berit H and Muller, David A and Raghu, Srinivas and Yu, Yijun and Hwang, Harold Y},
  journal={arXiv:2501.08022},
  year={2025},
  url={https://arxiv.org/abs/2501.08022}
}

@article{hu2025electronic,
  title={Electronic structures and multi-orbital models of {L}a$_3${N}i$_2${O}$_7$thin films},
  author={Hu, Xunwu and Qiu, Wenyuan and Chen, Cun-Qun and Luo, Zhihui and Yao, Dao-Xin},
  journal={arXiv:2503.17223},
  year={2025},
  url={https://arxiv.org/abs/2503.17223}
}

@article{geisler2024fermi,
  title={{F}ermi surface reconstruction in strained {L}a$_3${N}i$_2${O}$_7$ on {L}a{A}l{O}$_3$(001) and {S}r{T}i{O}$_3$(001)},
  author={Geisler, Benjamin and Hamlin, James J and Stewart, Gregory R and Hennig, Richard G and Hirschfeld, PJ},
  journal={arXiv:2411.14600},
  year={2024},
  url={https://arxiv.org/abs/2411.14600}
}

\end{document}